\documentclass[preprint,9pt]{elsarticle}

\usepackage{amssymb,url}
\usepackage{amsmath,mathtools}
\usepackage{booktabs}
\usepackage{subcaption}

\usepackage[normalem]{ulem}
\usepackage{comment}

\usepackage{multirow}
\usepackage{mleftright}
\usepackage{color,xcolor}
\usepackage{bm}
\usepackage{algorithm}
\usepackage[noend]{algpseudocode}
\usepackage{bbm}
\renewcommand*{\thesubfigure}{(\alph{subfigure})}

\newcommand{\E}{\mathbb{E}}
\newcommand{\R}{\mathbb{R}}
\newcommand{\C}{\mathbb{C}}
\newcommand{\adj}{\mathsf{H}}
\newcommand{\tr}{\mathsf{T}}
\DeclareMathOperator{\trace}{Tr}
\DeclareMathOperator{\rank}{rank}
\newcommand{\mat}{\mathbf}
\renewcommand{\vec}{\mathbf}
\newcommand{\xs}{\vec{x}_{\mathrm{s}}}
\newcommand{\xn}{\vec{x}_{\mathrm{n}}}
\newcommand{\hatxs}{\hat{\vec{x}}_{\mathrm{s}}}
\newcommand{\Ry}{\mat{R}_{\mathrm{y}}}
\newcommand{\Rs}{\mat{R}_{\mathrm{s}}}
\newcommand{\Rn}{\mat{R}_{\mathrm{n}}}
\newcommand{\stvec}{\vec{z}}
\newcommand{\unitvec}{\vec{u}}
\newcommand{\argmax}{\mathop{\rm arg~max}\limits}

\newcommand{\fs}{f_{\mathrm s}} % sampling frequency
\newcommand{\Tc}{T_{\mathrm{c}}}
\newcommand{\Tcin}{{T_{\mathrm{c}}^{\mathrm{wav}}}}
\newcommand{\Tcout}{{T_{\mathrm{c}}^{\mathrm{act}}}}
\newcommand{\Slocal}{S_{\mathrm{local}}} % sampling frequency
\newcommand{\thetaemb}{\theta_{\mathrm{emb}}}

\newcommand{\ditto}{---\char`\"---}
\renewcommand{\textcolor}[2]{#2}

\journal{Computer Speech \& Language}

\begin{document}

\begin{frontmatter}

%% Title, authors and addresses

%% use the tnoteref command within \title for footnotes;
%% use the tnotetext command for theassociated footnote;
%% use the fnref command within \author or \affiliation for footnotes;
%% use the fntext command for theassociated footnote;
%% use the corref command within \author for corresponding author footnotes;
%% use the cortext command for theassociated footnote;
%% use the ead command for the email address,
%% and the form \ead[url] for the home page:
%% \title{Title\tnoteref{label1}}
%% \tnotetext[label1]{}
%% \author{Name\corref{cor1}\fnref{label2}}
%% \ead{email address}
%% \ead[url]{home page}
%% \fntext[label2]{}
%% \cortext[cor1]{}
%% \affiliation{organization={},
%%             addressline={},
%%             city={},
%%             postcode={},
%%             state={},
%%             country={}}
%% \fntext[label3]{}

\title{Microphone Array geometry-independent Multi-Talker Distant ASR: NTT System for DASR Task of the CHiME-8 Challenge}

%% use optional labels to link authors explicitly to addresses:
%% \author[label1,label2]{}
%% \affiliation[label1]{organization={},
%%             addressline={},
%%             city={},
%%             postcode={},
%%             state={},
%%             country={}}
%%
%% \affiliation[label2]{organization={},
%%             addressline={},
%%             city={},
%%             postcode={},
%%             state={},
%%             country={}}

\author{Naoyuki Kamo$^*$, Naohiro Tawara$^*$, Atsushi Ando, Takatomo Kano, Hiroshi Sato, Rintaro Ikeshita, Takafumi Moriya, Shota Horiguchi, Kohei Matsuura, Atsunori Ogawa, Alexis Plaquet$^{**}$, Takanori Ashihara, Tsubasa Ochiai, Masato Mimura, Marc Delcroix, Tomohiro Nakatani, Taichi Asami, Shoko Araki} %% Author name

\cortext[cor1]{Equal contribution}
\cortext[cor2]{Alexis Plaquet is with IRIT, Université de Toulouse, CNRS, Toulouse INP, UT3, Toulouse, France, and a JSPS International Research Fellow. This work was done during an internship at NTT.}
%% Author affiliation
\affiliation{organization={NTT Corporation},%Department and Organization
            %addressline={}, 
            %city={},
            %postcode={}, 
            %state={},
            country={Japan}}

%% Abstract
\begin{abstract}
In this paper, we introduce a multi-talker distant automatic speech recognition (DASR) system we designed for the DASR task 1 of the CHiME-8 challenge. Our system performs speaker counting, diarization, and ASR. It handles a variety of recording conditions, from dinner parties to professional meetings and from two speakers to eight. We perform diarization first, followed by speech enhancement, and then ASR as the challenge baseline. However, we introduced several key refinements. First, we derived a powerful speaker diarization relying on end-to-end speaker diarization with vector clustering (EEND-VC), multi-channel speaker counting using enhanced embeddings from EEND-VC, and target-speaker voice activity detection (TS-VAD). For speech enhancement, we introduced a novel microphone selection rule to better select the most relevant microphones among those distributed microphones and investigated improvements to beamforming. Finally, for ASR, we developed several models exploiting Whisper and WavLM speech foundation models. In this paper, we present the original results we submitted to the challenge and updated results we obtained afterward. Our strongest system achieves a 63\% relative macro tcpWER improvement over the baseline and outperforms the challenge best results on the NOTSOFAR-1 meeting evaluation data among geometry-independent systems.
\end{abstract}

%%Graphical abstract
%\begin{graphicalabstract}
%\includegraphics{grabs}
%\end{graphicalabstract}

%%Research highlights
\begin{highlights}
\item A multi-talker ASR system achieving 63\% relative macro tcpWER improvement over the CHiME-8 DASR task baseline
\item A powerful diarization frontend combining EEND-VC, TS-VAD, and multi-channel speaker counting
\item Speech enhancement using improved microphone selection and SP-MWF beamformer
\item Four ASR backends exploiting speech foundation models (Whisper and WavLM)
\item  An extensive experimental study and ablation analyzing each component of our system
\end{highlights}

%% Keywords
\begin{keyword}
Robust ASR \sep multi-talker ASR \sep speaker diarization \sep CHiME-8 DASR
%% keywords here, in the form: keyword \sep keyword

%% PACS codes here, in the form: \PACS code \sep code

%% MSC codes here, in the form: \MSC code \sep code
%% or \MSC[2008] code \sep code (2000 is the default)

\end{keyword}

\end{frontmatter}

\section{Introduction}
Within the last decade, rapid progress in deep learning has led to dramatic improvements in the performance of automatic speech recognition (ASR). However, it remains challenging to recognize speech in multi-talker recordings, especially when using distant microphones \cite{umbach_proc_ieee21}. The problem of multi-talker distant ASR (DASR) consists of not only transcribing \emph{what} was spoken (i.e., ASR) but also correctly estimating \emph{who} uttered the utterances and \emph{when} (i.e., speaker diarization) \cite{cornell2023chime7,chime8-task1}. Moreover, it should be robust to overlapping speech, background noise, and reverberation, which are common distortions observed when recording with distant microphones in everyday living and working environments. This makes the multi-talker ASR problem extremely challenging, but solving it would have great implications, as it would enable broadening the range of ASR applications to automatic meeting minute generation, smart human-robot interactions, etc.

The CHiME challenge series has been a leading activity in triggering research on multi-talker distant ASR (DASR). It originated with the CHiME-1-2 challenges~\cite{chime1,chime2}, which involved DASR in living rooms. At that time, the problem consisted of recordings of simple commands or sentences read from a single speaker recorded by a distant microphone array in the presence of various noise sources that can appear in a living room, such as a vacuum cleaner, children's voices, etc. The CHiME-3-4 \cite{chime3} extended the problem to cover four different recording conditions, including bus, café, street, and pedestrian. The CHiME-5-6 \cite{chime5,chime6} increased the complexity by switching to multi-talker conversations in a dinner party setting recorded with distributed microphone arrays. This also added the requirement to perform speaker diarization. More recently, the DASR tasks of the CHiME-7-8 challenges (task 1) \cite{cornell2023chime7,chime8-task1} further increased the complexity by requiring the construction of a single array geometry-independent system that could perform multi-talker DASR for 1) diverse recording scenarios (dinner parties, interviews, meetings), 2) recordings with various distributed microphone array configurations, 3) recordings with the number of speakers varying from two to eight, 4) recording durations from a few minutes to more than one hour, and 5) limited amounts of in-domain training data. The CHiME-8 challenge also featured another DASR task called the NOTSOFAR-1 task (task 2)~\cite{vinnikov24_interspeech}, which focused on the DASR of meetings of four to eight speakers recorded with a single microphone array, with known array geometry, in relatively well-controlled conditions consisting of office meeting rooms. DASR and NOTSOFAR-1 ran as separate tasks, but the NOTSOFAR-1 data was also used as one of the datasets in the DASR task 1 evaluation.
In this paper, we describe the system we proposed for the DASR task of the CHiME-8 challenge~\cite{chime8-task1}. 

There are different ways to implement a multi-talker DASR system to perform both ASR and diarization while being robust to noise, reverberation, and overlapping speech. Here, we briefly review the two main approaches \cite{Raj2021}, \emph{separation-first} and \emph{diarization-first} approaches, which vary in terms of the processing order of the diarization, speech separation/enhancement, and ASR modules.

The continuous speech separation (CSS) pipeline \cite{chen_icassp20,yoshioka2019advances,Raj2021} is an example of a \emph{separation-first} approach, which performs speech separation on short time segments, then transcribes each segment, and finally performs diarization by clustering speaker embeddings computed on the separation outputs. The advantage of this approach is that ASR and diarization are greatly simplified if the speech signals can be well separated beforehand. For example, after separation, we can obtain speaker embedding less affected by overlapping speech, which can reduce the speaker confusion of diarization. In addition, diarization can exploit the ASR output in order to, e.g., refine speaker activity prediction \cite{von2024meeting}. Such a separation-first pipeline relies heavily on having good separation pre-processing, typically implemented with the combination of neural-network (NN) -based separation and a mask-based beamformer. The separation-first pipeline was utilized as a baseline for the NOTSOFAR-1 task 2 of the CHiME challenge. In that meeting recognition scenario, the recording conditions are relatively well-controlled, making NN-based separation very effective. It is, however, challenging to use such a separation-first pipeline for the domain- and microphone array geometry-independent DASR systems (which are required for the DASR task 1 of CHiME-8), where the recording conditions are much more diverse, making it difficult to train a robust NN-based separation frontend.

The NeMo baseline of the DASR task 1 of the CHiME-8 challenge is an example of a \emph{diarization-first} system \cite{chime8-task1}. The processing flow begins with diarization, followed by guided source separation (GSS) \cite{boeddecker18_chime} for speech enhancement (SE) and then ASR. The idea behind this approach is that it is easier to build a robust diarization system than separation because diarization requires estimating coarser information, i.e., the speech activity (discrete values) for diarization compared to waveforms for separation. Then, robust speech separation can be performed using GSS, which consists of an adaptive speech separation method such as a complex angular central Gaussian mixture model (cACGMM) \cite{ito2016complex} conditioned on the speaker activity prediction of the diarization pre-processing. Such a pipeline has been utilized by most systems proposed for the CHiME-6, 7, and 8 \cite{medennikov20_chime,chime7_ustc,chime7_stcon,kamo23_chime,stcon_chime8,ntt_chime8,niu24_chime} because it can be used with distributed microphone arrays and is relatively robust to the very challenging recording conditions for which NN-based separation remains challenging.

In this paper, we propose a modification of the diarization-first pipeline that can partially utilize the advantage of the separation-first pipeline. 
%modifying the \emph{diarization-first} pipeline, which we call segmentation-first.
Current state-of-the-art diarization systems, such as end-to-end diarization with vector clustering (EEND-VC) \cite{EEND-vector-clustering_Interspeech2021,EEND-vector-clustering_ICASSP2021,Bredin2021}, usually rely on a two-step processing, i.e., local segmentation and global clustering. The first step, local segmentation, consists of predicting the speech activity (i.e., \emph{when}) of each speaker in short audio segments of up to, e.g., 15--80 sec, assuming that the number of locally active speakers remains below a pre-set maximum number of local speakers, e.g., four. We then perform clustering of the speaker embeddings associated with each speaker activity in each segment to associate the local speaker activity with a recording-level speaker label (i.e., \emph{who}). This allows us to assemble the activity of each speaker together to form recording-level diarization results despite possible speaker ordering ambiguity between segments. It also enables handling a greater number of global speakers compared to the maximum number of local speakers.
%It can also handle a global number of speakers superior to the local maximum number of speakers. the as long as the number of locally active speakers 
%, i.e., more speakers than the maximum number of speakers in each local segment. 
In this paper, we exploit another advantage of the EEND-VC approach: name, since the segmentation and clustering can be performed independently, we propose performing GSS-based separation between these two steps, as GSS only requires local segmentation information. This approach, which we call \emph{segmentation-first}, can harness some of the benefits of the separation-first pipeline, i.e., extracting speaker embeddings on enhanced speech improves their quality, resulting in better clustering accuracy. 

\begin{figure}[t]
  \centering
\includegraphics[width=0.99\linewidth]{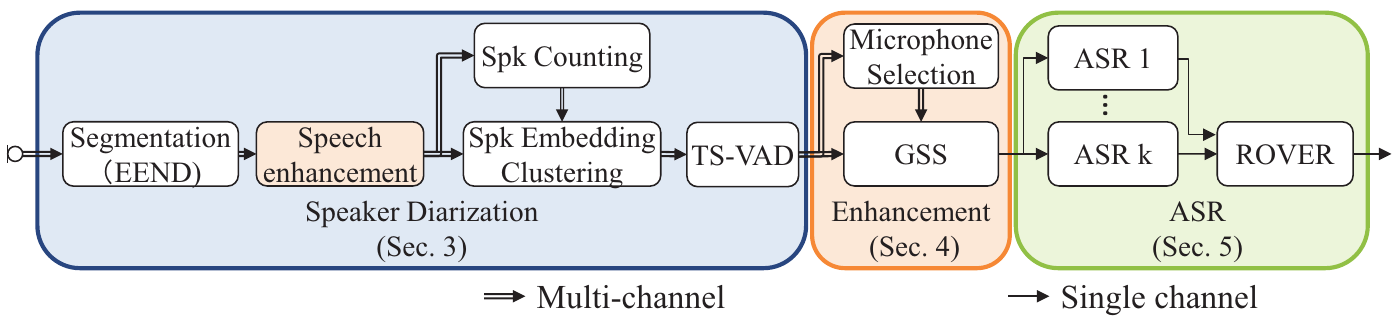}
  \caption{Overview of proposed multi-talker DASR system.}
  \label{fig:overview}
\end{figure}

Figure \ref{fig:overview} shows an overview of our proposed multi-talker DASR system. It consists of the modified EEND-VC frontend, which inserts GSS between segmentation and clustering. For clustering, we set the number of clusters to the number of speakers estimated using a novel multi-channel speaker counting approach. 
We then refine the diarization results using target speaker voice activity detection (TS-VAD) \cite{TS-VAD} implemented based on the memory-aware multi-speaker embedding with sequence-to-sequence architecture (NSD-MS2S) \cite{NSD-MS2S}, which achieved the top performance in the CHiME-7 challenge \cite{chime7_ustc}. For SE, we enhance the target speaker for each segment using GSS combined with weighted prediction error (WPE) \cite{yoshioka2012generalization,nakatani2010speech} and spatial-prediction multichannel Wiener filter (SP-MWF) beamformer \cite{benesty2008noncausal,cornelis2011performance}. In addition, we designed a novel microphone subset selection approach to reduce the influence of unreliable microphone signals. Finally, we perform ASR using four different backends based on Whisper, NeMo, and a WavLM-based RNN-T model. To obtain optimal results, we performed language model re-scoring and combined the output of the four systems using recognizer output voting error reduction (ROVER) -based system combination \cite{Fiscus1997ROVER}.

The proposed multi-talker DASR system addresses the requirements of CHiME-8 as follows:
\begin{enumerate}
\item \textbf{Robustness to recording conditions and speaking styles:} 
We exploit various speech foundation models to increase robustness and simplify system development, given the limited amount of in-domain data. Our diarization system exploits WavLM as feature extraction for segmentation and ECAPA-TDNN to obtain robust speaker embeddings for clustering and speaker counting. We also use several speech foundation models (WavLM, NeMo, and Whisper) to build various robust ASR models.
\item \textbf{Handling arbitrary and distributed microphone arrays:} 
We rely on a \emph{segmentation-first} pipeline, which allows us to use GSS that is robust to array geometry and distributed microphone configurations. For the diarization, we apply EEND-VC for each channel independently and combine the output using diarization output voting error reduction + overlap (DOVER-Lap) \cite{Raj2021Doverlap}. This approach allows us to exploit the multi-microphone recordings for diarization without building a multi-channel EEND system \cite{horiguchi2022multi}, which is particularly sensitive to training/test mismatch because of the difficulty of simulating realistic multi-channel speech mixtures. For SE, we designed a new microphone subset selection, which considers estimates of the reverberation condition at each microphone. Finally, we perform ASR on the single-channel output of the SE module.
\item \textbf{Handling various numbers of speakers and recording durations:} 
Our diarization system relies on EEND-VC, which can handle an arbitrary number of speakers. With EEND-VC, the result of the clustering of the speaker embeddings provides the speaker counts. However, the counting accuracy can worsen when speaker embeddings are unreliable or if there are not enough embeddings when the recordings are short. 
To improve speaker counting accuracy, we therefore propose decoupling the clustering for speaker counting from the clustering for assigning speaker labels. 
We improve the reliability of the estimated speaker counts by increasing the number of embeddings for clustering, computing them on shorter segments, and aggregating the results over the different microphones.
%The clustering for speaker counting is done on shorter segments (i.e. 15 sec) to provide more samples for clustering, and the results are aggregated over the different channels, providing reliable estimates. 
Then, we can perform speaker embedding clustering to assign speaker labels by fixing the number of clusters to the estimated speaker counts.
\item \textbf{Dealing with insufficient and unreliable transcriptions for training:}
One issue with the CHiME-8  DASR dataset is that the amount of reliable training data for ASR is limited. Using pre-trained foundation models can partially alleviate this issue, but the best performance is obtained when these models are fine-tuned on the target task. We propose two approaches to deal with data scarcity. First, we augment the training data with un-transcribed data from the VoxVeleb 1 and 2 datasets and apply contrastive data selection \cite{lu22_interspeech} to select data similar to the CHiME-8 task 1 datasets' conditions. Second, when retraining the Whisper model, we introduce a curriculum learning scheme, which filters out training samples with many recognition errors.
\end{enumerate}

This paper is an extention of our challenge system description paper \cite{ntt_chime8} and our conference papers describing the diarization part \cite{tawara_icassp24, tawara_icassp25}. Compared to our previous publications, we provide more exhaustive descriptions of all parts of our system, especially the diarization and SE frontends, and include extensive experimental analysis and ablation studies to clarify each module's contribution. Moreover, we provide new post-challenge results that further improved our performance, narrowing the gap with the top CHiME-8 system \cite{stcon_chime8} and surpassing it on the NOTSOFAR-1 data.

In the remainder of the paper, we provide an overall description of our proposed multi-talker DASR system in Section \ref{sec:overview}. 
We describe diarization and speaker counting in Section \ref{sec:diar}, SE in Section \ref{sec:se}, and the ASR backend in Section \ref{sec:asr}. 
We then discuss experimental results and analysis in Section \ref{sec:exp}, with extensive ablation for the different modules in \ref{sec:ablation}. We conclude the paper in Section \ref{sec:conclusion}.

As our aim is to provide an exhaustive description of our system, the overall paper is rather long. Readers interested in the details of each module can skip ahead to the relevant parts (diarization in Section \ref{sec:diar}, SE in Section \ref{sec:se} and ASR in Section \ref{sec:asr}), and those who are interested in the experiment analysis can read the overall description in Section \ref{sec:overview} and then proceed directly to the experiments in Section \ref{sec:exp}.

\section{Overview of proposed system}
\label{sec:overview}
The multi-talker DASR consists of transcribing a multi-talker conversation captured with distant microphones. Here, transcribing also includes assigning a speaker label and the boundaries for each utterance (start and end times).\footnote{Note that the system does not provide absolute speaker identities but should only associate a consistent speaker label for each speaker in a recording.}
We can model the signal received at a distant microphone:
\begin{align}
    \mathbf{y}_m = \sum_{s=1}^S \mathbf{x}_{m,s} + \mathbf{n}_m,
\end{align}
where $\mathbf{y}_m$, $\mathbf{x}_{m,s}$ and $\mathbf{n}_m  \in \R^T$ are the noisy observation, speech, and noise signals, respectively. $m$ and $s$ are the microphone and speaker indexes, respectively. $S$ represents the number of speakers and $T$ the signal duration.  For simplicity, we do not consider reverberation in the notations.
We define the set of all microphone signals as 
$\mathbf{Y} = \{\mathbf{y}_m \}_{1\leq m \leq M}$,
%$\mathbf{Y} = [\mathbf{y}_m^\tr]^\tr_{1\leq m \leq M} \in \R^{M \times T}$,
where $M$ is the number of microphones. In the CHiME-8 DASR task, the number of speakers, $S$, is considered unknown and varies from two to eight depending on the recordings.  The number of microphones, $M$, also varies depending on the datasets, and the array geometry is unknown.
%%%

The speech signal $\mathbf{x}_{m,s}$ consists of a long signal containing all utterances spoken by speaker $s$ in the recording and is zeros when the speaker is not active. 
The DASR system aims to output transcriptions and timings of all utterances of all speakers in the recording. Let us define the output $\mathbf{O}=\{\mathbf{O}_s\}_{1\leq s\leq S}$ as the sets of all utterances of all speakers, where $\mathbf{O}_s=\{\mathbf{o}_{s,j}\}_{1\leq j\leq J_s}$ represents the set of all $J_s$ utterances associated with speaker $s$. Here, $\mathbf{o}_{s,j} = (\mathbf{v}_{s,j}, \mathbf{b}_{s,j}) $ is the $j$-th utterance of speaker $s$, where $\mathbf{v}_{s,j}$ is the transcription or token sequence,  $\mathbf{b}_{s,j}= [t^{\text{start}}_{s,j}, t^{\text{end}}_{s,j}]$ represents the utterance boundaries, and $t^{\text{start}}_{s,j}$ and $ t^{\text{end}}_{s,j}$ are the start and end times of the utterance, respectively.

The DASR system estimates $\mathbf{O}$ given the microphone observations, as
\begin{align}
    \hat{\mathbf{O}} = \text{MT-DASR}(\mathbf{Y}),
\end{align}
where $\hat{\mathbf{O}}$ is the estimated utterances, including utterance timing and speaker attributions, and $\text{MT-DASR}(\cdot)$ is a function representing the multi-talker DASR system. The system performance can be evaluated in terms of time-constrained minimum-permutation word error rate (tcpWER) \cite{MeetEval23}, which is the WER accounting for the speaker and time assignments of each word.

We implement the proposed multi-talker DASR system as a cascade of diarization, SE, and ASR modules, as shown in Figure~\ref{fig:overview}. Below, we briefly define each module to demonstrate how they are interconnected. In the subsequent sections, we explain the details of each module. 

\textbf{Diarization:} The diarization module operates on the multi-channel signals of all recordings and predicts the speaker activity for each speaker in the recording. Here, by abuse of notations, we consider an estimated active speaker segment as an utterance, although it may not match an actual reference utterance. The diarization module thus estimates the utterance boundaries and assigns a speaker label to each estimated utterance, as
\begin{align}
    \hat{ \mathbf{B}} = \operatorname{DIAR}(\mathbf{Y}),
    \label{eq.diar}
\end{align}
where  $ \operatorname{DIAR}(\cdot)$ represents the diarization function, $\hat{ \mathbf{B}} = \{ \hat{\mathbf{B}}_{s} \}_{1\leq s \leq \hat{S}}$ is the set of utterance boundaries of all speakers, $\hat{\mathbf{B}}_{s}=\{ \hat{\mathbf{b}}_{s,j} \}_{1\leq j \leq \hat{J}_s}$ is the set of all estimated utterance boundaries for speaker $s$, $ \hat{\mathbf{b}}_{s,j}$, $\hat{J}_s $ is the number of estimated utterances, and $\hat{S}$ is the estimated number of speakers, which is also estimated by the diarization module. 
%Note here that the diarization does not provide absolute speaker labels but should only associate a consistent speaker label for each speaker in a recording. 
The speaker diarization module is composed of EEND-based segmentation, SE, speaker counting, speaker embedding clustering, TS-VAD refinement, and DOVER-Lap, as explained in Section \ref{sec:diar}.

\textbf{Speech enhancement:} The SE module aims to isolate the speech signal of the target speaker in each utterance, removing interference, noise, and reverberation. It operates on each utterance found by the diarization module and is conditioned on the estimated speakers' activity information. We can formulate its processing, as
\begin{align}
\label{eq:se:system}
    %\hat{\mathbf{X}}_{s,j} = \text{SE}(\mathbf{Y}, \hat{\mathbf{b}}_{s,j}, \bar{\mathcal{M}}),
    \hat{\mathbf{x}}_{s,j} = \operatorname{SE}(\mathbf{Y}, \hat{\mathbf{B}}, s, j),
\end{align}
where $\operatorname{SE}(\cdot)$ represents the SE function and $\hat{\mathbf{x}}_{s,j}$ is the single-channel estimated target speech for speaker $s$ associated with the $j$-th segment.\footnote{The SE module $\operatorname{SE}(\cdot)$ can also generate multi-channel outputs as explained in Section \ref{sec:se:chunkwiseSE}.}
It includes microphone subset selection, GSS, WPE, and beamforming, as explained in Section \ref{sec:se}.

\textbf{ASR:} Finally, we perform ASR on each enhanced signal channel speech signal, $\hat{\mathbf{x}}_{s,j}$, to obtain the transcriptions $\hat{\mathbf{v}}_{s,j}$ as
\begin{align}
    \hat{\mathbf{v}}_{s,j} = \text{ASR}(\hat{\mathbf{x}}_{s,j}),
    \label{eq:asr}
\end{align}
where $\text{ASR}(\cdot)$ represents the ASR module. To improve the performance, we perform ASR using several models and perform system-combination using ROVER. In addition, we also perform language model rescoring. We provide the details of the ASR module are provided in Section \ref{sec:asr}.

%=============================
\section{Speaker diarization module}
\label{sec:diar}
%=============================
%------------------------------------
\begin{figure}[t]
    \centering
    \includegraphics[width=1.00\linewidth]{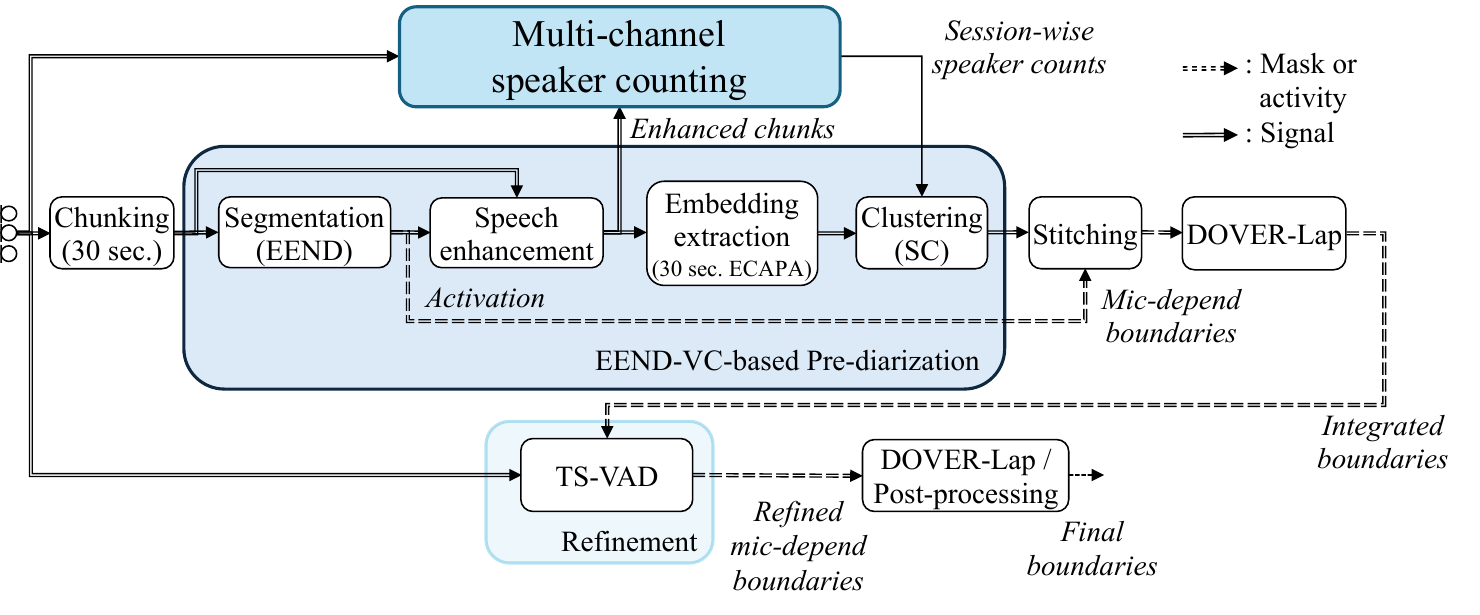}
    \caption{
        Schematic diagram of proposed speaker diarization system. Double-lined arrows represent multi-channel data flow, while single-lined arrows represent single-channel data flow.
        %\textcolor{red}{[TODO] Make more space in the figure to make it clearer. Make the name of the boxes consistent with the text and subsections.}
    }
    \label{fig:diar_overview}
\end{figure}
%------------------------------------
This section details our speaker diarization system defined in Eq.~\eqref{eq.diar}.
The entire diarization system, as shown in Figure \ref{fig:diar_overview}, consists of three main parts: i) EEND-VC-based pre-diarization, ii) multi-channel speaker counting, and iii) TS-VAD refinement.
TS-VAD is a very powerful approach for dealing with the challenging recording conditions of the CHiME challenge \cite{Medennikov2020TargetSpeakerVA, chime7_ustc}, but it requires initialization, which is usually performed using a simple speaker embedding clustering approach. For improved performance, we propose using a strong EEND-VC system instead.
We refer to the EEND-VC pre-diarization as DIA1 and its combination with TS-VAD refinement as DIA2. We also built a third diarization system, DIA3, which is the same as DIA2 but with a stronger TS-VAD system featuring more training data and longer training epochs.

%We introduce several modifications to the EEND-VC framework for the CHiME challenge. First, we perform SE using GSS on the local segments found by the EEND segmentation to improve the reliability of the speaker embeddings. Then, we decouple speaker counting from EEND-VC, which allows more precise speaker counting. Finally, we use DOVER-Lap to combine the diarization results obtained individually for each microphone, which we found an effective way to exploit the diversity of the distributed microphone recordings.

In the following subsections, we describe the EEND-VC pre-diarization in \ref{ssec:eend_vc}, the multi-channel speaker counting in \ref{ssec:speaker_counting}, TS-VAD refinement in \ref{ssec:tsvad}, and the details of the settings and training schemes in Section \ref{ssec:diar_settings}. %different components of the diarization system.

%====================================
\subsection{EEND-VC-based pre-diarization (DIA1)}
%====================================
\label{ssec:eend_vc}
%------------------------------------
\begin{figure}[t]
    \centering
    \includegraphics[width=0.99\linewidth]{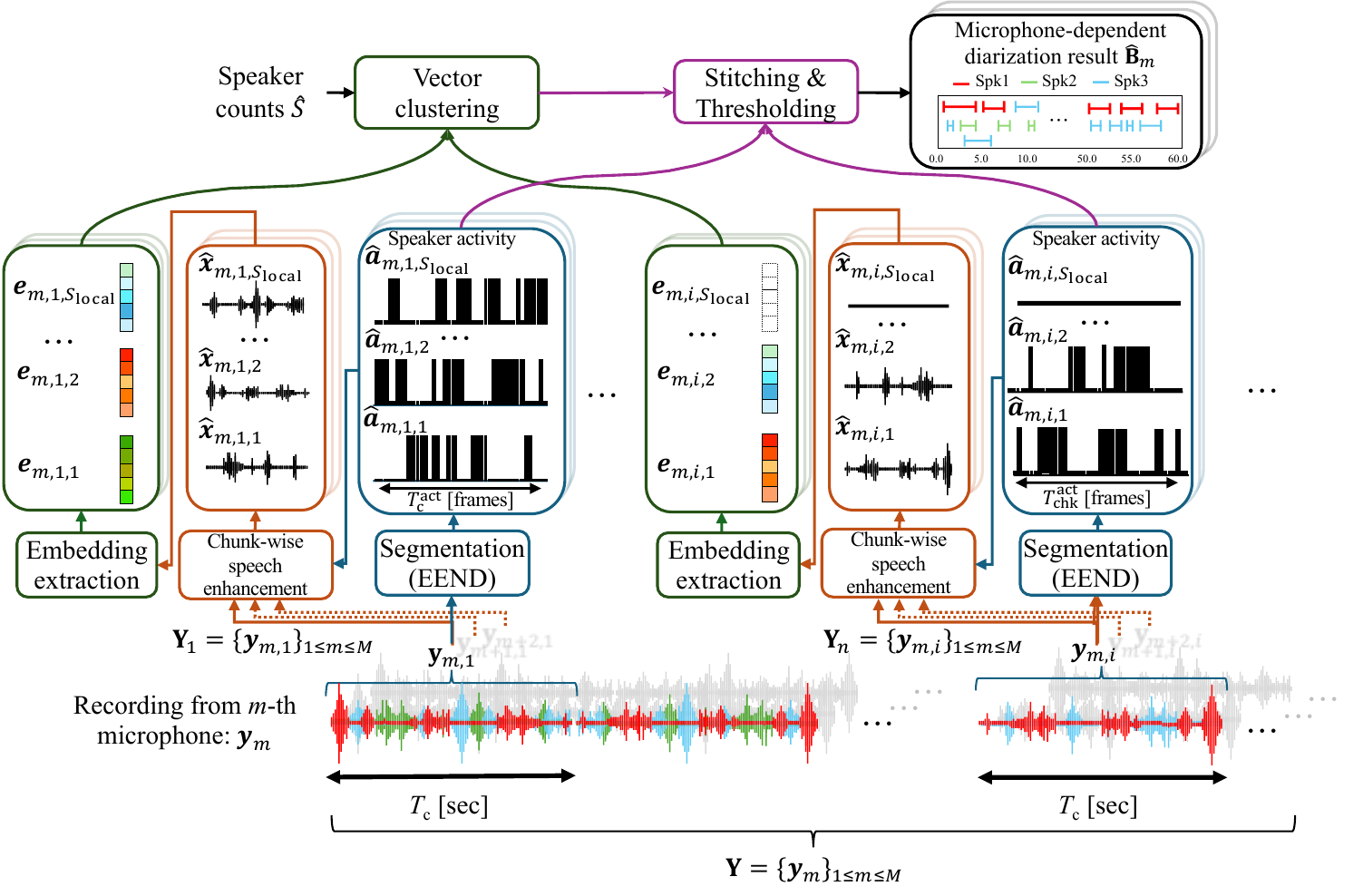}
    \caption{
        % Schematic diagram of the EEND-VC applied to the $m$-th channel audio containing three speakers in total. 
        % The first chunk $\mathbf{Y}_{1}$ contains three active speakers in total (red, green, and blue speakers shown in the waveform at the bottom), while the $i$-th chunk $\mathbf{Y}_{i}$ only contains two active speakers (blue and red speakers). 
        Schematic diagram of EEND-VC applied to $m$-th channel audio. For simplicity, this shows an example with three distinct speakers (represented by red, green, and blue waveforms in the bottom visualization). 
        The first chunk $\mathbf{y}_{m,1}$ contains three active speakers in total (red, green, and blue speakers), while the $i$-th chunk $\mathbf{y}_{m,i}$ only contains two active speakers (blue and red). 
    }
    \label{fig:eend-vc}
\end{figure}
%------------------------------------

% Figure \ref{fig:eend-vc} illustrates an example of data flow in our EEND-VC-based pre-diarization.
% For simplicity, the figure shows an example with three distinct speakers (represented by the red, green, and blue waveforms in the bottom visualization). The input consists of the set of the $M$ microphone signals, $\mat{Y}$.

% We introduce several modifications to the EEND-VC framework\cite{kinoshita21_interspeech} for the CHiME challenge. First, we perform EEND-VC per microphone and combine the results with DOVER-Lap, which we found an effective way to exploit the diversity of the distributed microphone recordings.
% Second, we perform SE using GSS on the local segments found by the EEND segmentation to improve the reliability of the speaker embeddings. Finally, we decouple speaker counting from EEND-VC, which allows more precise speaker counting. 
We use a modified version of the EEND-VC framework \cite{kinoshita21_interspeech} as the pre-diarization system, incorporating the following three changes. 
First, we perform SE using GSS on the local segments found by the EEND segmentation to improve the reliability of the speaker embeddings. 
Second, we perform EEND-VC for each microphone and combine the results with DOVER-Lap, which we found an effective way to exploit the diversity of the distributed microphone recordings.
Finally, we decouple speaker counting from EEND-VC, which allows more precise speaker counting. 

Figure \ref{fig:eend-vc} shows an example of the data flow in our EEND-VC-based pre-diarization.
The processing of the EEND-VC module consists of several steps, applied separately to each microphone, yielding microphone-dependent diarization results that consist of estimated utterance boundaries, $\hat{\mat{B}}_{m}=\{\hat{\mat{B}}_{m,s}\}_{1\leq s\leq \hat{S}}$.
First, the speech signal is divided into chunks and we use an NN to estimate the speaker activity for each chunk, which we refer to as EEND-based segmentation. 
We then perform chunk-wise SE to enhance the speech given to the speaker activities of each detected speaker in each segment, reducing the impact of noise, reverberation, and interfering speakers.
Next, we compute speaker embeddings from the enhanced speech signals. 
Finally, we perform clustering and stitching to assemble the global diarization output together. 
This process is performed for each channel independently, and the results are then combined using DOVER-Lap as explained in Section \ref{ssec:dover}.
%We explain the details of each of these steps in the next subsections.
%the estimated number of speakers, $\hat{S}$, denoted as $\hat{\mat{B}}_{m}=\{\hat{\mat{B}}_{m,s}\}_{1\leq s\leq \hat{S}}$, estimated using each $m$-th microphone.
%For simplicity, we omit the microphone index $m$ in the subsequent descriptions unless it is essential.

%=============================
\subsubsection{EEND-based local segmentation}
%=============================

First, we divide the microphone signal into chunks of $\Tc$ seconds, resulting in a set of $I$ chunks $\{{\vec{y}}_{m,i}\in\mathbb{R}^{\Tcin}\}_{1\leq i\leq I}$ for each microphone $m$, where each chunk contains $\Tcin = \fs\cdot\Tc$ samples and $\fs$ is the sampling frequency. In our system, we set $\Tc = 30$ seconds.
We then apply EEND separately to each channel of each audio chunk $\vec{y}_{m,i}$, yielding local speaker activity posteriors, as
%----------------------------
\begin{equation}
\hat{\mat{A}}_{m,i} = [\hat{\vec{a}}_{m,i,s=1},\dots, \hat{\vec{a}}_{m,i,s=\Slocal}] = 
  f_{\mathrm{EEND}}(\mathbf{y}_{m,i}),
  \label{eq.diar_eend}
\end{equation}
%----------------------------
where $\hat{\vec{a}}_{m,i,s}\coloneqq[\hat{a}_{m,i,s,t=1},\dots,\hat{a}_{m,i,s,t=\Tcout}]^{\tr} \in [0, 1]^{\Tcout}$ represents the posterior probabilities of the $s$-th local speaker’s speech activity at each frame index $t$ within the $i$-th chunk ($f_{\mathrm{EEND}}(\cdot)$ is the EEND function implemented with an NN).
Note that the length of speaker activity vectors, $\Tcout$, is shorter than $\Tcin$ due to the downsampling of EEND.
$\Slocal$ denotes the number of local speakers, a hyper-parameter of EEND that determines the maximum number of speakers present within each chunk.

After applying EEND, we obtain chunk-wise local segmentation results. 
To derive the global diarization results, we need to stitch together the local segmentation.
However, the order of speakers in the estimated local speaker activities can vary across chunks, and speakers who are not active in one chunk may be active in others.
% To solve this speaker permutation issue, we compute speaker embeddings and cluster them to assign session-wise (global) speaker labels to the local speakers.
% These steps are detailed in the next subsections.
To solve this speaker permutation issue, we compute speaker embeddings and cluster them to assign session-wise (global) speaker labels to the local speakers, as detailed in the following subsections.

%=============================
\subsubsection{Segment-wise speaker embedding computation}
%=============================

We calculate $D$-dimensional speaker embeddings $\{\vec{e}_{m,i,s} \in \mathbb{R}^{D}\}_{1\leq s\leq \Slocal}$ for each local speaker $s$ in each chunk $i$ from the $m$-th microphone given the local speaker activity $\hat{\mat{A}}_{m,i}$ obtained from Eq.~\eqref{eq.diar_eend}.
Most EEND-VC-based approaches, including our previous CHiME-7 system \cite{tawara_icassp24} and CHiME-7/8 baseline systems \cite{cornell2023chime7,chime8-task1}, extract speaker embeddings exclusively from non-overlapping intervals where only a single speaker is present, as
%----------------------------
\begin{align}
%f_{\mathrm{Emb}}(\vec{y}_{m,i}; \hat{\vec{a}}_{m,i,s}) 
%\coloneqq& 
\vec{e}_{m,i,s} = &f_{\mathrm{Emb}}(\vec{y}_{m,i} \odot \mathrm{upsample}(\tilde{\vec{a}}_{m,i,s})) \in \mathbb{R}^{D},
\label{eq.diar.emb2}
\end{align}
where
\begin{align}
%&\tilde{a}_{m,i,s,t} = \left\{
%\begin{array}{ll}
%    1 & \mathrm{if}~(\hat{a}_{m,i,s,t}\geq\thetaemb) \land (\hat{a}_{m,i,s',t} < \thetaemb), s'\neq s \\
%    0 & \mathrm{otherwise}
%\end{array}
%\right.,
\tilde{a}_{m,i,s,t} =&\mathbbm{1}\left((\hat{a}_{m,i,s,t}\geq\thetaemb) \land (\hat{a}_{m,i,s',t} < \thetaemb), s'\neq s\right),
\end{align}
%----------------------------
and $f_{\mathrm{Emb}}(\cdot)$ is a NN that maps a speech signal to a speaker embedding vector of dimension $D$, such as ECAPA-TDNN \cite{desplanques2020ecapa}. 
$\mathbbm{1}(\mathrm{condition})$ denotes an indicator function that returns 1 if the condition is satisfied and 0 otherwise $\tilde{\vec{a}}_{m,i,s}\coloneqq [\tilde{a}_{m,i,s,t=1},\dots,\tilde{a}_{m,i,s,t=\Tcout}]^{\tr}\in \{0,1\}^{\Tcout}$ is the frame-wise binary mask to select the non-overlapping frames, $\mathrm{upsample}(\cdot)$ is the linear interpolation used to upsample the length of speaker activity from $\Tcout$ to $\Tcin$, $\thetaemb$ is the speaker activity threshold for embedding extraction, and $\odot$ is the Hadamard product. This strategy efficiently captures speaker characteristics by focusing on single-speaker intervals.

However, it becomes challenging to extract speaker embeddings from highly overlapping segments that contain only a few brief single-speaker intervals.
To address this, our previous system used a very long chunk size (e.g., 80 seconds), setting the number of local speakers to the maximum anticipated for the use case scenario (four in the case of CHiME-7 \cite{kamo23_chime}).
However, when using long chunks, the number of local speakers in a chunk at inference time may exceed the anticipated maximum number of speakers. 
Moreover, using excessively long chunks can degrade EEND performance by increasing the risk of speaker permutation within chunks~\cite{tawara_icassp24}.

We therefore propose a more robust speaker embedding strategy that performs SE before speaker embedding extraction, thereby reducing interference from overlapping speakers.
Specifically, we apply a chunk-wise SE to enhance each local speaker's speech $\hat{\vec{x}}_{m,i,s}$ from the audio chunk, using their speaker activity $\hat{\vec{a}}_{m,i,s}$ and signals from all microphones $m\in\left\{1,\dots,M\right\}\eqqcolon\mathcal{M}$, $\mat{Y}_{i} = \{\vec{y}_{m,i}\}_{1\leq m\leq M}$.
The speaker embeddings are thus computed, as
%----------------------------
\begin{align}
\vec{e}_{m,i,s} = &
%f_{\mathrm{Emb}}(\vec{y}_{m,i}; \hat{\vec{a}}_{m,i,s}) 
%\coloneqq& 
f_{\mathrm{Emb}}(\hat{\vec{x}}_{m,i,s}),
\label{eq.diar_se1}
\\
\hat{\vec{x}}_{m,i,s}=& \textrm{ChunkWiseSE}(\vec{Y}_{i}, \mathbbm{1}(\hat{\vec{a}}_{m,i,s}\geq\thetaemb), \mathcal{M}),
\label{eq.diar_se2}
\end{align}
%----------------------------
where $\textrm{ChunkWiseSE}(\cdot)$ is the chunk-wise SE approach described in Section \ref{sec:se:chunkwiseSE}.

We analyze the effect of SE on EEND-VC performance in \ref{ssec:ablation_model}, showing that it can greatly reduce diarization errors.

%=======================SE on ======
\subsubsection{Vector clustering}
%=============================
\label{sec:diar:vector_clustering}

Next, vector clustering is applied to the speaker embeddings to assign each local speaker to a global speaker cluster as
%----------------------------
\begin{align}
\mat{\Phi}_{m} = 
\left\{\phi_{m,i,s}\right\}_{1\leq s\leq \Slocal, 1\leq i\leq I}
&= {\rm Clustering}\left(\left\{\vec{e}_{m,i,s}\right\}_{1\leq s\leq \Slocal, 1\leq i\leq I}\right),
\end{align}
%----------------------------
where $\phi_{m,i,s} \in \mathbb{Z}_{[1,\hat{S}]}$ denotes the mapping between the local speaker embeddings and the global speaker clusters, which represent the speaker identity.

Various different clustering approaches can be used here, such as k-means clustering, agglomerative hierarchical clustering \cite{ward1963hierarchical}, and spectral clustering \cite{shi2000sc}.
In this paper, we utilize spectral clustering with the Ng-Jordan-Weiss algorithm \cite{njw}.
We use a radial basis function kernel with $\gamma=1.0$, applied to the cosine similarity between speaker embeddings to compute the affinity matrix. Spectral clustering is implemented using scikit-learn==1.5.2, introducing a small modification to enforce the constraint that speaker embeddings from local activities within the same chunk should not be clustered into the same cluster \cite{kinoshita21_interspeech}.

Note that the number of speakers can be estimated as the number of remaining clusters estimated by the clustering method.
However, this tends to be inaccurate in practical situations where the session is extremely long or short, as reported in \cite{tawara_icassp24,chime7_stcon}.
Instead, we perform clustering by setting the number of clusters to the number of speakers estimated with the multi-channel speaker counting approach described in Section \ref{ssec:speaker_counting}.

%=============================
\subsubsection{Stitching and post-processing}
%=============================
\label{sec:diar:post_processing}

Finally, we stitch together the estimated microphone-dependent local speaker activities $\{\hat{\vec{a}}_{m,i,s}\}_{1\leq s\leq \Slocal}$ across $1\leq i\leq I$ chunks on each microphone based on the global speaker cluster labels provided by $\mat{\Phi}_{m}$, producing the global speaker activity $\{\hat{\vec{a}}^{\mathrm{global}}_{m,s}\}_{1\leq s\leq \hat{S}}$ for each microphone.

The global speaker activity is then converted into a set of utterance boundaries $\hat{\mat{B}}_{m}$ using the post-processing method from \cite{bredin2023pyannote}. 
Specifically, we apply a median filter with a kernel of 25 frames to smooth the global speaker activity and subsequently threshold it with $\theta_{\mathrm{eend}}$, which yields the binarized speaker activity $\{\tilde{\vec{a}}^{\mathrm{global}}_{m,s}\in\{0,1\}^{\Tcout}\}_{1\leq s\leq \hat{S}}$.
Each global speaker’s utterance boundaries are then identified by locating frame indices where the binarized speaker activity changes.
%, followed by converting these indices to the time domain.
%In the pre-diarization step, we do not apply padding boundary offsets or short-pause filling operations, as suggested in \cite{bredin2023pyannote,gelly2015minimum}, since these instantaneous activations may be advantageous for the subsequent diarization refinement step.

%====================================
\subsection{Multi-channel speaker counting}
%====================================
\label{ssec:speaker_counting}
%------------------------------------
\begin{figure}[t]
    \centering
    \includegraphics[width=1.00\linewidth]{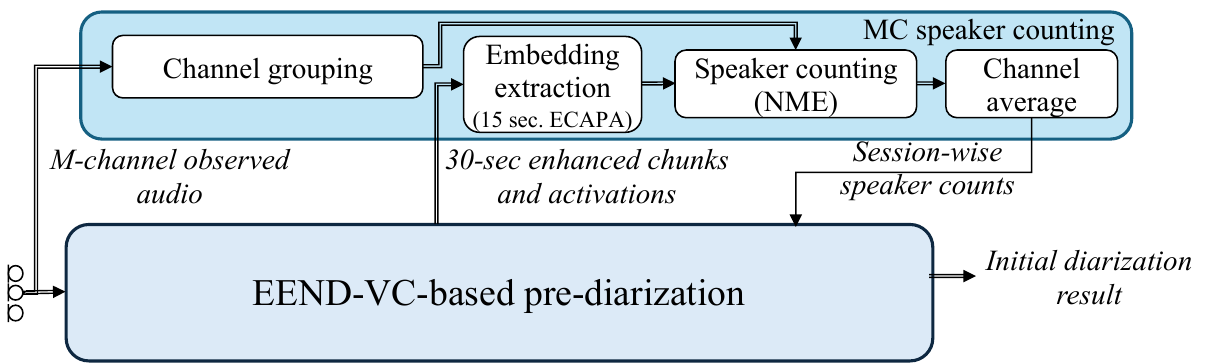}
    \caption{
        Schematic diagram of multi-channel speaker counting module.
    }
    \label{fig:diar_speaker_counting}
\end{figure}
%------------------------------------
%----------------------------------------------------------
\begin{algorithm}[t]
\renewcommand{\algorithmicfor}{\textbf{for each}}
\renewcommand{\algorithmicrequire}{\textbf{Input:}}
\renewcommand{\algorithmicensure}{\textbf{Output:}}
  \caption{Multi-channel speaker counting with microphone clustering}
    \small
    \begin{algorithmic}[1]
        \Require Set of observed $M$-channel audio $\{\mat{y}_{m}\}_{1\leq m\leq M}$, set of enhanced $M$-channel chunks $\{\hat{\mat{x}}_{m,i,s}\}_{1\leq m\leq M}$, and corresponding local speaker activities $\{\hat{\vec{a}}_{m,i,s}\}$, $\forall s\in\{1,\dots,\Slocal\}$ for each chunk $i\in\{1,\dots,I\}$
        \Ensure Session-wise speaker counting $\bar{s}$ 
        \For{$m$-th microphone, $i$-th chunk, $s$-th local speaker}
                \State $\left\{\hat{\vec{x}}^{\mathrm{subc}}_{m,i',s},
                         \vec{\hat{a}}^{\mathrm{subc}}_{m,i',s}
                       \right\}_{1\leq i'\leq I_{\mathrm{subc}}}
                       \leftarrow$
                       Split($\hat{\vec{x}}_{m,i,s},\vec{\hat{a}}_{m,i,s}$)
                \Comment{Split into subchunks}
                \For{$i'$-th subchunk}
                    \State $\vec{e}^{\mathrm{subc}}_{i,i',s}\leftarrow$ $f_{\mathrm{Emb}}(\hat{\vec{x}}^{\mathrm{subc}}_{m,i',s},\vec{\hat{a}}^{\mathrm{subc}}_{m,i',s}$) \Comment{Extract speaker embeddings}
                \EndFor
            %\EndFor
            %\textit{\# Extract speaker embedding}
        \EndFor
        %\State 
        \State Calculate $L$ using Eq. (\ref{eq.diar_sc1})\Comment{Microphone similarity matrix}
        \State $\{\mathcal{M}_{g}\}_{1\leq g \leq G} \leftarrow f_{\mathrm{ahc}}(L$)
        \Comment{Microphone grouping to obtain mic. groups}
        \For {$g\in\left\{1,\dots,G\right\}$}
        \Comment{Mic.-group-wise speaker counting}
        \State %s_{g}$ $\leftarrow$  
            %NME-SC$\left(\left\{\left\{e_{i,i',s}\right\}_{i'=1,\forall i\in\mathcal{M}_{g},\forall s}^{N_{\rm schk}}\right\}\right)$
            %$f_{\mathrm{NMC}}\left(\left\{\vec{e}_{m,i',s}'\right\}_{m \in \mathcal{M}_{g},1\leq i'\leq I_{\mathrm{subc}}, 1\leq s\leq S^{\mathrm{subc}}_{i'}
            %}\right)$ 
            % Estimate $s_{g}$ using Eq. \eqref{eq.diar.nmc}
            Estimate $s_{g}$ using NME on  $\mathcal{E}^{\mathrm{subc}}_{\mathrm{g}}$
        \EndFor
        \State Obtain session-wise speaker counts $\bar{s}$ using Eq.~(\ref{eq.diar_sc2})
    %\EndProcedure
  \end{algorithmic}
 \label{alg:diar.speaker_counting}
\end{algorithm}
%----------------------------------------------------------

Our preliminary experiments indicated that speaker counting tends to underestimate the number of speakers in short sessions.
We therefore propose augmenting the number of speaker embeddings used for counting by (1) considering shorter chunks and (2) exploiting the multiple microphone signals. 
By augmenting the number of speaker embeddings, we ensure stable counting performance even for short recordings. 

Figure \ref{fig:diar_speaker_counting} and Algorithm \ref{alg:diar.speaker_counting} detail the procedure of the proposed multi-channel speaker counting scheme.
The input of our speaker counting scheme is a set of $M$-channel observed signals $\{\vec{y}_{m}\}_{1\leq m\leq M}$, a set of $M$-channel enhanced speech signals divided in chunks $\{\hat{\vec{x}}_{m,i,s}\}_{1\leq m\leq M}$ estimated by Eq.(\ref{eq.diar_se2}) and the corresponding local speaker activities $\{\hat{\vec{a}}_{m,i,s}\}_{1\leq m\leq M}$ estimated by Eq.(\ref{eq.diar_eend}) for all local speakers and chunks.
We first divide these enhanced speech chunks and speaker activities into subchunks of $T_{\mathrm{subc}}$ seconds, producing $\hat{\vec{x}}^{\mathrm{subc}}_{m,i,s} \in \mathbb{R}^{T^{\mathrm{wav}}_{\mathrm{subc}}}$ and $\hat{\vec{a}}^{\mathrm{subc}}_{m,i,s}\in [0,1]^{T^{\mathrm{act}}_{\mathrm{subc}}}$ for $1\leq i\leq I_{\mathrm{subc}}$ (line 2 in Algorithm~\ref{alg:diar.speaker_counting}).
$T^{\mathrm{wav}}_{\mathrm{subc}}$ and $T^{\mathrm{act}}_{\mathrm{subc}}$ denote the number of samples and frames in each subchunk, respectively, and $I_{\mathrm{subc}}$ is the number of subchunks defined as $I_{\mathrm{subc}} = I \times \frac{T_{\mathrm{c}}}{T_{\mathrm{subc}}}$.

We then extract speaker embeddings $\{\vec{e}^{\mathrm{subc}}_{m,i',s} \in \mathbb{R}^{D}\}_{1\leq s \leq S_{\mathrm{local},i'}}$ for each $i'$-th subchunk and $m$-th microphone (lines 3--4 in Algorithm~\ref{alg:diar.speaker_counting}).
Note that we exclude the speaker embeddings extracted from subchunks with a total speech duration below a certain threshold $T_{\mathrm{min}}=0.75$ seconds to ensure that only reliable embeddings are used.
This results in $S^{\mathrm{subc}}_{i'}$ speaker embeddings in each $i'$-th subchunk. 
In all setups, we use $T_{\mathrm{subc}}=15$ second subchunks for speaker counting and $\Tc=30$ second chunks for vector clustering, which enables us to \textcolor{red}{approximately} double the number of speaker embeddings in the speaker counting step.
\textcolor{red}{These lengths are determined using the development set in the CHiME-8 challenge. We find that overly small subchunks result in deterioration during the subsequent clustering step, and subchunks with half the length of the chunks perform best across all scenarios.}
%We then further estimate the number of global speakers using the speaker embeddings obtained from all microphones to further increase the number of embeddings. 
%Here, we hypothesize that embeddings from a group of nearby microphones, where variations in transfer functions are minimized, can be effectively used for speaker counting.
Subsequently, we further increase the number of embeddings by exploiting the speaker embeddings obtained from all microphones.
There are various methods to utilize multi-channel data for speaker counting. One approach involves performing a single clustering step using speaker embeddings from all microphones. However, channel variations can result in inaccurate estimations of the number of speakers. 
Another approach is to conduct speaker counting for each microphone individually and then aggregate the results. However, this reduces the number of embeddings for clustering, leading to less reliable results. A compromise consists of performing speaker counting on groups of microphone signals that exhibit minimal variations in their transfer functions. This method allows the use of speaker embeddings from relatively similar channels, increasing the number of embeddings while limiting the impact of channel variability. The speaker counts from each microphone group can then be aggregated to achieve the final result.

This approach is implemented as follows. First, we apply a microphone grouping procedure similar to the one used for speech enhancement  \cite{park23_chime}. This approach is applied to the multi-microphone signals to identify groups of microphones with similar transfer characteristics. Speaker counting is then performed on each subchunk within these groups (lines 5–6).
Specifically, we first calculate a microphone similarity matrix $L=\left[l_{m_1,m_2}\right]\in\left[0,1\right]^{M\times M}$, where each element is the similarity between $m_1$-th and $m_2$-th microphone observed signals $\vec{y}_{m_1}, \vec{y}_{m_2}$.
The similarity is computed based on the inter-microphone correlations of the first $T_{\mathrm{corr}}$ samples of the signals, as
%-----------------------------------------------
\begin{align}
    l_{m_1,m_2} = \frac{1}{T_{\mathrm{corr}}} \sum_{\tau=1}^{T_{\mathrm{corr}}} \frac{(y_{m_1}(\tau)-\mu_{m_1}) (y_{m_2}(\tau)-\mu_{m_2})}{\sigma_{m_1} \sigma_{m_2}},
    \label{eq.diar_sc1}
\end{align}
%-----------------------------------------------
where $\tau$ is a time index and $\mu_{*}$ and $\sigma_{*}$ denote the mean and standard deviation of $\vec{y}_{*}$ for $1\leq \tau \leq T_{\mathrm{corr}}$, respectively.
We set $T_{\mathrm{corr}}=120$ seconds based on preliminary experimental results.
We then apply AHC to the microphones based on the computed similarity matrix, following Ward's method \cite{ward1963hierarchical}.
This approach determines the optimal number of microphone groups by recursively merging the closest pair of clusters that result in the smallest increase in total within-cluster variance, continuing until the minimum similarity between clusters falls below a predetermined threshold $\theta_{\mathrm{mic}}=0.05$.
This process groups microphones into $G$ groups, $\{\mathcal{M}_g\}_{1\leq g \leq G}$, of the microphones $\{1,\dots,M\}$.

For each microphone group $\mathcal{M}_{g}$, we estimate the number of speakers $s_g$ using the set of speaker embeddings within that group, as 
\begin{align}
\mathcal{E}^{\mathrm{subc}}_{\mathrm{g}}=\left\{\vec{e}^{\mathrm{subc}}_{m,i',s}\mid \forall m \in \mathcal{M}_{g}, 1\leq i' \leq I_{\mathrm{subc}}, 1\leq s\leq S^{\mathrm{subc}}_{i'} \right\}.
\end{align}
We determine the optimal number of speakers based on the approach proposed in \cite{park2020autotuning}, which finds the number of speakers from the normalized maximum eigengap (NME) of the Laplacian matrix computed over $\mathcal{E}^{\mathrm{subc}}_{\mathrm{g}}$ (lines 7--8).
% :
% %-----------------------------------------------
% \begin{align}
% s_{g} = f_{\mathrm{NMC}}\left(\mathcal{E}^{\mathrm{subc}}_{\mathrm{g}}\right),
% \label{eq.diar.nmc}
% \end{align}
% %-----------------------------------------------
% where $f_{\mathrm{NMC}}$ denotes the NMC-based speaker counting function.
%Note that in speaker counting, we exclude the speaker embeddings extracted from subchunks with a total speech duration below a certain threshold $T_{\mathrm{min}}$ to ensure that only reliable embeddings are used.

Finally, to further improve the speaker counting accuracy for each session, we integrate the estimated speaker counts from each microphone group to obtain session-level speaker counts by computing the weighted average of the estimated speaker counts $s_{g}$ for each microphone group $g$, as 
%-----------------------------------------------
\begin{align}
    \hat{S} = \left\lfloor \frac{1}{\sum_{g'=1}^G|\mathcal{E}^{\mathrm{subc}}_{g'}|}\sum_{g=1}^{G} |\mathcal{E}^{\mathrm{subc}}_{g}| s_{g} \right\rceil,
    \label{eq.diar_sc2}
\end{align}
%-----------------------------------------------
where $\lfloor\cdot\rceil$ is a rounding function to an integer.
% (line 10).
$|\mathcal{E}^{\mathrm{subc}}_{g}|$ represents the number of embeddings in the $g$-th microphone group, acting as a weight factor to emphasize estimations from groups with more microphones, thereby enhancing the reliability of those estimates.
The session-level speaker counts are utilized as the estimated number of speakers for each channel in the vector clustering described in Section \ref{sec:diar:vector_clustering}.

We perform an ablation study of the different components of the proposed multi-channel speaker counting in \ref{ssec:ablation_counting}.

%================================
\subsection{Multi-channel integration with DOVER-Lap}
%================================
\label{ssec:dover}
Various approaches have been proposed to integrate the results from multiple microphones\textcolor{red}{, including early fusion \cite{horiguchi2022multi,Ishiguro:11} and late fusion \cite{Raj2021Doverlap,Boeddeker:24}}.
The early fusion approach leverages fine spatial information, by \textcolor{red}{applying muti-channel speech enhancement before diarization \cite{yoshioka2022vararray,wang20c_interspeech}, spatial feature clustering \cite{Ishiguro:11}, or incorporating multi-channel features to the input of EEND \cite{horiguchi2022multi}, all of which could offer valuable insights for speaker discrimination.}
However, such methods may struggle with tracking speakers who move frequently and might be sensitive to the layout of the microphone array. An alternative is the late fusion approach, which combines diarization results from each microphone independently. \textcolor{red}{We can perform late fusion by using techniques such as posterior combination \cite{Boeddeker:24} or DOVER-Lap \cite{Raj2021Doverlap}.}
While the late fusion does not directly access the fine spatial details, its independent handling of each channel may offer greater robustness to speaker movement and variations in an array configuration.

In the proposed system, \textcolor{red}{motivated by prior studies \cite{Raj2021Doverlap} and our preliminary experiments,} we adopt the late fusion approach \textcolor{red}{using DOVER-LAP to merge the diarization results from all available microphones.}
%Specifically, we merge the diarization results from all available microphones with DOVER-Lap~\cite{Raj2021Doverlap}. 
For DOVER-Lap, we utilize the Hungarian algorithm to find the optimal speaker permutation between the diarization results across different microphones and integrate them using a voting scheme. \textcolor{red}{Specifically, we first aligns clustering labels across channels using the Hungarian algorithm and then aggregates the results using majority voting.}
This voting scheme can help mitigate the impact of outlier results, such as those from microphones far from the sources or affected by recording failures.
The effectiveness of DOVER-Lap-based microphone integration is demonstrated in the ablation study in \ref{sec:ablation:doverlap}.

%=====================================
\subsection{TS-VAD refinement (DIA2/DIA3)}
%=====================================
\label{ssec:tsvad}
%------------------------------------
\begin{figure}[t]
    \centering
    \includegraphics[width=0.7\linewidth]{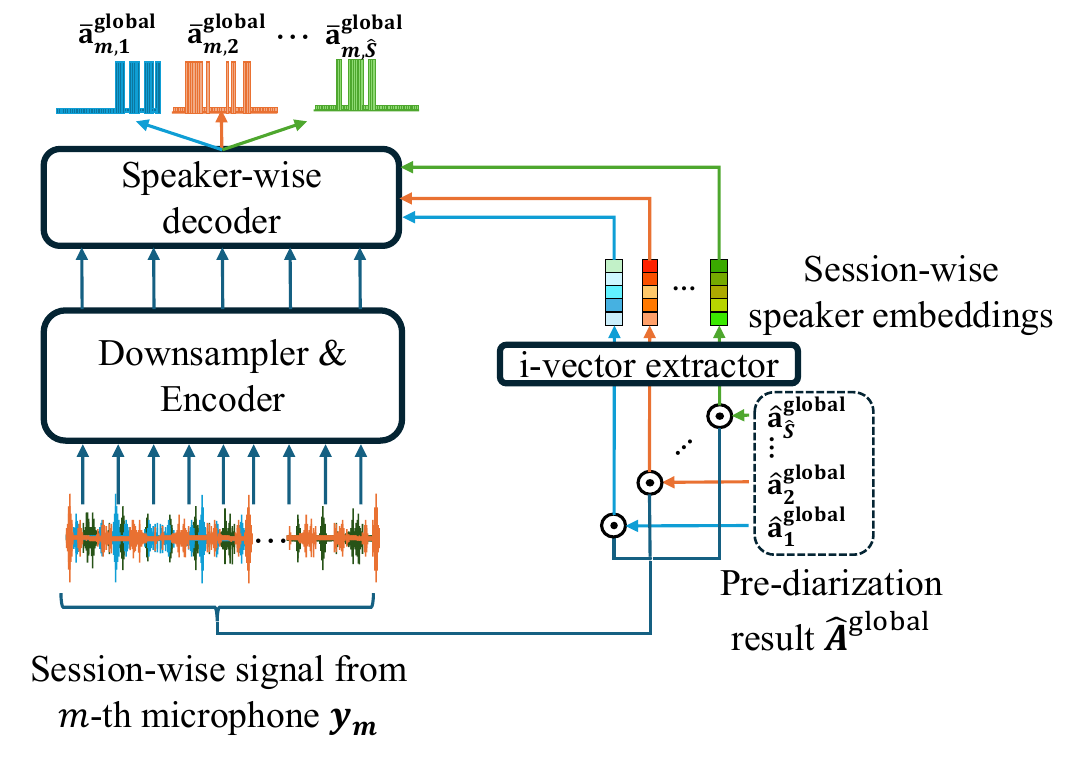}
    \caption{
    Outline of i-vector-based Seq2Seq TS-VAD.
    }
    \label{fig:ts-vad}
\end{figure}
%------------------------------------

After obtaining the initial diarization results using the EEND-VC module, we refine these results with target speaker voice activity detection (TS-VAD). 
Specifically, we adopt the sequence-to-sequence TS-VAD (Seq2Seq-TS-VAD)~\cite{NSD-MS2S}, applying it separately to each observed microphone signal to obtain microphone-dependent global speaker activity, as illustrated in Figure~\ref{fig:ts-vad}.

Seq2Seq-TS-VAD first extracts session-wise (global) speaker embeddings from the entire session, utilizing the global speaker activities estimated by the pre-diarization module. 
It then decodes each global speaker's activity using frame-wise features generated by the encoder module and the global speaker embeddings.
Note that the Seq2Seq-TS-VAD directly estimates the global speaker activity without permutation issues since the order of speakers in the output is determined by the global speaker embeddings.

For the encoder, we \textcolor{red}{utilize} Conformers consisting of a convolutional NN (CNN)-based downsampler followed by six Conformer blocks to transform the audio into frame-wise features.
For the decoder, we \textcolor{red}{used a six-layer Transformer-based} structure equipped with a memory-aware multi-speaker embedding (MA-MSE) module \cite{NSD-MS2S}.
%We apply memory-aware multi-speaker embedding with sequence-to-sequence architecture (NSD-MS2S) .
Session-level i-vectors are utilized as session-wise speaker embeddings, obtained by averaging the segment-level i-vectors belonging to the same speaker. 
These i-vectors are then combined with local speaker embeddings derived from the input mixture and the local segmentation information, which is implicitly performed within the speaker-wise decoder block in \textcolor{red}{Figure \ref{fig:ts-vad}}. 
These combined embeddings are then used to condition the Conformer-based TS-VAD module. 
Our system has almost the same model configuration as that in the top system of CHiME-7 \cite{chime7_ustc} but with a stronger initial diarization provided by EEND-VC. We confirm the importance of the initialization with EEND-VC in \ref{ssec:ablation_model}.

\begin{figure}[t]
    \centering
    \includegraphics[width=0.4\linewidth]{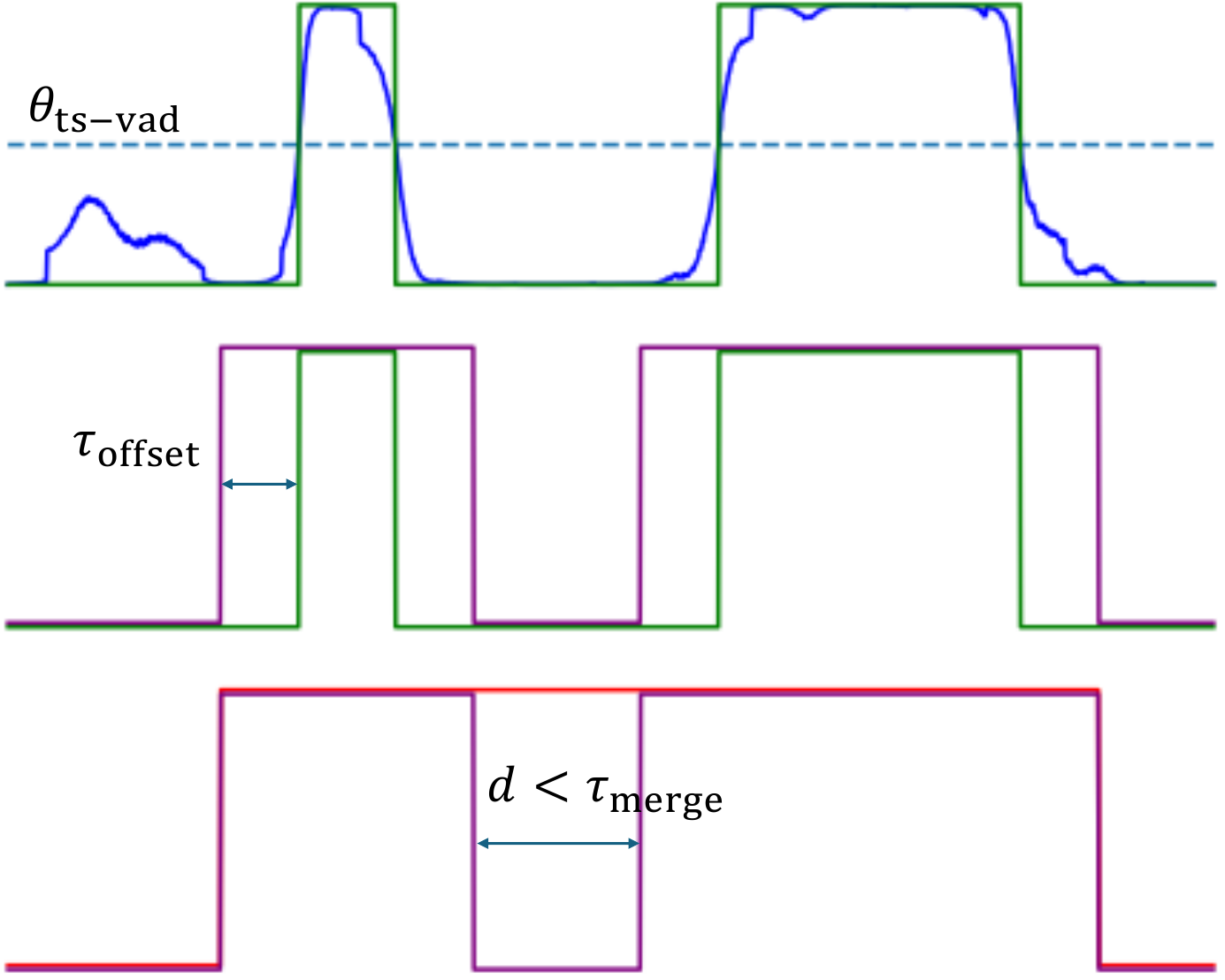}
    \caption{Postprocessing to obtain utterance boundaries from speaker activity. The blue plot represents the original activity from a speaker, the green plot shows the result after thresholding using $\theta_{\mathrm{ts-vad}}$, the purple plot represents the result after padding by an offset, $\tau_{\mathrm{offset}}$, and the red plot illustrates the activity obtained after filling pauses of duration $d$ shorter than $\tau_{\mathrm{merge}}$}
    %the pause intervals where the duration $d$ is shorter than $\tau_{\mathrm{merge}}$, which have been filled.}
    \label{fig:postprocessing}
\end{figure}

We obtain the diarization results from the speaker activity posteriors using the following post-processing steps. 
First, the global speaker activities are stitched together and thresholded using $\theta_{\textrm{ts-vad}}$. The boundaries of each utterance are then padded by a specified offset, $\tau_{\mathrm{offset}}$, and short pauses shorter than $\tau_{\mathrm{merge}}$ are filled with ones. As show in figure~\ref{fig:postprocessing}, these processes prevent the occurrence of spurious utterances while avoiding the omission of utterance boundaries and pauses, enabling better segmentation results for downstream ASR modules. These steps are performed on the activities obtained for each microphone and the activities are subsequently integrated into microphone-independent global speaker activities using DOVER-Lap, as in the pre-diarization step described in Section \ref{ssec:dover}. Note that boundary padding and short pause filling are only applied for ASR inputs, and the boundaries obtained by thresholding alone are used for GSS.

We report the effect of these post-processing steps in \ref{sec:ablation:asr_der_tuning}.
%The refined set of global speaker activities estimated for each microphone observation is stitched together and thresholded using $\theta_{\textrm{ts-vad}}$. The boundaries of each utterance are then padded by a specified offset, $\tau_{\mathrm{offset}}$, and short pauses shorter than $\tau_{\mathrm{merge}}$ are filled with ones for ASR. Figure~\ref{fig:postprocessing} illustrates how these three parameters work. These activities are subsequently integrated into microphone-independent global speaker activity using DOVER-Lap, as in the pre-diarization step. Note that these boundary padding and short pause filling are only applied for ASR inputs and the boundaries obtained by thresholding are used for GSS.

%===================================
\subsection{Settings}
%===================================
\label{ssec:diar_settings}
\subsubsection{Model configuration}
\textbf{EEND:} 
We utilized the pre-trained WavLM-large model as a feature extractor, following the approach used in \cite{WavLM}. 
Outputs from all WavLM Transformer layers were averaged using learnable weights, resulting in 1024-dimensional input speech features.
These features were then fed into the EEND module, following the same configurations as in our previous CHiME-7 submission~\cite{kamo23_chime}.
The EEND module consisted of six stacked Transformer encoder blocks with eight 256-dimensional attention heads, followed by a linear layer that projected the encoder’s output into four streams representing the frame-by-frame speaker activity probabilities for the four local speakers (i.e., $\Slocal=4$).
We applied thresholds of $\theta_{\mathrm{emb}} = 0.5$ and $\theta_{\mathrm{eend}} = 0.5$ to the outputs of the EEND modules to extract ECAPA-TDNN embeddings, as described in Eq. \eqref{eq.diar_se2}, and to obtain binarized speaker activity during the post-processing step, respectively.
 %where the EEND, GSS, and NME-SC hyperparameters were optimized to minimize the macro-average speaker clustering accuracy on the development set.
%which employs a pre-trained WavLM-large~\cite{wavlm} to obtain the input speech features, followed by six stacked Transformer-encoder blocks with eight 256-dimensional attention heads each.  The encoder's output is projected through a linear layer into four output streams, each representing frame-by-frame binary decisions for speaker activity. The EEND model was pre-trained using 50-second chunks derived from simulated conversation data~\cite{} and fine-tuned with 30-second chunks from a training subset of the CHiME-8 data.
%For inference with EEND, we used 30-second chunks for deriving activities used for GSS and speaker clustering and split them into 15-second subchunks for speaker counting.  The maximum number of speakers in a chunk was set to $N^{\text{max}} = 4$, which can correctly process at least ?? \% of 30-sec chunks in any scenario. 

\textbf{Vector clustering:} 
We used ECAPA-TDNN \cite{desplanques2020ecapa} for speaker embedding extraction, applying it to 30-second chunks ($\Tc = 30$) that were enhanced using a chunk-wise SE module.
We utilized SpeechBrain's ECAPA-TDNN~\cite{Speechbrain_ECAPA}, pre-trained on over 2000 hours of audio from more than 7000 speakers in VoxCeleb1 and VoxCeleb2~\cite{voxceleb}.
Subsequently, constrained spectral clustering was performed on these speaker embeddings, with the speaker count estimated using the speaker counting module.

\textbf{Speaker counting:} 
We determined the speaker count by identifying the value that minimizes the normalized maximum gap of the eigenvalues (NME) \cite{park2020autotuning} of the Laplacian matrix, computed from the cosine similarities between all pairs of speaker embeddings. 
To identify the NME, we used the default parameters of NME-SC from NeMo's CHiME-8 baseline \footnote{\url{https://github.com/chimechallenge/C8DASR-Baseline-NeMo/blob/main/nemo/collections/asr/parts/utils/offline_clustering.py}}, with one modification: we reduced the search range of $p$-values, which define the $p$-neighbor binarization of affinity matrix, from $\left[1, \lfloor\frac{N}{4}\rfloor \right]$ to $\left[1, \lfloor\frac{N}{10}\rfloor \right]$, where $N$ is the number of speaker embeddings, to mitigate the overestimation of the number of speakers.
This value was determined with the development set.

\textbf{TS-VAD:} 
The i-vector extractor was trained on VoxCeleb1 and VoxCeleb2 using the Kaldi recipe from egs/voxceleb/v1\footnote{\url{https://github.com/kaldi-asr/kaldi/tree/master/egs/voxceleb/v1}}.
We computed 40-dimensional Mel-frequency cepstral coefficients (MFCCs) calculated with a 25 ms window size and a 10 ms shift, followed by mean normalization using three-second sliding windows as acoustic features for i-vector extraction.
Silence regions and utterances shorter than 100 ms were removed based on local speaker activity estimated by the pre-diarization module.
Subsequently, i-vectors were extracted for each speaker’s utterance and averaged across the session to generate session-wise i-vectors for each speaker.
The implementation of Seq2Seq-TS-VAD relied on the publicly available implementation of NSD-MS2S \cite{NSD-MS2S,yang2024neural}.
The input to NSD-MS2S consisted of 40-dimensional filterbank features computed using a 25 ms window size and a 10 ms shift, followed by mean normalization with three-second sliding windows.
We used the default parameters implemented in NSD-MS2S code\footnote{\url{https://github.com/liyunlongaaa/NSD-MS2S}}, except that we used only a single deep interactive module block. 
%The model has 5.80 million parameters.

\textbf{Post-processing:}
To enable faster experimental turnaround, we tuned most parameters of the diarization module to achieve the best diarization error rate (DER) on the development set. However, DER is not well correlated with recognition performance. Therefore, after the initial tuning, we performed an extensive grid search on the optimal post-processing parameters after TS-VAD for ASR, including the activity threshold, boundary offset, and short-pause filling durations. After the tuning, we obtained $\theta_{\textrm{ts-vad}}=0.30$ , $\tau_{\mathrm{offset}}=0.0$ sec, and $\tau_{\mathrm{merge}}= 1.5$ sec as the best parameters. The effect of this tuning is reported in \ref{sec:ablation:asr_der_tuning}.

\subsubsection{Training data}
Each diarization model, including EEND and TS-VAD, was initially trained on simulated mixtures and subsequently fine-tuned using the CHiME-8 training set \cite{cornell2024chime}.
The simulated mixtures were primarily generated following a protocol designed to create natural utterance transitions \cite{yamashita2022improving}, with the following modifications to better match the real data statistics: i) long-form audio was first generated by incorporating turn-hold, turn-switch, and interruption patterns, with back-channels added afterward, ii) silence and overlap durations between utterances were sampled directly from the real data instead of from a fitted distribution, and iii) overlap durations in interruptions and back-channels were determined using absolute durations extracted from the real data, rather than relative ratios.
%We generated 1M/500 50-second 4-speaker mixtures using LibriSpeech for EEND-VC training/validation purposes, and 500k/500 80-second 4-speaker mixtures for TS-VAD, respectively.
We generated one million 50-second mixtures of four speakers using LibriSpeech \cite{panayotov2015librispeech} for EEND-VC training, and 500k 80-second mixtures of four speakers for TS-VAD in the DIA1 and DIA2 systems.
Additionally, we generated 1 million 50-second mixtures of four speakers to train a stronger TS-VAD for the DIA3 system.
We used 500 mixtures for validation in all cases.
Each mixture was augmented using the simulated room impulse responses \cite{ko2017study} and MUSAN noises \cite{snyder2015musan}.

\subsubsection{Training procedure}

\textbf{EEND:}
For the initial training, we froze the WavLM parameters and optimized the subsequent layer parameters to minimize the binary cross entropy between the model output and the ground truth labels, using permutation invariant training (PIT) \cite{Fujita:19}.
We trained the model for 25 epochs using the Adam optimizer with 25,000 warm-up steps, a learning rate of $10^{-3}$, and a batch size of 2048.
For fine-tuning, we updated the entire model using the CHiME-8 training set for three epochs with a learning rate of $10^{-5}$ and a batch size of one.
%The threshold value for the EEND output, $\theta_{\textrm{eend}}$, was set to 0.50.

\textbf{TS-VAD:}
For both the initial training and fine-tuning, we optimized all NSD-MS2S parameters to minimize the binary cross entropy between the model output and the ground truth labels.
We pretrained the model for six epochs on DIA2 and three epochs on DIA3.
We used Adam optimizer with a learning rate of $10^{-3}$ and a batch size of 256.
For fine-tuning, we again updated all NSD-MS2S parameters using the CHiME-8 training set for six epochs with a learning rate of $10^{-5}$ and a batch size of 1024.
Following the procedure used in \cite{NSD-MS2S}, we averaged the model parameters obtained over six epochs in the fine-tuning step.
%The threshold value for the TS-VAD output, $\theta_{\textrm{ts-vad}}$, was set to 0.45.
%-----------------------------------------------------------------------------
%-----------------------------------------------------------------------------
\section{Speech enhancement}
\label{sec:se}

\begin{figure}[t]
  \centering
\includegraphics[width=0.99\linewidth]{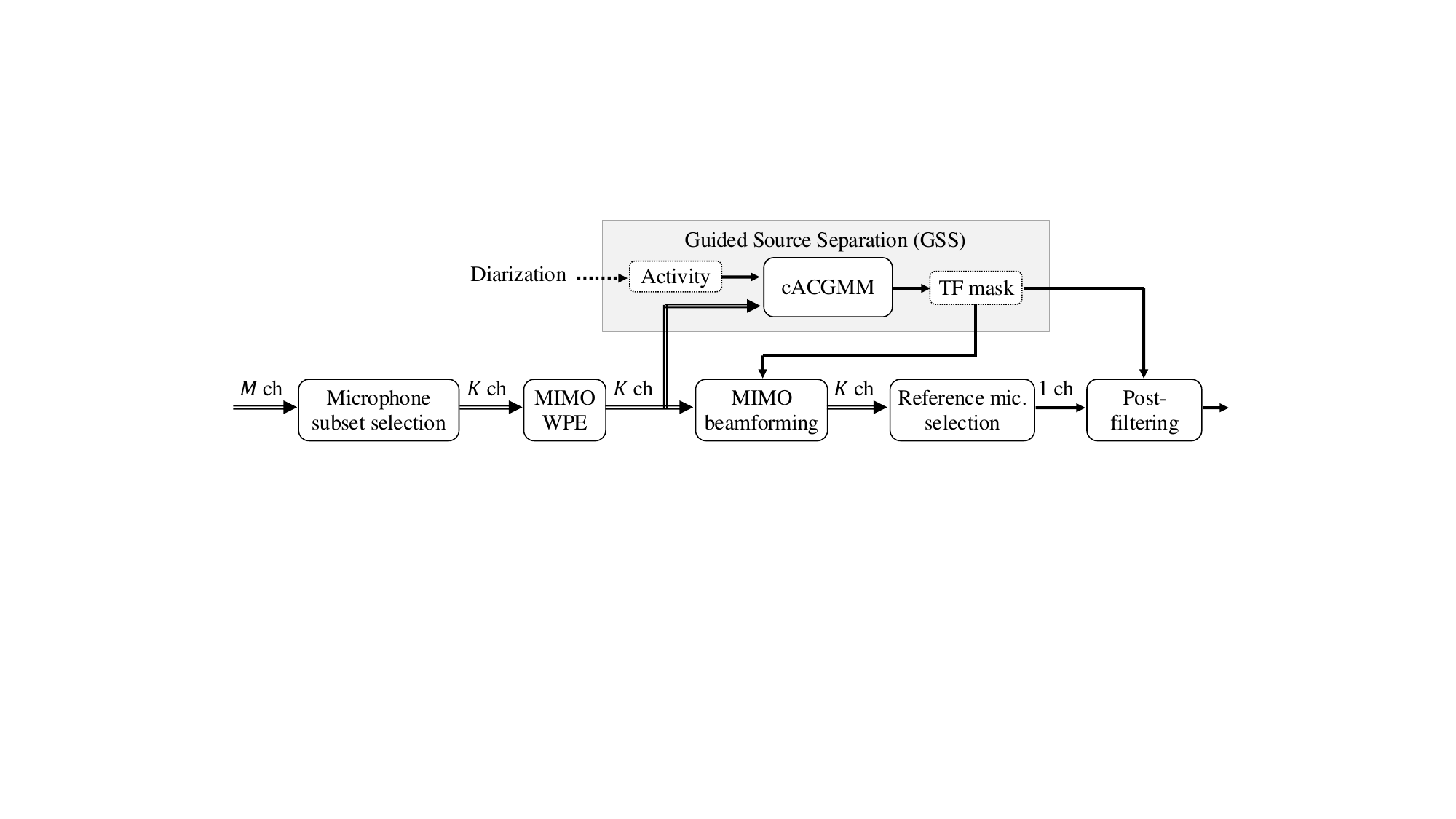}
  \caption{
    Processing flow of proposed SE system, which follows the same structure as the CHiME-8 DASR NeMo baseline system.
    While the baseline system uses R1-MWF for the MIMO beamforming module, our SE system utilizes SP-MWF (Section~\ref{sec:se:bf}).
    In addition, our microphone subset selection module differs from that of the baseline system (Section~\ref{sec:se:mic-selection}).
  }
  \label{fig:se}
\end{figure}

In this section, we explain how we implement our SE system described in Eq.~\eqref{eq:se:system}.
Figure~\ref{fig:se} shows the processing flow of our SE system, which follows the same structure as the official SE system~\cite{boeddecker18_chime} implemented in the CHiME-8 DASR NeMo Baseline~\cite{chime8-task1,park23_chime}.
These SE systems rely heavily on Guided Source Separation (GSS)~\cite{boeddecker18_chime} to robustly separate the target speaker with the help of speaker activity provided by the diarization system.
They also feature microphone subset selection and beamformer reference microphone selection to effectively handle distributed microphones and maximize their potential.
We briefly review the official SE system in Section \ref{sec:se:baseline}
and then describe our two key modifications to it in Sections \ref{sec:se:mic-selection} and \ref{sec:se:bf}.

\begin{figure}[t]
  \centering
\includegraphics[width=0.99\linewidth]{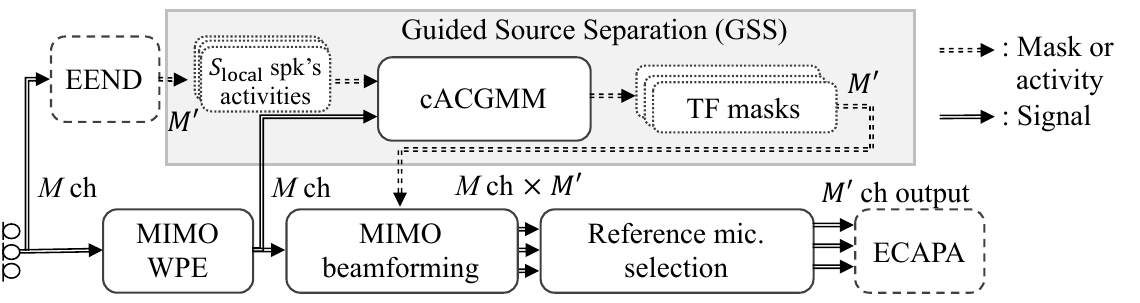}
  \caption{
  Processing flow of SE system used for speaker diarization. 
  Unlike the SE system, signals from all microphones are utilized for WPE and beamforming. 
  Additionally, GSS is applied based on local activity estimated on each of the $M'$ channels by the EEND module.
  }
  \label{fig:diar_se}
\end{figure}

\subsection{CHiME-8 DASR NeMo baseline}
\label{sec:se:baseline}

The processing flow of the baseline SE system is shown in Figure~\ref{fig:se}.
We here explain each module in Figure~\ref{fig:se} and how they are connected to each other.

In a distributed microphone scenario, some microphones may have a significantly lower signal-to-noise ratio (SNR) than others, and such microphones may not help or may even negatively affect the SE.
To ignore such uninformative microphones, the microphone subset selection module in Figure~\ref{fig:se} selects the top $K$ microphones among all $M$ microphones that are considered to be the most relevant to the subsequent SE processing.
Selecting fewer microphones also helps to reduce the computational load of the SE system.
To select the top $K$ microphones, the baseline system borrows a microphone ranking method from~\cite{WOLF2014170}, which is based on an audio feature called envelop variance (EV);
see Section~\ref{sec:se:mic-selection} for more details.
Note that the microphone subset selection is carried out independently for each utterance segment $\mathbf{b}_{s,j} = [t^{\text{start}}_{s,j}, t^{\text{end}}_{s,j}]$.

After selecting the top $K$ microphones for segment
$\mathbf{b}_{s,j} = [t^{\text{start}}_{s,j}, t^{\text{end}}_{s,j}]$,
the baseline system expands the segment to 
$\mathbf{b}_{s,j}^\Delta = [t^{\text{start}}_{s,j} - \Delta, t^{\text{end}}_{s,j} + \Delta]$, where $\Delta \geq 0$ is the context length and is set to $15$ seconds as a default.
Expanding the segment is primarily to enable the GSS module to estimate the time-frequency (TF) masks accurately.
After selecting $K$ microphones, we cut out the $K$-channel signal with expanded segment $\mathbf{b}_{s,j}^\Delta$,
transform the cut signal to the short-term Fourier transform (STFT) domain,
and apply dereverberation to it.
For dereverberation, the baseline system adopts a multiple-input multiple-output (MIMO) implementation~\cite{yoshioka2012generalization} of weighted prediction error (WPE)~\cite{nakatani2010speech}, which has served as an essential SE component in the past CHiME Challenge series.

The dereverberated $K$-channel signal with expanded segment $\mathbf{b}_{s,j}^\Delta$ is passed to the GSS module to estimate a single channel time-frequency (TF) mask for separating target speaker $s$ on segment $\mathbf{b}_{s,j}^\Delta$.
GSS~\cite{boeddecker18_chime} is a clustering-based TF mask estimation method built on a \textit{blind} method called the complex angular central Gaussian mixture model (cACGMM)~\cite{ito2016complex}.
GSS extends cACGMM to exploit the speaker activity of all sources, i.e., $\hat{\mathbf{B}}$ in Eq.~\eqref{eq:se:system}, which is provided by the diarization system. 
Guided by the speaker activity, GSS can estimate the TF masks for all sources more accurately than cACGMM and further associate each TF mask with the corresponding source, which is crucial for separating a target speaker $s$.
For more information on GSS, see~\cite{boeddecker18_chime}.

After GSS estimates the TF masks, the beamforming module cuts off the added context in $\mathbf{b}_{s,j}^\Delta$ from the $K$-channel dereverberated signal and optimizes an arbitrary MIMO beamformer for separating target speaker $s$ over original target segment $\mathbf{b}_{s,j}$
(not over $\mathbf{b}_{s,j}^\Delta$)
assuming that the $j$th utterance of target speaker $s$ is correctly included in segment $\mathbf{b}_{s,j}$.
The baseline system uses an effective implementation of the minimum variance distortionless response (MVDR) beamformer~\cite{benesty2008noncausal,souden2009optimal}, which is sometimes called Souden MVDR or the rank-1 multichannel Wiener filter (R1-MWF).
This beamformer avoids error-prone estimation of the target source steering vector and has been empirically proven effective when used as frontend SE systems for ASR through past CHiME challenges.
Section~\ref{sec:se:bf} explains how we obtain the MIMO beamformer using the TF masks.

The MIMO beamformer has $K$ separation filters, or $K$ separated signals, each designed to estimate a target clean signal as observed at each microphone.
The reference microphone selection module selects, among $K$ separated signals, the separated signal or the corresponding reference microphone that is considered the most effective in some sense.
For this purpose, the baseline system is based on the criterion of maximizing
\textcolor{red}{a certain signal quality measure}
%the output SNR
of the beamforming output signal~\cite{lawin2012reference,erdogan2016improved}.
See Section~\ref{sec:se:bf:refch} for a detailed description.

Finally, given the single-channel beamforming output signal,
the post-filtering module applies the blind analytic normalization (BAN)~\cite{warsitz2007blind} post-filter to it for adjusting the frequency-wise amplitude of the enhanced signal.
After that, using the TF mask for the target speaker provided by the GSS module, the post-filtering module also applies a TF masking to further enhance the signal.
This post-filtering module is discussed in Section~\ref{sec:se:bf:post-filter}.

\subsection{Microphone subset selection}
\label{sec:se:mic-selection}

As explained in Section~\ref{sec:se:baseline}, microphone subset selection is an essential component in SE systems that need to handle distributed microphones.
To select effective microphones, the conventional method implemented in the baseline system relies only on a single audio feature, i.e., EV (Section~\ref{sec:se:mic-selection:old}),
which may not be robust enough to deal with diverse recording scenarios, such as those in this CHiME-8 challenge.
We therefore propose leveraging multiple audio features to enhance robustness against environmental changes. 
We also propose imposing a lower bound on the number of microphones to be selected based on the assumption that the SE system can benefit from the spatial diversity of distributed microphones unless the number of microphones exceeds a certain threshold.
Section~\ref{sec:se:mic-selection:new} explains how these ideas are integrated into a single module.

\subsubsection{Conventional method based on envelope variance}
\label{sec:se:mic-selection:old}

To select the top $K$ most effective microphones out of $M$ microphones, where $K$ ($\leq M$) is specified by the user,
the conventional microphone subset selection module first calculates a signal quality measure called an EV \cite{WOLF2014170} of the input signal for each microphone.
The EV is a signal-level measure that is expected to have lower values when the input signal is more degraded by reverberation and background noise, and higher values when the signal is less degraded.
For the definition and detailed description of the EV, we refer readers to~\cite{WOLF2014170}.
The module then selects the $K$ microphones with the highest EV scores.
In the baseline system, $K$ is set to \textcolor{red}{$K = \lceil 0.8 \cdot M \rceil$} across all datasets and remains fixed, i.e., not varying depending on the dataset.
%Here $\lceil 0.8 \times M \rceil$ denotes the smallest integer that is greater than or equal to $0.8 \times M$.

\subsubsection{Proposed method based on EV and speech clarity index $C_{50}$}
\label{sec:se:mic-selection:new}

We extend the conventional microphone subset selection method in the following two aspects:
\begin{itemize}
  \item We select a set of microphones based not only on the EV but also on a speech quality indicator called \textit{speech clarity index $C_{50}$} (see, e.g.,~\cite{naylor2010speech}).
  \item We select at least $K_{\min} = 15$ microphones, if possible, and select all $M$ microphones when $M \leq K_{\min} = 15$.
\end{itemize}

First, we discuss the second point regarding the minimum number of microphones to select.
In the SE system, we assume that using at least $K_{\min}$ microphones always improves the quality of the enhanced signal in the SE system.
This is because using more microphones generally leads to a greater gain in the SNR of the beamforming output signal, motivating us to select as many microphones as possible.
On the other hand, the SNR improvement by beamforming tends to saturate as the number of microphones increases.
Moreover, selecting ineffective microphones, such as those with a very low SNR, could negatively impact the beamforming output signal.
The same argument for beamforming also applies to the TF mask estimation using GSS.
In the basis of on these observations, we hypothesize that discarding only ineffective microphones is more advantageous than selecting a small number of highly effective microphones.
We therefore impose a lower bound, $K_{\min}$, on the minimum number of microphones to select and pass to the subsequent module.
The value of $K_{\min} = 15$ is empirically determined using the development set.
For experimental results, see~\ref{sec:ablation:mic-selection}.

Next, we describe how the two audio features, EV and $C_{50}$, collaborate to solve microphone subset selection.
The speech clarity index $C_{50}$~\cite{naylor2010speech} is defined as the ratio of the energy in the early parts (0 to 50 ms) to that in the late parts (more than 50 ms) of the acoustic impulse response.
More formally, given an acoustic impulse response $h$ from a source to a microphone,
$C_{50}$ with respect to $h$ is defined as
\begin{align}
  C_{50} = 10 \log_{10}
  \mleft(
    \frac{
      \sum_{t=0}^{t_e} h(t)^2
    }{
      \sum_{t=t_e+1}^{\infty} h(t)^2
    }
  \mright),
\end{align}
where $t_e$ denotes the discrete time index corresponding to $50$ ms.
A microphone with a high $C_{50}$ value can be considered to capture a high-quality signal with less reverberation.
The clarity index $C_{50}$ for each microphone can be estimated using, e.g., the Brouhaha toolkit~\cite{lavechin2022brouhaha}.

If EV and $C_{50}$ can be estimated accurately, discarding low-quality microphones based on either measure would work well.
However, in diverse recording environments and distributed microphone setups, the estimates of EV and $C_{50}$ may not always be reliable, depending on the acoustic conditions.
Nevertheless, it is reasonable to assume that a microphone with both high EV and high $C_{50}$ values captures a high-quality signal and should therefore be selected.
To address this, while ensuring that at least $K_{\min} = 15$ microphones are selected, we propose the following selection method.

Let $\mathcal{K}_{\mathrm{EV}} \subseteq \{ 1, \ldots, M \}$ and $\mathcal{K}_{C_{50}} \subseteq \{ 1,\ldots, M \}$ be the sets of the top \textcolor{red}{$K_1 = \lceil 0.65 \cdot M \rceil$} microphones ranked by EV and $C_{50}$, respectively, where we determine the value $K_1$ empirically.
Let $\mathcal{K} = \mathcal{K}_{\mathrm{EV}} \cap \mathcal{K}_{C_{50}}$ denote the intersection of the two sets.
According to the definition, $|\mathcal{K}| \leq K_1 = | \mathcal{K}_{\mathrm{EV}} | = | \mathcal{K}_{C_{50}} |$.
We propose the following algorithm for selecting a set of microphones to pass to the next module in the SE system:%
\footnote{
    The algorithm presented in our workshop paper~\cite{ntt_chime8} contained a minor error in the fallback mechanism, which we have corrected in this paper.
    Note that we did not revise this fallback mechanism and we just used it in the CHiME-8 challenge.
}
\begin{itemize}
  \item If $M \leq K_{\min} = 15$, we do not apply microphone subset selection; in other words, we select all $M$ microphones.
  \item If $M > K_{\min}$ and $|\mathcal{K}| \geq K_{\min}$, we select $\mathcal{K} = \mathcal{K}_{\mathrm{EV}} \cap \mathcal{K}_{C_{50}}$.
  \item
    Otherwise, we rank the microphones in the complement of $\mathcal{K}$, i.e., $\mathcal{K}^c \coloneqq \{ 1,\ldots, M \} \setminus \mathcal{K}$, based on EV.
    We then select the top $K_{\min} - | \mathcal{K} |$ best microphones from $\mathcal{K}^c$, which we denote as $\mathcal{K}_{\mathrm{add}}$.
    Finally, we select $\mathcal{K} \cup \mathcal{K}_{\mathrm{add}}$ as the final output.
    Note that $| \mathcal{K} \cup \mathcal{K}_{\mathrm{add}} | = K_{\min}$.
  %\item If $M > K_{\min}$, $|\mathcal{K}| < K_{\min}$, and $| \mathcal{K}_{\mathrm{EV}} | \geq K_{\min}$, then we select $\mathcal{K}_{\mathrm{EV}}$.
  %\item If $M > K_{\min}$, $|\mathcal{K}| < K_{\min}$, and $| \mathcal{K}_{\mathrm{EV}} | < K_{\min}$, then we select the set of the top $K_{\min}$ microphones ranked by EV.
\end{itemize}
In this algorithm, we ensure that at least $K_{\min} = 15$ microphones are selected when $M \geq K_{\min}$.
The second step is based on the assumption that the microphones in the intersection $\mathcal{K} = \mathcal{K}_{\mathrm{EV}} \cap \mathcal{K}_{C_{50}}$ are highly reliable, and we shall select all of them.
The third step serves as a fallback mechanism,
using the baseline EV-based method to select the number of microphones needed to reach $K_{\min}$ from the microphones in $\mathcal{K}^c$.
%Note that, even in this case, we ensure that at least $K_{\min}$ microphones are selected, and that the highly reliable microphones in 
Note that even in this case the highly reliable microphones in 
$\mathcal{K} = \mathcal{K}_{\mathrm{EV}} \cap \mathcal{K}_{C_{50}}$ are included in the output.
The performance of this selection method is examined in~\ref{sec:ablation:mic-selection}.

%\begin{figure}[t]
%  \centering
%\includegraphics[width=0.5\linewidth]{fig/se/mic_selection.pdf}
%  \caption{Schematic diagram of the proposed microphone subset selection}
%  \label{fig:mic-selection}
%\end{figure}

\subsection{MIMO beamforming, reference microphone selection, and post-filtering}
\label{sec:se:bf}

This section describes in detail the three modules shown in Figure~\ref{fig:se}:
MIMO beamforming (Sections~\ref{sec:se:r1-mwf} and~\ref{sec:se:sp-mwf}),
reference microphone selection (Section~\ref{sec:se:bf:refch}),
and post-filtering (Section~\ref{sec:se:bf:post-filter}).
While the baseline system implements R1-MWF~\cite{souden2009optimal} for the MIMO beamforming, we propose replacing it with \textit{Spatial-Prediction MWF (SP-MWF)}~\cite{cornelis2011performance,benesty2008noncausal} as described in Section~\ref{sec:se:sp-mwf}.
As we will see, although this change to the MIMO beamformer only affects the frequency-wise amplitude of the MIMO beamforming output signal, it leads to the selection of a different reference microphone in the reference microphone selection module, which may be crucial in distributed microphone scenarios.
We discuss this in more detail in Section~\ref{sec:se:bf:discussion}.

\subsubsection{Signal model}
\label{sec:se:bf:signal}

With a slight abuse of notation, let
$\vec{y}(f,t) \in \C^K$ 
with frequency bin index $f$ and time frame index $t$
be the STFT representation of the $K$-channel dereverberated signal corresponding to
utterance segment $\mathbf{b}_{s,j} = [t^{\text{start}}_{s,j}, t^{\text{end}}_{s,j}]$
in the time domain.
%the output of the MIMO WPE module (cut out over segment $\mathbf{b}_{s,j} = [t^{\text{start}}_{s,j}, t^{\text{end}}_{s,j}]$), i.e.,
Suppose that the input signal $\vec{y}(f,t)$ is the superposition of
the source image of the target source $s$, denoted as $\xs(f,t)$,
and a noise-plus-interference component (hereafter noise), denoted as $\xn(f,t)$
\begin{align}
  \vec{y}(f,t) = \xs(f,t) + \xn(f,t) \in \C^K.
\end{align}
The goal of the MIMO beamforming is to estimate
$K$ separation filters, $\vec{w}_r(f) \in \C^K$ ($r = 1,\ldots,K$),
or
\begin{align}
  \mat{W}(f) = [\, \vec{w}_1(f), \ldots, \vec{w}_K(f) \,] \in \C^{K \times K}
\end{align}
in a matrix form so that
the $K$-channel separated signal
\begin{align}
  \hatxs(f,t) = \mat{W}(f)^\adj \xs(f,t) \in \C^K
\end{align}
is as close as possible to the target source component $\xs(f,t)$,
where $\empty^\adj$ represents the Hermitian transpose.
To estimate $\mat{W}(f)$, the beamforming module exploits two TF masks provided by the GSS module:
one for the target-plus-noise bins, denoted as $\alpha_{\mathrm{s}}(f,t)$,%
\footnote{
Here, strictly speaking, we cannot isolate the ambient noise from the target speech, and $s$ should be understood as target-plus-noise.
}
and the other for the noise-only bins, denoted as $\alpha_{\mathrm{n}}(f,t)$.
When considering the TF masks, we assume that the noise signal is present in all TF bins.
In what follows, we introduce several MIMO beamformers and explain how they are obtained using the TF masks.
We often omit the indices $f$ and $t$ to simplify the notations without causing confusion.

To explain MIMO beamformers, we first introduce some notation.
Let $\stvec(f) \in \C^K$ be the acoustic transfer function (ATF) or steering vector of the target source $s$ to the $K$ microphones,
and let
\begin{align}
  \Rs \equiv \E[ \xs \xs^\adj ] \in \C^{K \times K}
  \quad
  \text{and}
  \quad
  \Rn \equiv \E[ \xn \xn^\adj ] \in \C^{K \times K}
\end{align}
be the covariance matrices of the target source and the noise source, respectively.
In practice, two covariance matrices $\Rs$ and $\Rn$ are often estimated by weighted averaging using the TF masks:%
\footnote{
    Strictly speaking, $\Rs$ estimated by Eq.~\eqref{eq:Rs} is the covariance matrix of the target-plus-noise component.
    Nevertheless, both the baseline and our SE systems use this $\Rs$ as the target source covariance since this is common practice in the ASR community to avoid distortions in the estimated $\Rs$ at the expense of accurately estimating it.
    %For $\Rs$, both the baseline and our systems substitute $\Rs$ with $\Ry$, i.e., $\Rs \gets \Ry$.
}
\begin{align}
\label{eq:Rs}
    \mat{R}_{\mathrm{\nu}}(f) &=
    \frac{1}{ \sum_{t \in \mathcal{T}} \alpha_{\mathrm{\nu}}(f,t) }
    \sum_{t \in \mathcal{T}} \alpha_{\mathrm{\nu}}(f,t) \vec{x}(f,t) \vec{x}(f,t)^\adj,
    \quad
    \mathrm{\nu} \in \{ \mathrm{s}, \mathrm{n} \},
\end{align}
where $\mathcal{T}$ denotes the set of time frames corresponding to the considered segment,
$\mathbf{b}_{s,j} = [t^{\text{start}}_{s,j}, t^{\text{end}}_{s,j}]$, in the time domain.
In the ideal case, $\Rs$ is of rank 1 and has the following relation~\textcolor{red}{\cite{gannot2017consolidated}}:
\begin{align}
    \label{eq:Rs:rank1}
    \Rs = P_{\mathrm{s}} \stvec \stvec^\adj,
\end{align}
where $P_{\mathrm{s}} \in \R_{\geq 0}$ is the power spectral density of the target source.
\textcolor{red}{
Note, however, that in reverberant conditions where the window size of the STFT is shorter than the reverberation time, $\Rs$ can have a rank greater than 1.
Note also that
}
$\Rs$ estimated as in Eq.~\eqref{eq:Rs} is generally not of rank 1.

\subsubsection{MVDR and R1-MWF}
\label{sec:se:r1-mwf}

The MVDR beamformer with reference microphone $r \in \{ 1, \ldots, K \}$ is \textcolor{red}{
given as~\cite{gannot2017consolidated}
\begin{align}
    \notag
    \vec{w}_r
    &=
    \operatorname*{argmin}_{\vec{w}_r}
    \{
        \vec{w}_r^\adj \Rn \vec{w}_r
        \mid
        \vec{w}_r^\adj \vec{z} = z_r
    \}
    \\
    \label{def:mvdr}
    &=
    \frac{
        \Rn^{-1} \stvec
    }{
        \stvec^\adj \Rn^{-1} \stvec
    } z_r^*, 
\end{align}}
where $z_r \in \C$ denotes the $r$th element of $\stvec$ and $z_r^*$ is its complex conjugate.
This form of the MVDR beamformer estimates the target source as observed at the $r$th microphone.
If the relation in Eq.~\eqref{eq:Rs:rank1} holds, Eq.~\eqref{def:mvdr} is also expressed as \cite{benesty2008noncausal,souden2009optimal}
\begin{align}
  \label{def:mvdr:souden}
    \vec{w}_r
    =
    \frac{
        \Rn^{-1} \Rs \unitvec_r
    }{
        \trace(\Rn^{-1} \Rs)
    },
\end{align}
where $\trace(\cdot)$ is the trace of a matrix, and $\unitvec_r \in \C^K$ denotes the unit vector whose $r$-th element is one and all other elements are zero.
The latter expression of Eq.~\eqref{def:mvdr:souden} is beneficial since it allows us to directly use a general rank covariance matrix $\Rs$ instead of estimating ATF $\stvec$.

The MVDR formula in Eq.~\eqref{def:mvdr:souden} can also be derived from the so-called speech-distortion-weighted multichannel Wiener filter (SDW-MWF)~\cite{doclo2002gsvd,spriet2004spatially,doclo2007frequency}.
SDW-MWF is defined as a solution of the following optimization problem:
\begin{alignat}{2}
  &\operatorname*{minimize}_{ \vec{w}_r } & \quad &
  \E[\, | \textcolor{red}{ \vec{w}_r^\adj \xs - \unitvec_r^\top \xs } |^2 \,]
  + \gamma \E[\, | \vec{w}_r^\adj \xn |^2 \,].
\end{alignat}
A hyperparameter $\gamma \geq 0$ is introduced to balance the trade-off between speech distortion and noise reduction in the separated signal.
With the assumption that $\xs$ and $\xn$ are uncorrelated, i.e., $\E[\, \xs \xn^\adj \,] = 0$,
the problem is solved as
\begin{align}
  \label{eq:sdw-mwf}
  \mathbf{w}_r = ( \Rs + \gamma \Rn) ^{-1} \Rs \unitvec_r.
\end{align}
Souden, Benesty, and Affes~\cite{souden2009optimal} pointed out that, when $\Rs$ is of rank 1,
\eqref{eq:sdw-mwf} becomes equivalant to
\begin{align}
  \label{def:r1-mwf}
  \vec{w}_r
  =
  \frac{
      \Rn^{-1} \Rs \unitvec_r
  }{
      \gamma + \trace(\Rn^{-1} \Rs)
  }.
\end{align}
This beamforming formula in Eq.~\eqref{def:r1-mwf} is referred to as 
the rank-1 MWF (R1-MWF)
or
the parameterized MWF (PMWF).
When $\gamma = 0$, R1-MWF reduces to the MVDR beamformer as implemented in~\eqref{def:mvdr:souden}.
The baseline SE system implements R1-MWF (or PMWF) in the MIMO beamforming module with the default setting of $\gamma = 0$.

\subsubsection{Spatial-prediction MWF (SP-MWF)}
\label{sec:se:sp-mwf}

SP-MWF~\cite{cornelis2011performance}, which we implement in our MIMO beamforming module, is equivalent to R1-MWF and SDW-MWF when $\rank \Rs = 1$.
It is defined as
\begin{align}
  \label{def:sp-mwf}
  \mathbf{w}_r
  = 
  \frac{
      (\unitvec_r^\top \Rs \unitvec_r)
      \Rn^{-1} \Rs \unitvec_r
  }{
      \gamma \unitvec_r^\top \Rs \unitvec_r
      +
      \unitvec_r^\top \Rs \Rn^{-1} \Rs \unitvec_r
  }
  =
  \frac{
      \Rn^{-1} \Rs \unitvec_r
  }{
      \gamma
      +
      \unitvec_r^\top \Rs \Rn^{-1} \Rs \unitvec_r
      /
      \unitvec_r^\top \Rs \unitvec_r
  }.
\end{align}
The equivalence under $\rank \Rs = 1$ or $\Rs = P_{\mathrm{s}} \stvec \stvec^\adj$ can be confirmed by checking that
the second term of the denominator in Eq.~\eqref{def:sp-mwf}
is equivalent to that of R1-MWF in Eq.~\eqref{def:r1-mwf}, as
\begin{align}
  \frac{
    \unitvec_r^\top \Rs \Rn^{-1} \Rs \unitvec_r
  }{
    \unitvec_r^\top \Rs \unitvec_r
  }
  = \frac{ P_{\mathrm{s}}^2 | z_r |^2 \stvec^\adj \Rn^{-1} \stvec }{ P_{\mathrm{s}} | z_r |^2 }
  = P_{\mathrm{s}} \stvec^\adj \Rn^{-1} \stvec
  = \trace(\Rn^{-1} \Rs).
\end{align}
Clearly, SP-MWF differs from R1-MWF in Eq.~\eqref{def:r1-mwf} when $\rank \Rs \geq 2$, but even in this case,
they are proportional, meaning that they are the same except for the scale, or norm $\| \vec{w}_r(f) \|_2$, at each frequency bin,
as confirmed by checking that the difference in Eqs.~\eqref{def:r1-mwf} and \eqref{def:sp-mwf} is only in the scalar denominator.

To understand how the scale of SP-MWF is determined, let us illustrate an essential relationship between SP-MWF and a certain implementation of the MVDR beamformer.
Since $x_{\mathrm{s},r} \coloneqq \unitvec_r^\top \xs$ is the clean source signal as observed at microphone $r$,
the \textit{relative transfer function (RTF)} of the target source can be estimated as the solution of
\begin{align}
  \label{P:sp:sv}
  \operatorname*{minimize}_{\stvec \in \C^K }
  \quad
  \E \mleft[\, \| \xs - ( \unitvec_r^\top \xs ) \stvec \|_2^2 \,\mright],
\end{align}
which is solved as
\begin{align}
  \label{def:sp:sv}
  \stvec =
  \frac{
      \Rs \unitvec_r 
  }{
      \unitvec_r^\top \Rs \unitvec_r 
  }
  \in \C^K.
\end{align}
%Observe that it satisfies $z_r \coloneqq \unitvec_r^\top \stvec = 1$.
Then, substituting $\stvec$ in Eq.~\eqref{def:sp:sv} into the MVDR formula of Eq.~\eqref{def:mvdr}, we obtain
\begin{align}
  \label{def:distortionless-MWF}
  \mathbf{w}_r
  =
  \frac{
    (\unitvec_r^\top \Rs \unitvec_r)
    \Rn^{-1} \Rs \unitvec_r
  }{
    \unitvec_r^\top \Rs \Rn^{-1} \Rs \unitvec_r
  }
  =
  \frac{
    \Rn^{-1} \Rs \unitvec_r
  }{
    \unitvec_r^\top \Rs \Rn^{-1} \Rs \unitvec_r
    /
    \unitvec_r^\top \Rs \unitvec_r
  },
\end{align}
which is nothing but SP-MWF with $\gamma = 0$.
The beamformer of Eq.~\eqref{def:distortionless-MWF}, referred to as Distortionless MWF, was developed in \cite{benesty2008noncausal} and extended to SP-MWF in \cite{cornelis2011performance}, where the hyperparameter $\gamma \geq 0$ was introduced as in R1-MWF.

As we mentioned, R1-MWF and SP-MWF are identical when $\rank \Rs = 1$ and proportional even when $\rank \Rs \geq 2$.
Although they differ only in the norm $\| \vec{w}_r(f) \|_2$ at each frequency bin, this difference may significantly affect the output of the reference microphone selection module.
We will return to this topic in Section~\ref{sec:se:bf:discussion} after reviewing the reference microphone selection and post-filtering modules in Sections~\ref{sec:se:bf:refch} and~\ref{sec:se:bf:post-filter}, respectively.

\subsubsection{Reference microphone selection}
\label{sec:se:bf:refch}

After optimizing an MIMO beamformer $\mat{W}(f)$, or $\vec{w}_r(f)$ for each microphone $r \in \{ 1,\ldots,K \}$,
a reference microphone $r_\star$, considered to be the most effective, is chosen as the one that maximizes the output SNR of the beamforming output signal \cite{lawin2012reference,erdogan2016improved}:
\begin{align}
    \label{eq:se:refch}
    r_\star = \operatorname*{argmax}_{ r \in \{ 1,\ldots,K \} }
    \frac{
        \sum_f \vec{w}_r(f)^\adj \Rs(f) \vec{w}_r(f) 
    }{
        \sum_f \vec{w}_r(f)^\adj \Rn(f) \vec{w}_r(f) 
    }.
\end{align}
The output of the reference microphone selection module is
\begin{align}
  \label{eq:bf-output}
  \hat{x}_{\mathrm{s}, r_\star}(f,t) = \vec{w}_{r_\star}(f)^\adj \vec{y}(f,t) \in \C,
\end{align}
which is then passed to the final post-filtering module.

\subsubsection{post-filtering}
\label{sec:se:bf:post-filter}

The post-filtering module in the baseline system receives $\hat{x}_{\mathrm{s},r_\star}(f,t)$ in Eq.~\eqref{eq:bf-output} and applies the BAN post-filter~\cite{warsitz2007blind} to it:
\begin{align}
  \hat{x}_{\mathrm{s}, r_\star}(f,t)
  \gets
  \frac{
    \sqrt{ \vec{w}_{r_\star}(f)^\adj \Rn(f) \Rn(f) \vec{w}_{r_\star}(f) }
  }{
    \vec{w}_{r_\star}(f)^\adj \Rn(f) \vec{w}_{r_\star}(f)
  }
  \hat{x}_{\mathrm{s}, r_\star}(f,t).
\end{align}
Note that, unlike the baseline system, our SE system does not apply the BAN post-filter, as it is not effective when used in conjunction with SP-MWF.
We refer readers to~\ref{sec:ablation:bf} for the corresponding ablation study.

After that, the post-filtering module applies TF masking to further enhance the signal:
\begin{align}
\label{eq:se:output}
  \hat{x}_{\mathrm{s}, r_\star}(f,t)
  \gets
  \max \{ \alpha_{\mathrm{s}}(f,t), \delta \}
  \hat{x}_{\mathrm{s}, r_\star}(f,t).
\end{align}
Here, the TF mask $\alpha_{\mathrm{s}}(f,t)$ is floored to prevent excessive enhancement that could cause speech distortion.
The flooring coefficient $0 \leq \delta \leq 1$ is set to
$- 9$ (dB), or $\delta \approx 0.355$, as a default.

\subsubsection{SP-MWF vs.\ R1-MWF in the SE system}
\label{sec:se:bf:discussion}

We explain the key difference between SP-MWF and R1-MWF when they are used in the SE system shown in Figure~\ref{fig:se}.
As described in Section~\ref{sec:se:sp-mwf}, these two beamformers are proportional and differ only in the norm $\| \vec{w}_r(f) \|_2$ at each frequency bin.
This difference affects the following two aspects in the final output of the SE system, $\hat{x}_{s, r_\star}(f,t)$ at Eq.~\eqref{eq:se:output}:
\begin{itemize}
  \item The reference microphone $r_\star \in \{ 1,\ldots, K \}$, selected based on Eq.~\eqref{eq:se:refch}.
  \item The amplitude variation of the separated signal $\hat{x}_{\mathrm{s},r^\star}(f,t)$ across frequency bins, even when the same reference microphone is chosen for both SP-MWF and R1-MWF.
\end{itemize}
In the reference microphone selection rule in Eq.~\eqref{eq:se:refch},
the norm of the filter, $\| \vec{w}_r(f) \|_2^2$, determines the frequency-wise weight of the separated components when calculating the output SNR.
As a result, the selected microphone $r_\star$ may frequently differ depending on whether SP-MWF or R1-MWF is used for the MIMO beamformer $\vec{w}_r(f)$.
In a distributed microphone scenario, such as the one considered in this CHiME challenge, selecting an effective microphone, e.g., a microphone near the target source with a high SNR, can substantially improve the quality of the separated signal compared to selecting an ineffective microphone.
Although determining which beamformer is theoretically more effective for reference microphone selection remains an open question, we empirically demonstrate in \ref{sec:ablation:bf} that SP-MWF outperforms R1-MWF on the development set of this challenge.

It is worth noting that SP-MWF and R1-MWF differ only in the reference microphone selection if we apply the BAN post-filter to both.
%To better understand how we can compare SP-MWF and R1-MWF for reference microphone selection,
To see this,
let us consider the case where the same reference microphone $r^\star$ is selected for both beamformers, and the BAN post-filter is applied in the post-filtering module.
In this case, the output of the SE system is expressed as
\begin{align}
  \hat{x}_{\mathrm{s},r_\star}(f,t)
  =
  \underbrace{
    \max \{ \alpha_{\mathrm{s}}(f,t), \delta \}
    \vphantom{
      \frac{
        \sqrt{ \vec{w}_{r_\star}(f)^\adj \Rn(f) \Rn(f) \vec{w}_{r_\star}(f) }
      }{
        \vec{w}_{r_\star}(f)^\adj \Rn(f) \vec{w}_{r_\star}(f)
      }
    }
  }_{\text{TF masking}}
  \cdot
  \underbrace{
    \frac{
      \sqrt{ \vec{w}_{r_\star}(f)^\adj \Rn(f) \Rn(f) \vec{w}_{r_\star}(f) }
    }{
      \vec{w}_{r_\star}(f)^\adj \Rn(f) \vec{w}_{r_\star}(f)
    }
  }_{\text{BAN post-filtering}}
  \cdot
  \underbrace{
    \vec{w}_{r_\star}(f)^\adj \vec{y}(f,t)
    \vphantom{
      \frac{
        \sqrt{ \vec{w}_{r_\star}(f)^\adj \Rn(f) \Rn(f) \vec{w}_{r_\star}(f) }
      }{
        \vec{w}_{r_\star}(f)^\adj \Rn(f) \vec{w}_{r_\star}(f)
      }
    }
  }_{\text{beamforming}},
\end{align}
which is independent of the scale of the beamformer $\vec{w}_{r_\star}(f)$ due to the BAN post-filter.
This implies that when the BAN post-filter is applied, the performance difference between SP-MWF and R1-MWF arises solely from the difference in reference microphone selection. 
We will demonstrate this in the ablation study in \ref{sec:ablation:bf}.

\subsection{Chunk-wise SE diarization}
\label{sec:se:chunkwiseSE}
%\textcolor{red}{\textbf
%{Written by Marc. Need to double-check by Tawara-san and Ikeshita-san}
The SE processing we used within the diarization module to obtain better speaker embeddings, as explained in Section \ref{ssec:eend_vc}, is shown in Figure~\ref{fig:diar_se}. It differs slightly from the SE processing explained above and shown in Figure~\ref{fig:se}. There are two main differences, GSS is performed chunk-wise, i.e., it is given the local speaker activity computed by the EEND model from each chunk, instead of utterance-wise, and therefore relies on estimated utterance boundaries as described in Eq.~\eqref{eq:se:system}. Note that no context is used (i.e. $\Delta =0$), making it different from the utterance-wise SE approach.
The other difference is that we perform GSS $M$-times, which each time given the chunk activity estimated from a different microphone. This results in having $M$ TF-masks. Then, MIMO beamforming and reference channel selection (without post-filtering) are performed $M$ times independently, each time with a different mask. 
%We used the same hyperparameters as in SE processing, except that no flooring was applied to the TF mask in Eq.~\eqref{eq:se:output}.
We utilized the same hyperparameters as those used by the proposed SE system $\operatorname{SE}(\cdot)$,
except for $\delta = 1$ in Eq.~\eqref{eq:se:output}, meaning that no masking was applied to the beamforming output signal $\hat{x}_{\mathrm{s},r_\star}$ given by Eq.~\eqref{eq:bf-output}.
%}

The output of the chunkwise SE module thus consists of $M$ enhanced signals, which allows the computation of enhanced speaker embeddings for each local speaker and each microphone. We analyze the effect of SE on diarization and speaker counting in Appendices \ref{ssec:ablation_model} and \ref{ssec:ablation_counting}.

%------------------------------------------------------
%------------------------------------------------------
%%% ASR %%%%
\section{Speech recognition backend}
\label{sec:asr}

\begin{figure}[t]
  \centering
\includegraphics[width=0.75\linewidth]{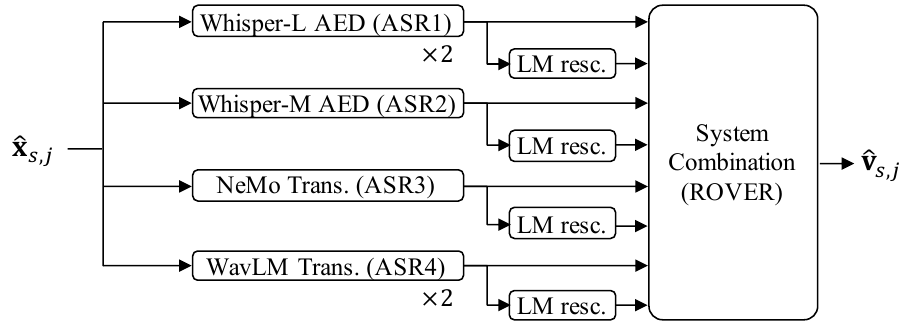}
  %\vspace{-5mm}
  \caption{Proposed ASR backend.}
  %\vspace{-5mm}
  \label{fig:asr}
\end{figure}

In this section, we describe how we implemented our ASR backend (``$\text{ASR}(\cdot)$'' in Eq. (\ref{eq:asr})). Figure~\ref{fig:asr} shows the schematic diagram of our ASR backend, which uses four different ASR models, performs language model (LM) for restoring, and combines their output with system combination. The details of the ASR models, the training data used for their training, the language model (LM) used for rescoring, and the system combination approach are described below.

\subsection{Training data for ASR backend}
We used 70 hours of CHiME-8 training data \textcolor{red}{released by the challenge organizers\cite{cornell2024chime} (including data from CHiME-6, DiPCo, Mixer 6, and NOTSOFAR-1 (only the real recordings))} processed with GSS for the Oracle segmentation, which is commonly used for building ASR models (ASR1-4).
We did not use train\_call and train\_intv in Mixer 6 at this time because preliminary experiments showed minimal improvements with them.
For pre-training of the transducer-based ASR model with WavLM Large (ASR4), we also used LibriSpeech \cite{panayotov2015librispeech} and VoxCeleb1+2 \cite{voxceleb} datasets in addition to the CHiME-8 described above. 
The development set was utilized for measuring token accuracy or training loss for early stopping.

\subsection{ASR models}
\label{ssec:asr_models}
We implemented four ASR models: two attentional encoder-decoder (AED) models~\cite{vaswani2017att} and two transducer models~\cite{graves2012rnnt}. For building strong ASR systems, we adopt several pre-trained models: 1) two different sizes of Whisper-based AED models (ASR1/ASR2), 2) the official pre-trained transducer model based on the CHiME-8 DASR NeMo baseline system (ASR3), and 3) the self-supervised learning (SSL) model, i.e., WavLM~\cite{WavLM}, for the encoder of the transducer model (ASR4). The following describes the detailed model structure and training pipeline of each ASR model.

\subsubsection{Whisper Large v3 (ASR1)}
\label{sssec:ASR1}
Whisper Large v3 model~\cite{radford2022whisper} is a strong ASR model trained on a large amount of data. It consists of  1540M parameters, i.e., a 32-layer Transformer encoder and decoder, each with 8 attention heads and a hidden dimension of 1280. It utilizes the byte-level byte-pair encoding (BPE)~\cite{sennrich2016bpe} tokenizer from  GPT-2~\cite{radford2019language} and has a vocabulary size of 51,864 tokens.
We fine-tuned all parameters of the Whisper Large v3 model using the CHiME-8 training data, with an early stopping set to a patience of 5,000 steps.

To obtain an optimal performance on the CHiME-8 task, we fine-tuned the Whisper Large v3 model on the CHiME-8 training data. However, since this data is very noisy and part of it is unreliable for fine-tuning,  we also adopted a curriculum learning scheme. Our idea is to filter out utterances that have a high character error rate (CER) and can thus be considered unreliable.
Here, we computed the CER on the fly during training. This means that we can gradually increase the number of training utterances as fine-tuning progresses, the model becomes stronger and CER decreases.

To implement this curriculum learning practically, we changed the transcription reference to the self-generated decoding result when the CER exceeds 30~\%, and multiplied the loss by a small value of $10^{-3}$  to reduce its impact. 

In preliminary experiments, we found that the proposed scheme could stabilize training. However, we could not investigate its impact deeply due to the limited time required to develop the challenge system and the cost of fine-tuning the Whisper Large model.
This model turned out to be our best ASR model, as shown in the ablation study in \ref{sec:ablation:asr}(Table \ref{tab:asr_ablation}).

\subsubsection{Whisper Medium (ASR2)}
\label{sssec:ASR2}
The Whisper Medium English model\cite{radford2022whisper}  is another strong model available for the challenge. It has significantly fewer parameters than Whisper Large v3 but is trained only on English data. Therefore, it is also potentially strong for CHiME-8.

The Whisper Medium English model has approximately 770M parameters and consists of a 24-layer Transformer encoder and decoder, each with 8 attention heads and a model width of 1024. The large model utilizes the byte-level BPE tokenizer of GPT-2 and has the same vocabulary size of 51,864 tokens.

We fined-tuned all model parameters using the CHiME-8 training data, with an early stopping set to a patience of 5,000 steps.
Note that this model was created in the early development of our system without the curriculum learning scheme of ASR1. As shown in our ablation study (\ref{sec:ablation:asr}), it performed worse than ASR1 but was still effective for system combination.

\subsubsection{NeMo Transducer (ASR3)} 
\label{sssec:ASR3}
The CHiME baseline utilizes a NeMo transducer model \cite{nemo} for the ASR backend. This model has 644M parameters, i.e., two-layer 2D convolutional CNNs followed by 24 fast Conformer blocks \cite{anmol2020conformer,rekesh2023fastconformer}.  The prediction and joint networks consist of a 640-dimensional long short-term memory (LSTM) and a 640-dimensional feed-forward network.
The vocabulary consists of 1025 BPE tokens. 

The NeMo model was trained on a large dataset.
Similar to ASR1 and ASR2 models, we fine-tuned all parameters of the NeMo transducer on the CHiME-8 training data, with an early stopping set to a patience of 5 epochs.

\subsubsection{WavLM Transducer (ASR4)} 
\label{sssec:ASR4}
For the fourth model, we developed a WavLM-based transducer model.
The input feature of the model consists of the weighted sum of the output of the Transformer layers of the WavLM Large model.
For the encoder, we use two 2D-CNN layers followed by 18 branchformer blocks \cite{kim2022branchformer}. The prediction and joint networks consist of two-layer 640-dimensional LSTMs and 512-dimensional feed-forward layers, respectively. The vocabulary size consists of 500 BPE tokens. This model has the smallest size among our ASR models and consists of approximately 422M parameters.

Although the other ASR systems (ASR1-3) and the WavLM pre-encoder are pre-trained using a large amount of data, ASR4 contains randomly initialized parameters, i.e., the branchformer encoder, prediction, and joint networks, which are trained from scratch. These network parameters should also be trained on a large amount of data. For this model training, we used supervised data from CHiME-8 and LibriSpeech, as well as unsupervised data from the vast amount of VoxCeleb1+2 dataset~\cite{voxceleb}. We utilized VoxCeleb1+2, transcribed by a publicly available pre-trained NeMo CTC model, as semi-supervised training data. To discard noisy data from the large VoxCeleb1+2 dataset, we performed data selection for efficient training. We also conducted \textit{multi-step training}, which dramatically improved the performance of the integration of the WavLM and transducer model. \textcolor{red}{ We describe the \textit{data selection} and \textit{multi-step training} processes below.}

\begin{itemize}
    \item \textbf{Data selection} The contrastive data selection algorithm~\cite{lu22_interspeech,kamo23_chime} is applied to the VoxCeleb1+2 data. This algorithm utilizes a general LM and a target LM to select data with high domain relevance scores on the target domain and low scores on the general domain. Each LM consists of a 5-gram LM with Kneser-Ney smoothing, trained using discrete acoustic units on the general (VoxCeleb1+2) and target (CHiME-8) domain datasets, respectively. In this work, to generate the acoustic units, we discretized the 21st WavLM layer output instead of using the w2v-BERT codebook~\cite{lu22_interspeech}. For the quantization, we run the k-means algorithm with 1024 clusters. We calculated the domain relevance score by measuring the probability differences between the target- and general-domain LMs. 

    \item \textbf{Multi-step training:} In previous work~\cite{kamo23_chime}, we found a novel training scheme that consists of multiple training steps. We conducted three-step training to build the ASR4 system sequentially: 1) partial parameters were trained using the CHiME \& LibriSpeech \& VoxCeleb1+2 datasets while freezing the WavLM frontend, 2) all network parameters (including WavLM) were fine-tuned using the same data from the first step, and 3) we fine-tuned it using only the CHiME-8 data. Note that we used early stopping with patience of 5 epochs for all training steps.
\end{itemize}

As described in~\cite{kamo23_chime}, applying \textit{data selection} reduced the VoxCeleb1+2 dataset to a quarter through experimentation using a threshold hyperparameter, while simultaneously enhancing recognition performance. \textit{Multi-step training} achieved recognition performance improvements compared to single-step training~\cite{kamo23_chime}.

The ASR4 model achieved a competitive performance with the larger ASR1 model, as shown in our experiments in \ref{sec:ablation:asr}.
\subsection{LM rescoring and system combination (ASR5)} 
\label{sec:lm_rescoring}
In the decoding step, we generated N-best hypotheses for each ASR system and performed LM rescoring using external LM. For each ASR system, we chose the hypothesis with the highest score after beam search decoding as the best hypothesis. 
Note that both the beam size and N-best size are set to 4.
Each hypothesis of the $j$-th utterance of speaker $s$, generated from each ASR backend with LM rescoring, is determined as follows:
\begin{eqnarray}
%  \vspace{-0.2cm}
  \hat{\mathbf{v}}_{s,j} = \argmax_{\mathbf{v}_{s,j}} \Big\{ \text{log} \ p_{\text{ASR}}(\mathbf{v}_{s,j}| \hat{\mathbf{x}}_{s,j}) +  \beta_{1} \ \text{log} \ p_{\text{LM}}(\mathbf{v}_{s,j}) + \beta_{2} \ |\mathbf{v}_{s,j}| \Big\},
  \label{eq:hypilm}
%  \vspace{-0.2cm}
\end{eqnarray}
where $\hat{\mathbf{x}}_{s,j}$ is the input speech obtained from the SE module, $\hat{\mathbf{v}}_{s,j}$ is the hypothesis, $\text{log} \ p_{\text{ASR}}(\mathbf{v}_{s,j}|\hat{\mathbf{x}}_{s,j})$ is the ASR model score for $\mathbf{v}_{s,j}$ given $\hat{\mathbf{x}}_{s,j}$, which corresponds to Eq. (\ref{eq:asr}). 
$\text{log} \ p_{\text{LM}}(\mathbf{v}_{s,j})$ indicates the LM score for $\mathbf{v}_{s,j}$, $\beta_1$ is the weight for the LM,
$\beta_2$ is the reward given proportional to the length of $\mathbf{v}_{s,j}$
(the weight $\beta_1$ and reward $\beta_2$ were tuned using the development set),
and $\hat{\mathbf{v}}_{s,j}$ is the best hypothesis.

We built a Transformer-LM consisting of 35M parameters for LM rescoring. The LM has the vocabulary of 1000 BPE tokens. We pre-trained the LM using 1/10 of the LibriSpeech text dataset and then fine-tuned using the CHiME-8 train text dataset.
At the inference,
%the LM uses
we used
256 past rescored (re-ranked) 1-best tokens as the context
%(i.e., context carry-over)
to rescore the current hypotheses
(i.e., context carry-over)
\cite{Ogawa_ICASSP2022,Ogawa_arXiv2024}. 

We built six different ASR models (the four models described above and two versions of ASR1 and ASR4 with different training setups). These systems vary in architecture, training procedure, and data. Therefore, they show different error patterns. To improve performance, we performed the system combination of the original 1-best hypotheses of the four ASR systems (ASR5) and their rescored versions by an external LM, using ROVER \cite{Fiscus1997ROVER}, as shown in Figure~\ref{fig:asr}. We discuss the effectiveness of the system combination in the experiments in \ref{sec:ablation:asr}.

%ROVER~\cite{Fiscus1997ROVER} is indeed a method used for combining recognition outputs from multiple speech recognizers to enhance overall accuracy. In this process, multiple hypotheses generated by different recognizers are aligned using dynamic programming, where insertion or deletion positions are marked with a blank symbol ``*'', to ensure temporal alignment across all hypotheses. After alignment, a voting mechanism is applied at each time step, selecting the word that occurs most frequently among the recognizer outputs as the final ``answer.'' This consensus-based approach typically yields higher accuracy by mitigating the weakness of each recognizer to recover individual errors.

\section{Related works}
We briefly summarize the STCON CHiME-8 system \cite{stcon_chime8}, which achieved the top performance on the DASR task.
STCON follows a diarization-first pipeline and also uses TS-VAD.

\textbf{Diarization:}
For pre-diarization, they use VAD to find speech activity combined with a more advanced speaker embedding extractor, which provides as many embeddings as speakers in a chunk. These embeddings are combined with ECAPA-TDNN embeddings. For clustering, they used GMM-based clustering, which is also utilized for speaker counting. To further refine speaker counting, they developed a Wav2Vec2.0-based classifier to detect if speech from a cluster corresponds to non-speech and should thus be discarded. Microphone selection is performed before diarization.

For TS-VAD training, they generated room impulse responses with similar characteristics as the CHiME-8 data and generated 2500 hours of training data. They also used refined utterance boundaries obtained by forced alignment on the CHiME-8 training data as a target.

They generated 12 different diarization pipelines for each microphone, with variations of the pre-diarization, and fused their results. They also created seven variations of the TS-VAD model, applied it to WPE-processed and non-processed speech, and combined their outputs.

\textbf{SE:} In addition to GSS, they also developed a target speech extraction model, which was optimized first with a scale-invariant signal-to-distortion ratio (SiSDR) loss and then fine-tuned jointly with an ASR backend. The TSE module uses the TF mask obtained from GSS as a clue to determine the target speaker, and processing is performed on the GSS output.

%\textbf{ASR:} They developed several backends, including some DNN-HMM hybrid ASR models trained with Kaldi \cite{povey2011kaldi}, and at least three end-to-end ASR models using WavLM as input features, including Uconv-Conformer \cite{andrusenko2023uconvformer} and an E-Branchformer \cite{kim2022branchformer} trained with ESPnet \cite{watanabe2018espnet} and a ZipFormer \cite{yao2024zipformer} trained with K2 \footnote{\url{https://github.com/k2-fsa}}. They performed ASR on two versions of the enhanced speech using several backends and combined their results. Besides, they also used session-wise unsupervised adaptation.

\textbf{ASR:}: Several backends has been developed, including 1) TDNN-based acoustic models within DNN-HMM hybrid ASR systems trained using Kaldi \cite{povey2011kaldi}; 2) three AED models, specifically Uconv-Conformer \cite{andrusenko2023uconvformer}, E-Branchformer \cite{kim2022branchformer}, and Whisper, all trained with ESPnet \cite{watanabe2018espnet}; and 3) a Transducer model utilizing the Zipformer encoder \cite{yao2024zipformer}, trained with K2\footnote{\url{https://github.com/k2-fsa}}. All of these ASR systems utilized WavLM Large as input features. ASR was performed on two versions of the enhanced speech using these backends, and their results were combined. They also used session-wise unsupervised adaptation.

To summarize, the major difference of our system compared to the above is its use of more advanced speaker embedding extraction, which may achieve better speaker counting. The SE part uses an additional neural TSE frontend. Finally, they combine many diarization systems and many ASR systems to obtain their final results.

% \textcolor{red}{Maybe we could discuss the difference between our paper and the best system for DASR and NOTSOFAR-1 in the experiments.\\
% If needed, we could also describe the baseline systems here if it makes it easier for the explanations of SE part, eg.}

%%%%%%%%%%%%%
%------------------------------------------------
%------------------------------------------------
\section{Experiments}
\label{sec:exp}
In this section, we report the results on the CHiME-8 development and evaluation sets used during the challenge. The test set comprises four datasets: CHiME-6, DiPCo, Mixer 6, and NOTSOFAR-1. 
CHiME-6 and DiPCo consist of four-speaker home-party recordings of about 2.5 hours and 30 minutes, respectively. Mixer 6 data are 2-speaker interviews that are 15 minutes long, and NOTSOFAR-1 data are short 6-minute meetings of 4 to 8 participants recorded in office meeting rooms. We refer to the challenge overview paper for more details about the data \cite{cornell2024chime}.

\subsection{Evaluated systems}
We report the results with different versions of our system, which we compare to the NeMo baseline and the STCON system \cite{stcon_chime8}.

For diarization, we tested a system using EEND-VC without TS-VAD refinement (DIA1), described in Section \ref{ssec:eend_vc}, and a system with TS-VAD refinement (DIA2), described in section \ref{ssec:tsvad}, which were developed for the challenge. We also experimented with a third diarization system, that uses a TS-VAD model trained with more data (DIA3). In addition, we also explored the impact of speaker counting errors by comparing the results using DIA3 with Oracle speaker counts in our diarization pipeline, which we refer to as DIA4. 

For SE, except when explicitly mentioned, all our systems utilized the same SE module described in Section \ref{sec:se}, with improved microphone subset selection and beamformer.

For ASR, we report the results using a WavLM-based transducer (ASR4) (see Section \ref{sssec:ASR4}), the lightest yet competitive ASR model; Whisper-L (ASR1), described in Section \ref{sssec:ASR4}, a larger and more powerful system; and the system combination of the output of six ASR systems (ASR1 and ASR2 plus two variants of ASR1 and ASR4 with different training hyperparameters) with and without LM rescoring (ASR5), amounting to a total of 12 systems, as described in Section \ref{sec:lm_rescoring}.

\subsection{Evaluation metrics}
The systems submitted to the CHiME-8 challenge were evaluated in terms of tcpWER \cite{MeetEval23}, which measures the WER between the reference transcription of each speaker and the hypotheses of the system, including speaker and time allocation, but allowing a 5-second margin between the reference and hypotheses.

To further analyze the results, we also report the diarization performance in terms of DER and speaker counting performance. For the latter, we report the speaker counting accuracy (SCA) and the mean absolute speaker counting error (SCE), which is the mean of the absolute difference between the estimated and actual number of speakers.

%\newpage
\subsection{Overall results}
\begin{table}[t]
  \caption{Overall results on the development set in terms of DER [\%]  ($\downarrow$) SCE ($\downarrow$) and tcpWER [\%] ($\downarrow$). DER is computed using md-eval with a collar of 0.25 sec. The best performances among all systems are indicated in \textbf{bold fonts}, and the best performances among our systems are \uline{underlined}. \ditto ~ denotes a ditto mark. All our systems use the SE frontend described in Section \ref{sec:se}.}
  %\vspace{-3mm}
  \label{tab:overall_results_dev}
  \centering
  \resizebox{\linewidth}{!}{%
  \begin{tabular}{@{}l@{\hspace{0.1cm}}l@{\hspace{0.2cm}}l@{\hspace{0.2cm}}c@{\hspace{0.1cm}}c@{\hspace{0.1cm}}cc@{\hspace{0.1cm}}c@{\hspace{0.1cm}}cc@{\hspace{0.1cm}}c@{\hspace{0.1cm}}cc@{\hspace{0.1cm}}c@{\hspace{0.1cm}}cc@{\hspace{0.1cm}}c@{\hspace{0.1cm}}c@{}}
  %{@{}llc@{\hspace{0.2cm}}c@{\hspace{0.2cm}}c@{\hspace{0.2cm}}c@{\hspace{0.2cm}c@{}}
    \toprule
    && & \multicolumn{3}{c}{CHiME-6} &\multicolumn{3}{c}{DiPCo} & \multicolumn{3}{c}{Mixer 6} &\multicolumn{3}{c}{NOTSOFAR-1} & \multicolumn{3}{c}{Macro}\\
    \cmidrule(l{0pt}r{\tabcolsep}){4-6}\cmidrule(l{\tabcolsep}r{\tabcolsep}){7-9}\cmidrule(l{\tabcolsep}r{\tabcolsep}){10-12}\cmidrule(l{\tabcolsep}r{\tabcolsep}){13-15}\cmidrule(l{\tabcolsep}){16-18}
    & &&DER & SCE & tcpWER & DER & SCE & tcpWER& DER & SCE & tcpWER& DER & SCE & tcpWER& DER & SCE & tcpWER \\
    \midrule
0 & \multicolumn{2}{l}{Baseline (NeMo)}& 45.65 & 1.00 & 49.29 &  45.92 & 3.00 & 78.87 &  25.16 & \textbf{0.00} & 15.75  &  38.05 & 1.71 & 56.21 &  38.70  & 1.43 & 50.03 \\  
%%%0' & \multicolumn{3}{l}{STCON(constrained-lm)} &24.55 & 0.00 &wer&22.26 & 0.00 &wer& 6.58& 0.03 &wer& 18.63& 0.48 &wer& 18.01& 0.13 & 20.2 \\ 
0' & \multicolumn{2}{l}{STCON (sys2)} & 24.55 & \textbf{0.00} & \textbf{22.47} &  \textbf{22.26} & \textbf{0.00} & \textbf{28.41} & \textbf{6.58}& 0.03 & 9.84 & 18.63& \textbf{0.48} & 18.71 &  18.01& \textbf{0.13} & 19.86 \\ 
%0' & USTC (NOTSOFAR) &\\
\midrule
\multicolumn{18}{c}{\textit{Results submitted to the challenge}} \\
\midrule
1 & DIA1  & ASR4 & 25.05 & \textbf{0.00} & 30.14 & 23.73 & \textbf{0.00} & 35.86 & 7.66 & \textbf{0.00} & 10.94 & 15.88 & 0.51 & 23.85 & 18.08 & \textbf{0.13} & 25.20 \\
2 & DIA2  & ASR1 & 25.14 & \ditto & 28.21 & 25.50 & \ditto & 35.32 & 8.90 & \ditto & 10.66 & 14.91 & \ditto & 20.41 & 18.61 & \ditto & 23.65\\
3 & DIA2  & ASR5 %(LM+ROVER) 
& \ditto & \ditto & 25.49 & \ditto & \ditto & 31.25 & \ditto & \ditto & 9.63 & \ditto & \ditto & 18.79 & \ditto & \ditto & 21.29 \\ 
\midrule
\multicolumn{18}{c}{\textit{Post-challenge results}} \\
\midrule
4 & DIA3  & ASR1 & \textbf{\uline{22.31}} & \ditto & 25.45 & \uline{22.67} & \ditto & 32.69 & \uline{7.11} & \ditto & 10.50 & \textbf{\uline{11.45}} & \ditto & 18.88 & \textbf{\uline{15.89}} & \ditto & 21.88 \\
%%% 5 & DIA3 & SE & ASR5(ROVER) & \ditto & \ditto & 23.25 & \ditto & \ditto & 29.30 & \ditto & \ditto & 9.47 & \ditto & \ditto & 17.61 & \ditto & \ditto & 19.91 \\ %%% Move to different table
5 & DIA3  & ASR5%(LM+ROVER) 
& \ditto & \ditto & \uline{22.99} & \ditto & \ditto & \uline{29.18} & \ditto & \ditto & \textbf{\uline{9.26}} & \ditto & \ditto & \textbf{\uline{17.55}} & \ditto & \ditto & \textbf{\uline{19.74}} \\
\midrule
\multicolumn{18}{c}{\textit{Results with Oracle number of speakers}} \\
\midrule
\textit{6} & \textit{DIA4} &  \textit{ASR1} & \textit{22.31} & - & \textit{25.45} & \textit{22.67} & - & \textit{32.69} & \textit{7.11} & - & \textit{10.50} & \textit{11.02} & - & \textit{18.33} & \textit{15.78} & - & \textit{21.74} \\
\textit{7} & \textit{DIA4}  & \textit{ASR5 } & \ditto & \textit{-} & \textit{22.99} & \ditto & \textit{-} & \textit{29.18} & \ditto & \textit{-} & \textit{9.26} & \textit{\ditto} & \textit{-} & \textit{16.87} & \ditto & \textit{-} & \textit{19.58} \\
%
%\textit{7} & \textit{DIA4} & \textit{SE} & \textit{ASR1} & \textit{22.28} & - & \textit{25.37} & \textit{22.60} & - & \textit{32.95} & \textit{7.23} & - & \textit{10.35} & \textit{11.02} & - & \textit{18.33} & \textit{15.78} & - & \textit{21.75} \\
%\textit{8} & \textit{DIA4} & \textit{SE} & \textit{ASR5 }%(LM+ROVER) 
%& \ditto & \textit{-} & \textit{23.04} & \ditto & \textit{-} & \textit{29.35} & \ditto & \textit{-} & \textit{9.27} & \textit{\ditto} & \textit{-} & \textit{16.87} & \ditto & \textit{-} & \textit{19.63} \\
      \bottomrule
  \end{tabular}}%
  %\vspace{-3mm}
\end{table}

\begin{table}[t]
  \caption{Overall results on the evaluation set in terms of DER [\%]  ($\downarrow$) SCE ($\downarrow$) and tcpWER [\%] ($\downarrow$). DER is computed using md-eval with a collar of 0.25 sec. The best performances among all systems are indicated in \textbf{bold fonts}, and the best performances among our systems are \uline{underlined}. \ditto ~ denotes a ditto mark. All our systems use the SE frontend described in Section \ref{sec:se}.}
  %\vspace{-3mm}
  \label{tab:overall_results_eval}
  \centering
  \resizebox{\linewidth}{!}{%
  \begin{tabular}{@{}l@{\hspace{0.1cm}}l@{\hspace{0.2cm}}l@{\hspace{0.2cm}}c@{\hspace{0.1cm}}c@{\hspace{0.1cm}}cc@{\hspace{0.1cm}}c@{\hspace{0.1cm}}cc@{\hspace{0.1cm}}c@{\hspace{0.1cm}}cc@{\hspace{0.1cm}}c@{\hspace{0.1cm}}cc@{\hspace{0.1cm}}c@{\hspace{0.1cm}}c@{}}
  %{@{}llc@{\hspace{0.2cm}}c@{\hspace{0.2cm}}c@{\hspace{0.2cm}}c@{\hspace{0.2cm}c@{}}
    \toprule
    & && \multicolumn{3}{c}{CHiME-6} &\multicolumn{3}{c}{DiPCo} & \multicolumn{3}{c}{Mixer 6} &\multicolumn{3}{c}{NOTSOFAR-1} & \multicolumn{3}{c}{Macro}\\
    \cmidrule(l{0pt}r{\tabcolsep}){4-6}\cmidrule(l{\tabcolsep}r{\tabcolsep}){7-9}\cmidrule(l{\tabcolsep}r{\tabcolsep}){10-12}\cmidrule(l{\tabcolsep}r{\tabcolsep}){13-15}\cmidrule(l{\tabcolsep}){16-18}
    & &&DER & SCE & tcpWER & DER & SCE & tcpWER& DER & SCE & tcpWER& DER & SCE & tcpWER& DER & SCE & tcpWER \\
    \midrule
0 & \multicolumn{2}{l}{Baseline (NeMo)}&  63.25 & 2.00 & 73.80  &  41.73 & 1.40 &  57.10  &  16.13 & 0.00 & 23.20  &  51.36 & 2.00 & 72.00  &  43.12 & 1.35 & 56.50 \\
%%%0' & \multicolumn{3}{l}{STCON(constained lm)}    &  & 33.6 & 20.4 & 0.60 &  20.2 &	6.8 & 0.00 & 11.0 & 18.9 & 0.09 & 14.8 & 20.8& 0.17 & 19.9     \\ 
0' & \multicolumn{2}{l}{STCON (sys2)} & 36.97 & \textbf{0.00} & \textbf{33.12}  & 20.38 & 0.60 &  \textbf{19.88}  &	\textbf{6.81} & \textbf{0.00}  & \textbf{10.87}  & 18.93 & \textbf{0.09} & 14.61 & 20.77 & \textbf{0.17} & \textbf{19.62}  \\
%0' & USTC (NOTSOFAR-1) &\\
\midrule
\multicolumn{18}{c}{\textit{Results submitted to the challenge}} \\
\midrule
1 & DIA1  & ASR4 & 39.71  & 0.50 & 44.80 &  20.61 & \textbf{\uline{0.40}} & 26.20  &  \uline{8.24} & \uline{0.09} & 15.60 &  16.37 & \uline{0.15} & 22.10 &  21.23 & 0.29 & 27.20\\
2 & DIA2  & ASR1 & 38.46 & \ditto & 38.70  &  22.93 & \ditto & 25.00  &  8.59 & \ditto & 14.90  & 15.83 & \ditto & 18.30  &  21.46 & \ditto & 24.30\\
3 & DIA2  & ASR5 & \ditto & \ditto & 35.30  & \ditto  & \ditto & 22.40  & \ditto & \ditto & \uline{13.50}  & \ditto  & \ditto & 16.80  & \ditto & \ditto & 22.00\\ 
\midrule
\multicolumn{18}{c}{\textit{Post-challenge results}} \\
\midrule
4 & DIA3  & ASR1 & \textbf{\uline{32.59}} & \ditto & 37.40 & \textbf{\uline{20.33}} & \ditto & 25.02 & 8.26 & \ditto & 15.32 & \textbf{\uline{10.92}} & \ditto & 15.03 & \textbf{\uline{18.03}} & \ditto & 23.19 \\
%5 & DIA3 & SE & ASR5(ROVER) & \ditto & \ditto & 33.94 & \ditto & \ditto & 21.76 & \ditto & \ditto & 13.99 & \ditto & \ditto & 13.75 & \ditto & \ditto & 20.67 \\
5 & DIA3  & ASR5 & \ditto & \ditto & \uline{33.96} & \ditto & \ditto & \uline{21.60} & \ditto & \ditto & 13.83 & \ditto & \ditto & \textbf{\uline{13.73}} & \ditto & \ditto & \uline{20.78} \\
\midrule
\multicolumn{18}{c}{\textit{Results with Oracle number of speakers}} \\
\midrule
\textit{6} & \textit{DIA4}  & \textit{ASR1} & \textit{30.29} & - & \textit{34.92} & \textit{19.52} & - & \textit{24.11} & \textit{6.41} & - & \textit{12.04} & \textit{10.40} & - & \textit{14.22} & \textit{16.66} & - & \textit{21.32} \\
\textit{7} & \textit{DIA4}  & \textit{ASR5} & \ditto & - & \textit{31.99} & \ditto & - & \textit{21.20} & \ditto & - & \textit{10.77} & \ditto & - & \textit{12.87} & \ditto & - & \textit{19.21} \\
      \bottomrule
  \end{tabular}}%
  %\vspace{-3mm}
\end{table}
We first report the overall results in terms of DER, SCE, and tcpWER computed on the different datasets and their macro average. Ablation studies on the different modules are presented in \ref{sec:ablation}.

Tables \ref{tab:overall_results_dev} and \ref{tab:overall_results_eval} show the results on the development and evaluation sets, respectively.
We compare our results with the challenge's NeMo baseline \cite{chime8-task1} (system 0) and STCON (unconstrained\_lm)\footnote{STCON (unconstrained\_lm) system was submitted to the unconstrained LM track of the DASR task, which allows the use of large language models (LLMs). It utilizes a Llama-2-7B7 LLM for LM restoring as in \cite{Ogawa_arXiv2024}. LM rescoring with LLM improved the performance by 0.3 points in terms of Macro tcpWER on the evaluation set. Our systems do not utilize such LLM-based LM rescoring.} system \cite{stcon_chime8}(system 0’) that achieved the best tcpWER on the task at the challenge.\footnote{\url{https://www.chimechallenge.org/challenges/chime8/task1/results}} We report the results of the three systems we submitted to the challenge: system 1, which is the most lightweight model and uses only the simpler DIA1 and ASR4 models,  system 2, which utilizes our best diarization (DIA2)\footnote{At the time of the challenge submission.} and ASR (ASR1) models, and system 3, which combines the recognition hypotheses obtained with several ASR backends (ASR5). 

We also report the post-challenge results (systems 4 and 5), where we replaced the diarization module (DIA2) of systems 2 and 3 with a stronger diarization system (DIA3), which uses a TS-VAD model trained with more data. The EEND model, speaker counting, SE, and ASR modules remain the same. Note that all our systems rely on the same speaker counting module, and consequently, the estimated number of speakers is the same.

We also show results with the Oracle number of speakers (systems 6 and 7), where we replaced the estimated speaker counts in DIA3 of systems 4 and 5 with the Oracle ones (DIA4) (all other components remain the same). We believe that for some practical applications such as meeting transcriptions, it is reasonable to assume that we can know in advance the total number of speakers. Besides, it helps us understand the impact of speaker counting errors on the overall performance. %For systems 6 and 7, we used our stronger diarization system (DIA3) with the Oracle number of speakers, which we call DIA4.

Among our systems (without Oracle information), the most advanced systems, 3 and 5, naturally achieve the best macro DER and tcpWER for the challenge and post-challenge, respectively. Our systems greatly reduce the tcpWER over the challenge baseline for all test conditions and achieve performance close to the STCON system. Notably, our stronger post-challenge diarization system improved our challenge macro DER by more than three points, outperforming all challenge submissions. This resulted in a reduction of more than one point reduction of the macro tcpWER results, also achieving the best performance on the development and evaluation sets of NOTSOFAR-1 set. Comparing the results on the development and evaluation sets, we observe that our system has a higher SCE on the evaluation set. This may indicate over-tuning of the parameters on the development set. 

\begin{figure}[t]
\centering
%\begin{subfigure}{0.45\linewidth}
\centering
 \includegraphics[width=0.99\linewidth]{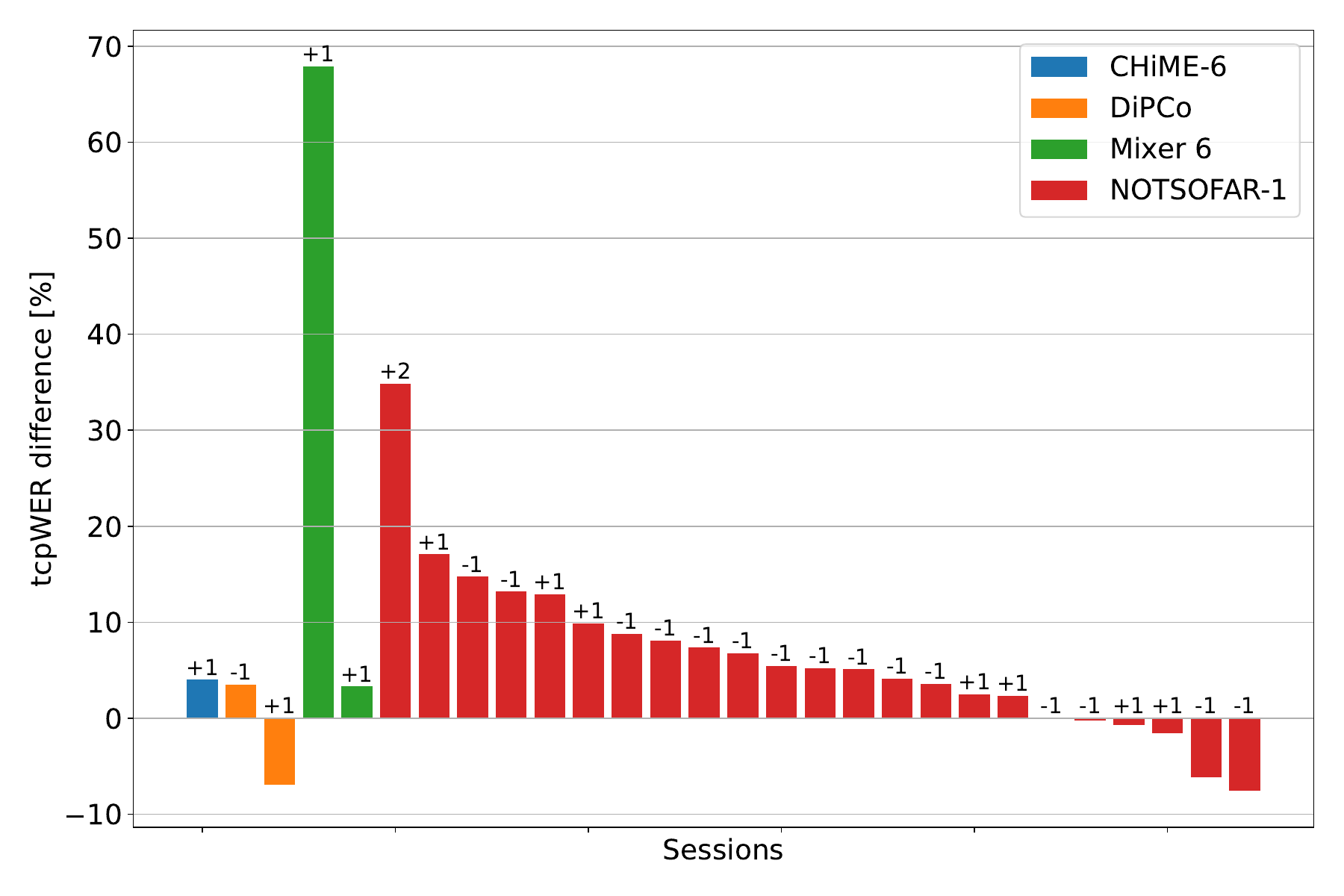}
   \caption{Impact of speaker count errors on tcpWER of the different sessions of the evaluation set. Vertical axis shows the difference between the tcpWER of systems using estimated and Oracle speaker counts, i.e., systems 5 (DIA3, SE, ASR5) and 7 (DIA4(Oracle speaker nb) SE ASR5), respectively. We report results only for sessions with speaker counting errors. The number above each bar indicates the speaker counting error. Positive values represent false alarms, while negative values represent missed speakers.}
    \label{fig:diff_tcpwer}
\end{figure}

Figure \ref{fig:diff_tcpwer} shows the impact of speaker counting errors on tcpWER by comparing the results without and with Oracle speaker counts for all sessions of the evaluation set. We observe that for some sessions, wrongly estimating the number of speakers can have dramatic consequences in terms of tcpWER. Indeed, if a speaker is split into two speakers, it will lead to a large number of deletion and insertion errors. For example, for Mixer 6, we only overestimated the number of speakers for two of the 23 sessions in the eval set. However, for one, it led to a drastic increase of the tcpWER reaching about 85 \% because one of the speakers was split into two speakers, but the speech was well transcribed. The error on this session alone explains the relatively high tcpWER we obtained on that dataset compared to when using the Oracle number of speakers.  

In contrast, for several sessions, speaker errors degrade the tcpWER by less than 5 points and can sometimes even lead to slightly better performance,  as seen in Figure~\ref{fig:diff_tcpwer}. This indicates that SCEs do not always have a high impact on performance. This is probably the case when the amount of speech allocated to the wrong speaker is small.

The results with the Oracle number of speakers show that further improvement to the speaker counting accuracy is crucial to achieving better performance with our system. In addition, although the tcpWER level on tasks such as Mixer 6 and NOTSOFAR-1 are already relatively low, they remain above 20 \% on tasks such as CHiME-6 and DiPCo. This indicates that further progress in diarization, SE, and ASR would be needed to bring down the tcpWER in challenging home-party recording scenarios.

\subsection{ASR and diarization error distribution}
\begin{figure}[t]
\centering
%\begin{subfigure}{0.45\linewidth}
\centering
 \includegraphics[width=0.99\linewidth]{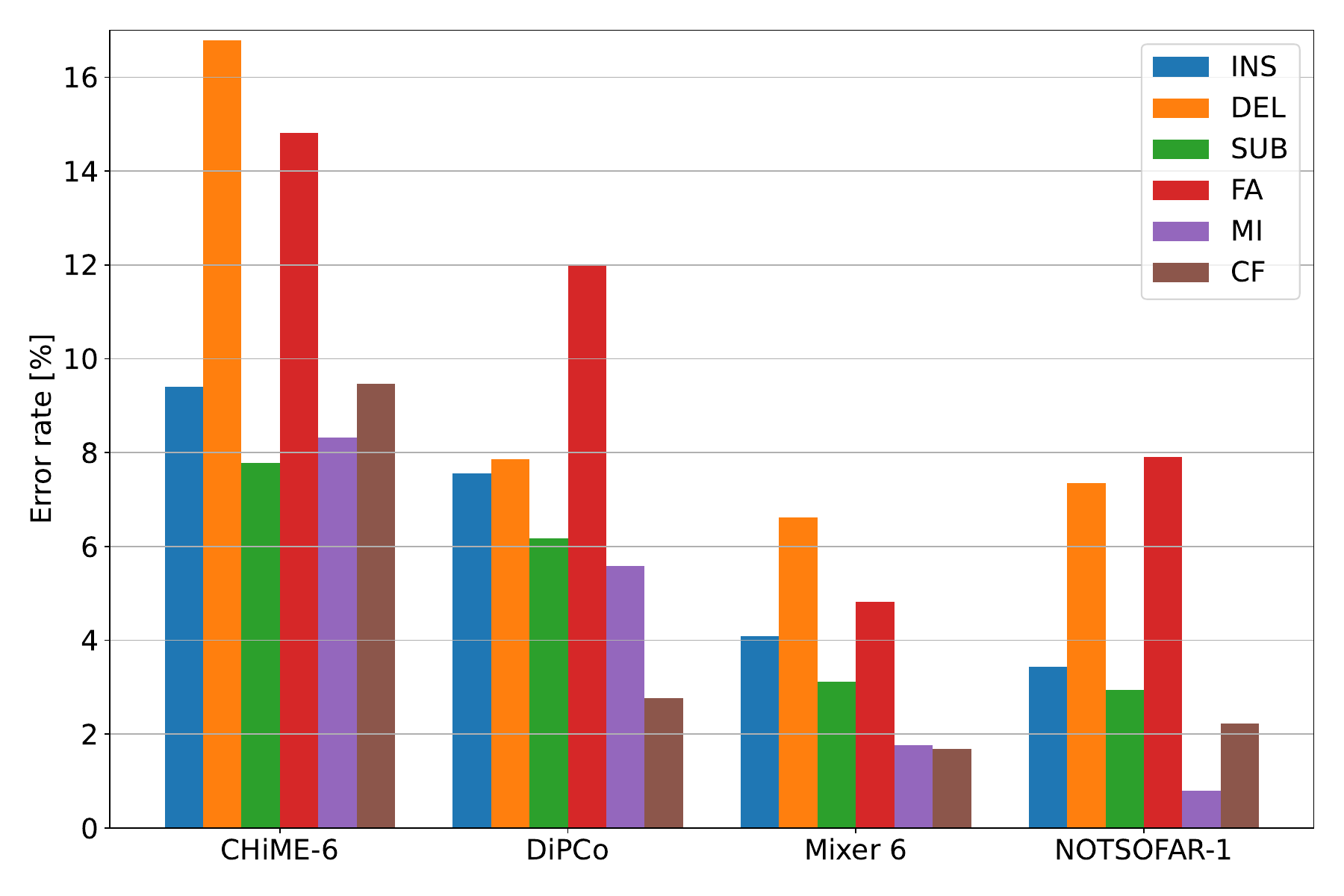}
   \caption{Distribution of ASR and diarization errors for different datasets of the evaluation set of our best system (DIA3 SE ASR5). ASR errors are measured in terms of insertions (INS), deletions (DEL), and substitutions (SUB). Diarization errors are measured in terms of missed speech (MI), false alarm (FA), and confusion (CF).}
   \label{fig:overall_result_details_2}
\end{figure}

Figure \ref{fig:overall_result_details_2} shows the distribution of ASR and diarization errors for the different datasets obtained with our top-performing system (system 5 in Table \ref{tab:overall_results_eval}). We observe that deletion errors are the main cause of errors for tcpWER and false alarms for diarization. The number of confusion errors is also much higher for CHiME-6 than for the other datasets. 

False alarms dominate diarization errors, but insertions do not dominate recognition errors, suggesting a complex relationship between diarization and recognition errors. In fact, as discussed in \ref{sec:ablation:asr_der_tuning}, short-pause filling, which increases false alarm, reduces tcpWER as the ASR backends can benefit from longer segments and ignore short pauses in-between.

%The relation between diarization and recognition errors is complex. For example, there are many false alarms partly because we use short-pause filling, which is shown to reduce tcpWER as the ASR backends can benefit from longer segments and ignore short pauses in-between (see ablation in Appendix \ref{sec:ablation:asr_der_tuning}). This may explain why the large number of false alarms does not imply that insertion errors dominate. 

In addition, we should not forget that speaker confusion errors of diarization translate into deletion and insertion errors in terms of tcpWER because the speech can be recognized but is allocated to the wrong speaker. This may explain the large number of deletions for the CHiME-6 data and suggests that reducing confusion errors on this data would be important to further improve performance by, e.g., using more discriminative speaker embeddings for EEND-VC. 

\subsection{Comparison of NOTSOFAR-1 results with CHiME-8 task 2 systems}
For the NOTSOFAR-1 data, we can compare the results of our microphone array geometry-independent systems with those submitted to the CHiME-8 task 2, which allowed developing systems dedicated only to the NOTSOFAR-1 data. In addition to the fact that the domain is limited to office meetings, the task 2 participants could take advantage of the fact that all NOTSOFAR-1 recordings were performed with a microphone array having the same known array geometry. 

The results of this task were evaluated by session-wise macro tcpWER, i.e., the macro average of all NOTSOFAR-1 sessions. Therefore, the numbers reported here and in the task 2 results \footnote{\url{https://www.chimechallenge.org/challenges/chime8/task2/results}} slightly differ from the other numbers in this paper and task 1 results.\footnote{\url{https://www.chimechallenge.org/challenges/chime8/task1/results}} We report the session-wise macro tcpWER on the NOTSOFAR-1 data in Table \ref{tab:notsofar_results} to allow comparison of our general system with NOTSOFAR-1-specific systems. 

Our best challenge system was five points behind the top system on task 2. With our post-challenge system, we could reduce this gap to two points, showing the possibility of achieving a high-performing microphone array geometry-independent multi-talker DASR system. %This gap could be further narrowed down if we use the Oracle number of speaker information.

\begin{table}[t]
    \caption{Comparison of task 2 (geometry-dependent) and task 1 (geometry-independent) systems of the CHiME-8 challenge on the NOTSOFAR-1 data in terms of session-wise macro tcpWER on the evaluation set. The best performance among all systems is indicated in \textbf{bold fonts}. The best performance among the array geometry-independent systems is \uline{underlined}.}
    \centering
    \begin{tabular}{@{}lcc@{}}
    \toprule
          &	Geometry  &	tcpWER  \\
    System     &	 independent &	(\%) \\
         \midrule
%\midrule
% \multicolumn{3}{c}{\textit{Challenge results}}		\\
% \midrule
%\midrule
NAIST \cite{hirano24_chime}	& &	23.00\\
NPU-TEA \cite{huang24b_chime}	& &	18.70\\
USTC-NERCSLIP \cite{niu24_chime} & &	\textbf{10.80} \\
\midrule
STCON \cite{stcon_chime8}	&\checkmark &	13.80\\
Ours (DIA2 SE ASR5)	&\checkmark &	15.90\\
%\midrule
%\midrule
%\multicolumn{3}{c}{\textit{Post-challenge results}}		\\
%\midrule
%\midrule
Ours (DIA3 SE ASR5) (Post-challenge)	&	\checkmark &\uline{12.95} \\
%\textit{Our (DIA3(Oracle speaker nb) SE ASR5)}	& \checkmark  &	\textit{12.04}\\
\bottomrule
    \end{tabular}
    \label{tab:notsofar_results}
\end{table}

\subsection{System complexity}
\begin{table}[t]
    \centering
    \caption{Comparison of RTF and tcpWER for three proposed systems on development set of NOTSOFAR-1.}
    \label{tab:rtf}
    \begin{tabular}{@{}lccccc@{}}
    \toprule
    ID & Diar & SE & ASR & RTF & tcpWER\\
    \midrule
    1 & DIA1 & SE & ASR4 & 2.46 & 23.85\\
    2 & DIA2 & SE & ASR1 & 3.14 & 20.41 \\
    3 & DIA2 & SE & ASR5 & 4.03 & 18.79 \\
\bottomrule
    \end{tabular}
\end{table}

The three systems we proposed have very different computational complexities. To provide an idea of the complexities, we calculated the real-time factor (RTF). Table \ref{tab:rtf} reports the RTF and the tcpWER of the three proposed systems, computed on one A6000 GPU on the NOTSOFAR-1 dataset. 

Note that the RTF values are only indicative, as we did not optimize our systems for computational efficiency. These numbers show the increased complexity coming from using more advanced diarization and ASR systems. 
Our lightest system achieves an RTF of 2.5, which is still relatively high. Future works should focus on further reducing the computational complexity of the simpler system while improving its performance.

Note also that the post-challenge system DIA3 has the same complexity as DIA2 since it uses the same EEND-VC module, and the model configurations of the TS-VAD models are the same; only the training data differ.

\section{Conclusion}
\label{sec:conclusion}
We have introduced the microphone array geometry-independent multi-talker DASR system we submitted to the DASR task of CHiME-8. We have provided a detailed description of all parts of our system and performed a detailed analysis and ablation study of the effect of each component. 

Our best post-challenge system significantly improved the macro tcpWER performance over the baseline by up to 63 \% and outperformed the top challenge system in terms of DER. Moreover, on the NOTSOFAR-1 data, we achieved new state-of-the-art results for the array-geometry agnostic system, narrowing the gap to about two points behind the best array-geometry-dependent system in the challenge. 

Our analysis has also identified several important future research directions. First, we should improve speaker counting further, as small errors can lead to large performance degradation. Moreover, we should tackle the high confusion error rate on the CHiME-6 data by developing more robust speaker embedding extractor and clustering approaches. Finally, high performance comes at the price of high system and computational complexity. Simplifying our system and reducing its computational complexity while maintaining performance will also be part of our future research directions.

%% The Appendices part is started with the command \appendix;
%% appendix sections are then done as normal sections
%\newpage
\appendix
\section{Ablation analysis}
\label{sec:ablation}

We performed ablation studies on the diarization, SE, and ASR modules to clarify the effect of our implementation choices.

\subsection{Ablation experiments on the diarization module}
In this section, we report the DER values optimized for diarization and not for ASR. Therefore, there is a slight discrepancy between the results in Tables \ref{tab:overall_results_dev} and those in this section.

\subsubsection{Diairization model configuration}
\label{ssec:ablation_model}

\begin{table}[t]
  \caption{DER [\%] $ (\downarrow$) on the development set for different configurations of the diarization system. The DER was computed using md-eval with a collar of 0.25 sec.}
  %\vspace{-3mm}
  \label{tab:diarization_results}
  \centering
  %\resizebox{\linewidth}{!}{%
  \begin{tabular}{@{}llc@{\hspace{0.2cm}}c@{\hspace{0.2cm}}c@{\hspace{0.2cm}}c@{\hspace{0.2cm}}c@{}}
    \toprule
     & Model & CHiME-6 & DiPCo & Mixer 6 & NOTSOFAR-1 & Macro\\
    \midrule
1 & EEND-VC w/o GSS &39.70&33.06&10.37&13.00&24.03\\
2 & EEND-VC w/ GSS  &28.52&24.38&9.69&10.67&18.32\\
\midrule %\\
%EEND-VC + TS-VAD &&23.97&21.01&6.11&9.72&15.20\\
3 & EEND-VC + TS-VAD  &21.58&19.35&5.79&8.66&13.85\\
4 & NeMo + TS-VAD &27.96&40.34&7.93&26.77&25.75\\
%DIA0 & Baseline (NeMo)&  45.65 	&	45.92 	&	25.16 & 38.05 & 38.70 \\
%\midrule
%DIA1 & EEND-VC w/ ECAPA & 28.52 	&	24.38 	&	9.69 	&	10.67 	&	18.32 \\ % system 74
%DIA2 & DIA1 + TS-VAD & 23.97 	&	21.01 	&	6.11 	&	9.72 	&	15.20  \\ % system 74'
      \bottomrule
  \end{tabular}
  %}%
  %\vspace{-3mm}
\end{table}
Table \ref{tab:diarization_results} shows the DER for different configurations of the EEND-VC pre-diarization and the impact of pre-diarization on TS-VAD performance as discussed in Section \ref{sec:diar}.
First, comparing rows 1 and 2, we see that using SE before computing speaker embeddings as explained in Section \ref{ssec:eend_vc} has a significant impact on the performance as it reduces the macro DER by close to six points. This indicates that by using SE, we can obtain more discriminative speaker embeddings, thereby improving the accuracy of the speaker embedding clustering. Using SE also impacts the speaker counting performance, as shown in \ref{ssec:ablation_counting}. The effect of SE is consistent across all datasets.

Next, we confirm the importance of a good initialization for TS-VAD by comparing the results with TS-VAD using our proposed EEND-VC (row 3) and NeMo baseline (row 4) as pre-diarization. These results demonstrate that TS-VAD is much more effective and achieves a significantly lower DER when using EEND-VC.

\subsubsection{Speaker Counting}
\label{ssec:ablation_counting}
\begin{table*}[t]
  \caption{Speaker counting results in speaker counting accuracy (SCA), speaker counting error (SCE), and average speaker counts (\#SPK).}
  %\vspace{-3mm}
  \label{tab:counting_results}
  \centering
  \scriptsize%
  \resizebox{\linewidth}{!}{%
  \setlength{\tabcolsep}{1mm}
  \begin{tabular}{@{}ll@{}cccccccccccccc@{}}  \toprule
       & & \multicolumn{3}{c}{CHiME-6} & \multicolumn{3}{c}{DiPCo} 
       & \multicolumn{3}{c}{Mixer 6} & \multicolumn{3}{c}{NOTSOFAR-1} 
       & \multicolumn{2}{c}{Macro Avg.} \\
        &  & \multicolumn{3}{c}{(\#SPK=4.0)} & \multicolumn{3}{c}{(\#SPK=4.0)} 
       & \multicolumn{3}{c}{(\#SPK=2.0)} & \multicolumn{3}{c}{(\#SPK=4.7)} 
       & \multicolumn{2}{c}{} 
       \\ \cmidrule(l{0pt}r{\tabcolsep}){3-5}\cmidrule(l{\tabcolsep}r{\tabcolsep}){6-8}\cmidrule(l{\tabcolsep}r{\tabcolsep}){9-11}\cmidrule(l{\tabcolsep}r{\tabcolsep}){12-14}\cmidrule(l{\tabcolsep}){15-16} 
    & & \#SPK &SCA$\uparrow $ & SCE$\downarrow $ 
          & \#SPK &SCA$\uparrow $ & SCE$\downarrow $ 
          & \#SPK &SCA$\uparrow $ & SCE$\downarrow $ 
          & \#SPK &SCA$\uparrow $ & SCE$\downarrow $ 
          & SCA$\uparrow $ & SCE$\downarrow $  \\
    \midrule
%\multirow{8.5}{*}{\begin{tabular}{@{}c@{}}\rotatebox[origin=c]{90}{\textbf{Development set}}\end{tabular}}&
0& 
NeMo  Baseline 
& 5.0 & 50.0 & 1.00 
& 7.0 & 0.0 & 3.00 
& 2.0 & \textbf{100.0} & \textbf{0.00} 
& 3.0 & 13.8 & 1.71 
& 41.0 & 1.43 \\
\midrule
%\multicolumn{2}{l}{\textbf{Our baseline}}\\
1& Channel-wise  
& 3.3 & 36.4 & 0.86 
& 3.2 & 30.0 & 0.90 
& 2.1 & 92.8 & 0.09 
& 4.2 & 51.4 & 0.68 
& 52.7 & 0.63  \\
2& ~+  Using SE   
& 4.1 & 93.2 & 0.07 
& 4.0 & 95.7 & 0.04 
& 2.0 & 98.3 & 0.02 
& 4.3 & 56.8 & 0.56 
& 86.0 & 0.17  \\ 
\midrule
%&\multicolumn{2}{l}{\textbf{Proposed speaker counting}}\\
3& Mic-group-wise   
& 3.2 & 21.1 & 1.11  
& 2.9 & 10.0 & 1.10 
& 2.1 & 92.9 & 0.08
& 4.2 & 53.7 & 0.67 
& 44.4 & 0.74 \\
4& ~+  Using SE  
& 4.0 & \textbf{100.0} & \textbf{0.00} 
& 4.1 & 90.0 & 0.10 
& 2.0 & 98.5  & 0.02 
& 4.4 & \textbf{59.7} & \textbf{0.49} 
& 87.1 & 0.15 \\
5& ~+  Group avg 
& 4.0 & \textbf{100.0} & \textbf{0.00} 
& 4.0 & \textbf{100.0} & \textbf{0.00} 
& 2.0 & \textbf{100.0} & \textbf{0.00} 
& 4.4 & 58.5 & 0.51  
& \textbf{89.6} & \textbf{0.13}  \\
%\\\midrule\midrule
% \multirow{8.5}{*}{\begin{tabular}{@{}c@{}}\rotatebox[origin=c]{90}{\textbf{Evaluation set}}\end{tabular}}&(B)& Baseline (NeMo) & 
%  & 6.0 & 0.0 & 2.00 
%  & 5.4 & 20.0 & 1.40 
%  & 2.1 & \textbf{100.0} & \textbf{0.00}
%  & 2.9 & 17.5 & 2.00 
%  & 34.4 & 1.35 \\
% \cmidrule(l{\tabcolsep}){2-18}
% &\multicolumn{2}{l}{\textbf{Our baseline}}\\
% &(C1)& Channel-wise counting & 
%  & 3.7 & 13.6 & 0.95 
%  & 4.0 & \textbf{92.0} & \textbf{0.08}
%  & 2.1 & 90.4 & 0.14 
%  & 4.3 & 61.6 & 0.58
%  & 64.4 & 0.44  \\
% &(C2)& ~+ Using enhanced signals & \checkmark  
%  & 3.9 & 36.4 & 0.70 
%  & 4.4 & 61.7 & 0.47 
%  & 2.2 & 82.0 & 0.21 
%  & 4.6 & 81.2 & 0.23 
%  & 65.3 & 0.40  \\ 
% \cmidrule(l{\tabcolsep}){2-18}
% & \multicolumn{2}{l}{\textbf{Proposed speaker counting}}\\
% &(M1)& Microphone-group-wise counting & 
%  & 3.5 & 15.0 & 0.95 
%  & 3.9 & 76.0 & 0.24 
%  & 2.1 & 91.8 & 0.11 
%  & 4.4 & 66.3  & 0.47 
%  & 62.3 & 0.44 \\
% &(M2)& ~+  Using enhanced signals & \checkmark 
%  & 4.0 & 40.0 & 0.65 
%  & 4.4 & 68.0 & 0.52 
%  & 2.1 & 82.5  & 0.21 
%  & 4.6 & \textbf{85.6} & \textbf{0.17} 
%  & 68.7 & 0.39 \\
% &(M3)& ~~++  Group averaging & \checkmark 
%  & 4.0 & \textbf{100.0} & \textbf{0.00} 
%  & 4.4 & 80.0 & 0.40 
%  & 2.1 & 91.3 & 0.09
%  & 4.6 & \textbf{85.6} & \textbf{0.17} 
%  & \textbf{89.2} & \textbf{0.15} \\
      \bottomrule
  \end{tabular}}
  %\vspace{-6mm}
\end{table*}
%-------------------------------------

Table \ref{tab:counting_results} shows the speaker counting accuracy and error on the development set for the different configurations explained in Section \ref{ssec:speaker_counting}. % Note that systems 1 and 2 present the performance per channel, systems 3 and 4 present it per microphone group, and systems 0 and 5 present it per session. 
System 0 presents the results from the NeMo-based baseline system, which failed to correctly estimate the number of speakers in most scenarios except for Mixer 6.
It tended to overestimate the number of speakers for CHiME-6 and DiPCo, probably because it uses short segments of up to three seconds. This caused significant variations within a speaker’s voice in highly overlapped chunks, leading to an overestimation of the number of speakers. %It tended to overestimate the number of speakers for CHiME-6 and DiPCo, probably because it uses short segments of up to three seconds, causing significant variations within a speaker’s voice in highly overlapped chunks, leading to overestimating the number of speakers.
Conversely, on NOTSOFAR-1, the baseline underestimated the number of speakers because the sessions were shorter. This resulted in less reliable clustering because of the limited number of speaker embeddings available.

%. This may be due to the baseline system performing speaker counting on relatively short segments, up to three seconds, which could cause significant variations within a speaker’s voice in highly overlapped chunks, leading to an over-count of speakers. Conversely, the baseline underestimated the number of speakers on NOTSOFAR-1, likely because it consisted of shorter sessions with a limited number of available speaker embeddings, making it challenging for the clustering process to accurately estimate the number of speakers due to the scarcity of samples.

The remaining lines show the results of our speaker counting system. Systems 1 and 2 show the results of performing speaker counting on each channel% with the integration into session-level speaker counts
, while systems 3, 4, and 5 display the results using microphone groups.
Systems 1 and 3 utilized speaker embeddings derived from the original signals, while systems 2, 4, and 5 utilized embeddings from enhanced signals. It is important to note that system 1 refers to our previous CHiME-7 system \cite{kamo23_chime}, with the key difference being the use of a shorter chunk length (reduced from 80 seconds to 30 seconds).

In Table \ref{tab:counting_results}, we can see that our baseline (system 1) outperforms the NeMo-based baseline in terms of the macro average computed across all scenarios. The improved performance can be attributed to using longer subchunks, i.e., 15 vs. three seconds for the NeMo baseline. However, system 1 tends to underestimate the number of speakers (except for Mixer 6), %probably because the longer subchunks led to a reduced number of embeddings, which contributed to the underestimation of speaker counts. System 2, which performs speaker counting on embeddings from enhanced signals, significantly improves the performance.
probably because, in longer subchunks, errors in estimating speaker activation often result in extracting speaker embeddings that include speech from other speakers or noise segments. As a result, the variance of the speaker embeddings increases, causing different clusters to be counted as a single cluster. In contrast, system 2 significantly improves the performance, as the interfering speakers' speech and noise are suppressed, resulting in less variation in the embeddings.

Using microphone group-wise counting (system 3) and enhanced signals (system 4) improved the overall counting performance. Notably, the performance improved for the shorter sessions of NOTSOFAR-1, likely because group-wise counting helped increase the number of embeddings, thereby enhancing performance. 
Moreover, applying group averaging (system 5) further improved the counting accuracy, achieving the best macro-average performance across all scenarios (89.6 in SCA of Macro Avg.). These findings suggest that our proposed group-wise speaker counting on embeddings from enhanced signals, followed by aggregation through group averaging, demonstrates a robust performance across various situations.
\textcolor{red}{
%[TODO: ADDITIONAL COMMENTS FOR THE NOTSOFAR-1 RESULT] 
The overall SCA in NOTSOFAR-1 was relatively lower than in the other scenarios, possibly because NOTSOFAR-1 is a unique scenario involving more speakers (up to eight) in short sessions (averaging six minutes) compared to the other scenarios \cite{vinnikov24_interspeech}. This discrepancy may cause the hyper-parameters,  including chunk and subchunk lengths, to overfit the other three scenarios. Note that although the SCA is low, the SCE on NOTSOFAR-1 remains reasonable (0.51), and the relative speaker error is moderate considering the higher number of speakers in the sessions of this dataset.}
%Future research should explore a more generalized approach to address this gap.}
% it tends to underestimate speaker counts, except for Mixer 6. This underestimation is likely due to our diarization pipeline producing more discriminative speaker embeddings using longer 15-second subchunks, in contrast to the NeMo-based baseline's shorter chunks. However, the longer subchunks led to a reduced number of chunks, which contributed to the underestimation of speaker counts.

%The system 3 result shows that microphone group-wise counting yielded similar performance to channel-wise counting but slightly improved SCA and SCE for NOTSOFAR-1. This improvement is likely due to the shorter session lengths in NOTSOFAR-1, where group-wise counting helped increase the number of embeddings, thereby enhancing performance.

%Finally, the results of systems 2, 4, and 5 indicate that performing speaker counting on embeddings from enhanced signals significantly improved the performance of both channel-wise and group-wise approaches. Specifically, applying group averaging over the group-wise counting results led to further improvements, achieving the best macro-average performance across all scenarios (89.6 in SCA of Macro Avg.). These findings suggest that our proposed group-wise speaker counting on embeddings from enhanced signals, followed by aggregation through group averaging—demonstrates robust performance across various situations.

\subsubsection{Effect of DOVER-Lap for combining channel-wise diarization results}
\label{sec:ablation:doverlap}
%------------------------------------
\begin{figure}[t]
    \centering
    \includegraphics[width=0.95\linewidth]{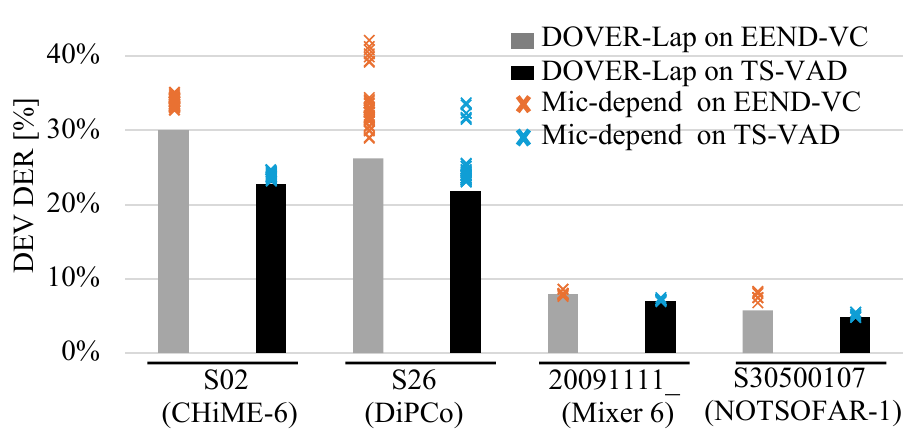}
    \caption{
    Effect of DOVER-Lap for combining the channel-wise diarization results. We report DER values for four sessions of the development set (one for each dataset). Orange and blue cross-marks show the channel-wise results using EEND-VC (DIA1) and TS-VAD (DIA2), respectively. Gray and black bars show the results obtained after using DOVER-Lap. %and t DERs on DOVER-Lap, both with and without EEND-VC-based pre-diarization and TS-VAD results. Individual plots represent microphone-dependent results, while the bars are the DOVER-Lap results.
    }
    \label{fig:dir_abr_doverlap}
\end{figure}
%------------------------------------
We perform diarization for each microphone and combine the results with DOVER-Lap as explained in Section \ref{ssec:dover}.
Figure \ref{fig:dir_abr_doverlap} shows the DER on one session of the development set for each dataset with and without TS-VAD refinement and before and after using DOVER-Lap.
We observe that before applying DOVER-LAP, the DER depends greatly on the microphone used. This tendency is especially visible with the DiPCo data, where the microphones are more widely distributed compared to those in other datasets.
DOVER-Lap effectively combines the results obtained with each microphone and achieves the lowest DER, which is usually lower than the best microphone output.
These findings confirm the effectiveness of using DOVER-Lap to combine the diarization output of all the microphones in a distributed microphone array setting.

\subsubsection{Tuning diarization hyperparameters for ASR}
\label{sec:ablation:asr_der_tuning}

\begin{figure}[tb]
\centering
\begin{subfigure}{0.45\linewidth}
\centering
\includegraphics[width=\linewidth]{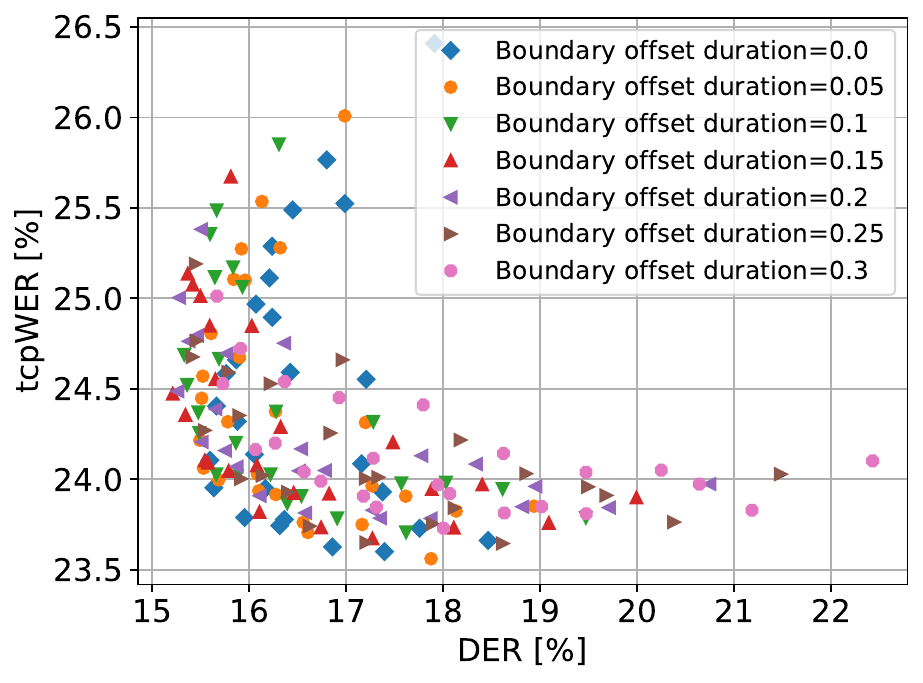}
\caption{Boundary offset duration}
\label{fig:diar_asr_tuning_pad}
\end{subfigure}
\hfill
\begin{subfigure}{0.45\linewidth}
\centering
\includegraphics[width=\linewidth]{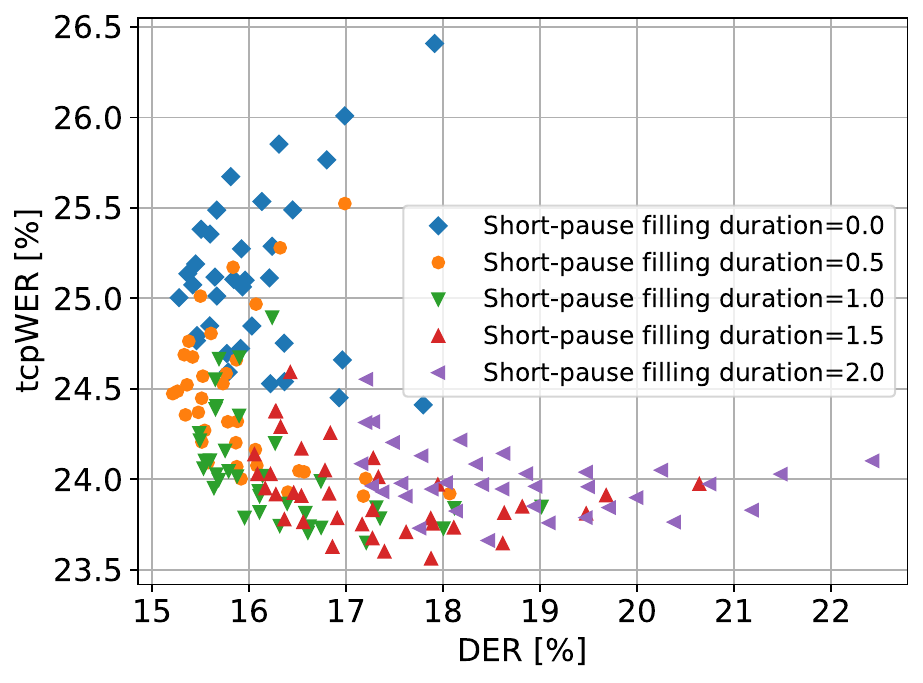}
\caption{Short-pause filling duration}
\label{fig:diar_asr_tuning_merge}
\end{subfigure}
\begin{subfigure}{0.45\linewidth}
\centering
\includegraphics[width=\linewidth]{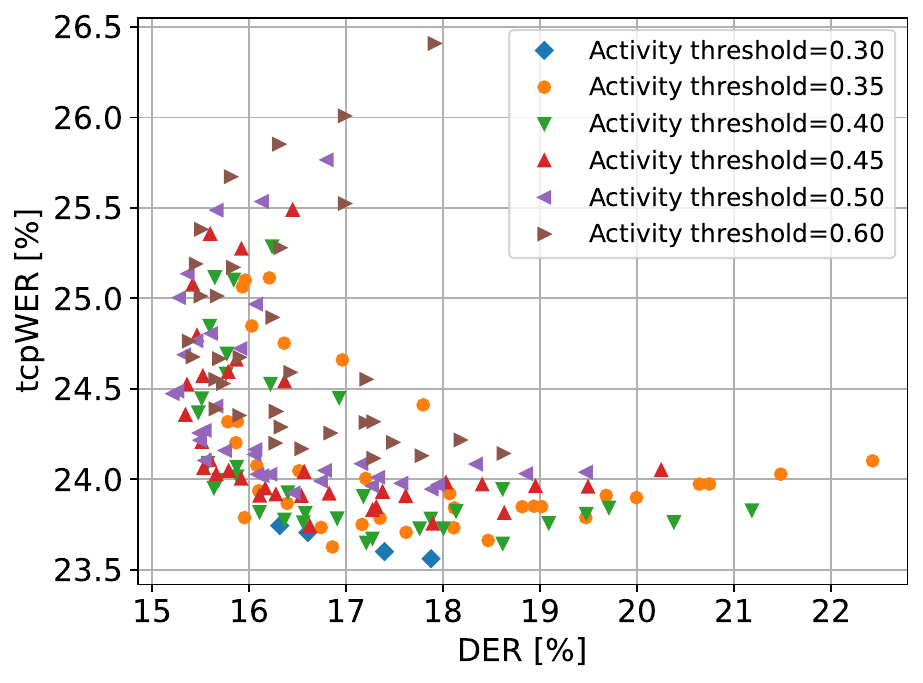}
\caption{Activity threshold}
\label{fig:diar_asr_tuning_threshold}
\end{subfigure}
\caption{Relation between macro averaged tcpWER and DER of the development set for different tuning parameters using DIA2 and ASR1.}
\label{fig:diar_asr_tuning}
\end{figure}

Diarization post-processing is utilized to convert the speaker activity posteriors into utterances boundaries, as explained in Sections \ref{sec:diar:post_processing} and \ref{ssec:tsvad}. In this section, we investigate the impact of the post-processing hyperparameters on DER and tcpWER, including the boundary offset duration $\theta_{\textrm{offset}}$, the short-pause filling duration $\theta_{\textrm{merge}}$, and the activity threshold $\theta_{\textrm{ts-vad}}$.

%The hyperparameters that achieve the best DER may not achieve the best tcpWER. For example, the diarization system may tend to output short segments cut at every short pause. However, for ASR, short pauses may not matter much; instead, it may be beneficial to process longer segments to exploit speech context.

%To obtain optimal tcpWER, we chose to perform an exhaustive search on the diarization post-processing hyper-parameters described in Section \ref{sec:diar:post_processing}, including the activity threshold, boundary offset duration, and short-pause filling duration.

Figure \ref{fig:diar_asr_tuning} shows the relationship between tcpWER and DER for the different tuning parameters when using DIA2 and ASR1. The three graphs are identical and differ only by their labels, indicating different values of the hyper-parameters. 
The optimal hyperparameters are different for diarization and ASR. 
We observe that the configuration achieving the best tcpWER (about 23.5 \%) is close to 3 points behind the configuration achieving the best DER (about 15.2 \%). 

Looking into each parameter in more detail, we can see from Figure~\ref{fig:diar_asr_tuning_pad} that the offset does not have a major impact, and setting it to a small value ($\theta_{\textrm{offset}} =0.0$ or $0.05$) achieves the best tcpWER. 

Looking at Figure~\ref{fig:diar_asr_tuning_merge}, we see that merging segments together by filling short pauses has a larger impact on tcpWER, and the best performance is achieved with a filling duration of $\theta_{\textrm{merge}} =1.5$ sec. These results suggest that the diarization system may tend to output short segments by cutting speech at every short pause. However, for ASR, short pauses may not matter much; instead, it may be beneficial to process longer segments to exploit speech context.

Finally, from Figure~\ref{fig:diar_asr_tuning_threshold}, we also observe that the activity threshold plays an important role. The optimal threshold for tcpWER ($\theta_{\textrm{ts-vad}}=0.30$) tends to be lower than that for DER ($\theta_{\textrm{ts-vad}}=0.50$), which indicates that the ASR system is less affected by false alarms than missed speech.

\subsection{Ablation experiments on SE module}
\subsubsection{MIMO beamforming, reference microphone selection, and post-filtering}
\label{sec:ablation:bf}

We tested whether SP-MWF was more effective than the R1-MWF utilized by the baseline.
%in the MIMO beamforming module in the SE system.
We also investigated whether the BAN post-filter was effective in conjunction with SP-MWF.
The hyperparameter $\gamma$ was set to $\gamma = 0$ for both SP-MWF and R1-MWF as in the baseline.

Table~\ref{tab:ablation:bf} shows the tcpWER on the development set when the beamforming and post-filters are varied in the official baseline SE system with Oracle diarization.
To evaluate the tcpWER, we used the baseline NeMo ASR system.
The number of microphones selected in the microphone subset selection module was tested for $K = M$ and $K = \lceil 0.8 \textcolor{red}{\cdot} M \rceil$.
As discussed in Section~\ref{sec:se:bf:discussion}, when the BAN post-filter is applied, the performance difference between R1-MWF and SP-MWF is attributed solely to their effectiveness of the reference microphone selection.
Thus, by comparing lines (1) with (3) and (5) with (7) in Table \ref{tab:ablation:bf}, we observed that SP-MWF selected a more effective reference microphone than R1-MWF.
This trend was only evident in the CHiME-6 and DiPCo datasets, likely because in these datasets the microphone arrays in these datasets are more widely distributed, making the selection of an effective reference microphone more important than in the other two datasets.

Next, by comparing lines (3) with (4) and (7) with (8), we observed that the BAN post-filter was not effective with SP-MWF.
Applying the BAN post-filter is independent of the choice of the reference microphone and only alters the frequency-wise amplitude of the separated signal.
Therefore, it turned out that the scale of SP-MWF was more appropriate than the scale defined by the BAN post-filter, although the performance improvement was small.

\begin{table}[t]
    \caption{tcpWER [\%] ($\downarrow$) on the development set when varying the beamformers and post-filters in the SE system. The results were obtained with Oracle diarization and the NeMo ASR system.}
    \label{tab:ablation:bf}
    \centering
    \resizebox{\linewidth}{!}{%
        \begin{tabular}{@{}lcccccccc@{}}
        \toprule
        ID & \#mics ($K$) & Beamformer & post-filter & CHiME-6 & DiPCo & Mixer 6 & NOTSOFAR-1 & Macro \\
        \midrule
        (1) & \multirow{4}{*}{$M$} & R1-MWF & BAN
            & 20.87 & 31.52 & 10.64 & 12.07 & 18.77 \\
        (2) & & R1-MWF & - 
            & 21.79 & 36.09 & 11.58 & 12.84 & 20.58 \\
        (3) & & SP-MWF & BAN
            & \textbf{20.46} & 31.16 & \textbf{10.59} & 12.00 & 18.55 \\
        (4) & & SP-MWF & - 
            & 20.62 & \textbf{30.83} & 10.70 & \textbf{11.85} & \textbf{18.50} \\
        \midrule
        (5) & \multirow{4}{*}{$\lceil 0.8 \textcolor{red}{\cdot} M \rceil$} & R1-MWF & BAN
            & 20.01 & 31.52 & \textbf{10.87} & 13.64 & 19.01 \\
        (6) & & R1-MWF & - 
            & 21.00 & 34.73 & 11.85 & 14.29 & 20.47 \\
        (7) & & SP-MWF & BAN
            & 19.74 & 30.65 & 10.89 & 13.61 & 18.72 \\
        (8) & & SP-MWF & - 
            & \textbf{19.72} & \textbf{30.00} & 10.96 & \textbf{13.44} & \textbf{18.53} \\
        \bottomrule
        \end{tabular}%
    }
\end{table}

\subsubsection{Microphone subset selection}
\label{sec:ablation:mic-selection}

We evaluate the effect of introducing $K_{\min}$, the minimum number of microphones selected in microphone subset selection (see Section~\ref{sec:se:mic-selection:new}).
We also investigate the effectiveness of exploiting two acoustic features, EV and $C_{50}$.

Table~\ref{tab:ablation:mic-subset-selection:1} shows the tcpWER on the development set as we vary the value of $K_{\min}$ among $\{ 0, 5, 10, 15, 20 \}$.
To evaluate the tcpWER, we used the ASR system, WavLM Transducer (ASR4 in Table~\ref{tab:asr_ablation}).
Except for $K_{\min}$, we used our proposed full SE system with Oracle diarization.
%, the same system used in Table ????.
Since there are only $M = 10$ microphones in Mixer 6 and $M = 7$ in NOTSOFAR-1,
all the microphones are selected for these datasets when $K_{\min} \geq 10$,
resulting in the same tcpWER (note that the tiny difference of 0.01 \% in tcpWER may be caused by numerical errors).
For all datasets, the result with $K_{\min} \leq 5$ is worse than that with $K_{\min} \geq 10$, particularly for Mixer 6 and NOTSOFAR-1, demonstrating the effectiveness of the mechanism that selects at least $K_{\min}$ microphones.
For CHiME-6, the result with $K_{\min} = 20$ is significantly worse than that with $K_{\min} = 15$.
Since CHiME-6 has only 20 or 24 microphones, when $K_{\min} = 20$, all or almost all of the microphones are selected, including those that may be problematic for the SE system.
By setting $K_{\min}$ to 15, we can discard these problematic microphones while still selecting a sufficient number of microphones to improve the accuracy of multichannel SE methods, such as beamforming and GSS.
On the basis of these observations, we conclude that the $K_{\min}$ mechanism is essential for CHiME-8.

Next, Table~\ref{tab:ablation:mic-subset-selection:2} shows the tcpWER on the development set when using either or both of the EV and $C_{50}$ in the proposed SE system with Oracle diarization.
To evaluate the tcpWER, we used the baseline NeMo ASR system.
Note that the results for Mixer 6 and NOTSOFAR-1 are the same across all methods since we select all $M$ microphones due to $K_{\min} = 15$.
The results for CHiME-6 and DiPCo show that relying on two features is more advantageous than relying on just a single feature, clearly demonstrating the effectiveness of the proposed approach.
In future work, we plan to develop a method that can handle more than two audio features and explore more sophisticated yet simple combinations of these features.

\begin{table}[t]
    \caption{tcpWER [\%] ($\downarrow$) on the development set when varying the minimum number of microphones selected ($K_{\min}$) in the microphone subset selection module. The results were obtained with Oracle diarization and our ASR4 model.}
    \label{tab:ablation:mic-subset-selection:1}
    \centering
    \resizebox{\linewidth}{!}{%
        \begin{tabular}{@{}lcccccc@{}}
        \toprule
        %& $M$ & 20 or 24 & 35 & 10 & 7 \\
        %\midrule
        &  & CHiME-6 & DiPCo & Mixer 6 & NOTSOFAR-1 & Macro \\
        ID & $K_{\min}$ & ($M= 20$ or $24$) & ($M=35$) & ($M=10$) & ($M=7$) \\
        \midrule
        (1) & 0
            & 20.66 & 27.68 & 11.55 & 15.92 & 18.95 \\
        (2) & 5
            & 20.68 & 27.66 & 11.45 & 13.94 & 18.43 \\
        (3) & 10
            & 20.29 & 27.63 & \textbf{10.77} & \textbf{13.23} & 17.98 \\
        (4) & 15
            & \textbf{19.77} & \textbf{27.52} & \textbf{10.78} & \textbf{13.24} & \textbf{17.83} \\
        (5) & 20
            & 20.60 & 27.58 & \textbf{10.78} & \textbf{13.23} & 18.05 \\
        \bottomrule
        \end{tabular}%
    }
\end{table}

\begin{table}[t]
    \caption{tcpWER [\%] ($\downarrow$) on the development set when using either or both of the EV and $C_{50}$ for microphone subset selection. The results were obtained with Oracle diarization and the NeMo ASR system.}
    \setlength{\tabcolsep}{4pt}
    \label{tab:ablation:mic-subset-selection:2}
    \centering
    \resizebox{\linewidth}{!}{%
        \begin{tabular}{@{}lccclccccc@{}}
        \toprule
        ID & $K_{\min}$ & EV & $C_{50}$
        & \multicolumn{1}{c}{$\mathcal{K}$, $\mathcal{K}_{\mathrm{EV}}$, $\mathcal{K}_{C_{50}}$ in \S\ref{sec:se:mic-selection:new}}
        & CHiME-6 & DiPCo & Mixer 6 & NOTSOFAR-1 & Macro \\
        \midrule
        (1) & 15 & $\checkmark$ & - 
            & $\mathcal{K} = \mathcal{K}_{\mathrm{EV}}$, $\mathcal{K}_{C_{50}} = \{ 1,\ldots,M \}$
            & 19.64 & 29.47 & 10.94 & 15.10 & 18.79 \\
        (2) & 15 & - & $\checkmark$
            & $\mathcal{K} = \mathcal{K}_{C_{50}}$, $\mathcal{K}_{\mathrm{EV}} = \{ 1,\ldots,M \}$
            & 19.77 & 29.12 & 10.93 & 15.10 & 18.73 \\
        (3) & 15 & $\checkmark$ & $\checkmark$
            & $\mathcal{K} = \mathcal{K}_{\mathrm{EV}} \cap \mathcal{K}_{C_{50}}$
            & 19.34 & 28.98 & 10.93 & 15.09 & 18.59 \\
        \bottomrule
        \end{tabular}%
    }
\end{table}

%\newpage
\subsection{Ablation experiments on the ASR module}
\label{sec:ablation:asr}
\subsubsection{evaluation of ASR models with Oracle diarization}

\begin{table}[tb]
  \caption{tcpWER [\%] ($\downarrow$) on the development set with Oracle diarization and SE frontend.}
  %\vspace{-3mm}
  \label{tab:asr_ablation}
  \centering
  \resizebox{\linewidth}{!}{%
  \begin{tabular}{@{}l@{\hspace{0.2cm}}lc@{\hspace{0.2cm}}c@{\hspace{0.2cm}}c@{\hspace{0.2cm}}cc@{}}
    \toprule
     
    ID & Model & CHiME-6 & DiPCo & Mixer 6 & NOTSOFAR-1 & Macro \\
    \midrule
    ASR0 & NeMo Trans. (Baseline) &  19.78 & 31.01 & 10.61 & 17.95 & 19.84 \\ % Combined system Line 102
    \midrule
    ASR1 & Whisper-L AED & 17.80 & 26.29 &	10.43 & 13.05 & 16.89 \\ % Whisper L: Combined system Line 107
    ASR1'     & \quad w/o Curriculum Learning & 19.30 &  27.20 & 11.39 & 13.19 & 17.77 \\ % Whisper L: Combined system Line 107
    ASR2 & Whisper-M AED  & 19.81 & 27.15 & 11.16 & 13.57 & 17.92 \\ % Whisper M: Combined system Line 106
    ASR3 & NeMo Trans.    & 20.30 & 28.33 & 11.25 & 14.33 & 18.55 \\ % NeMo: Combined system Line 105
    ASR4 & WavLM Trans.   & 19.76 &	27.52 &	10.79 &	13.23 & 17.82 \\ % WavLM: Combined system Line 104
    ASR4'     &\quad w/o SpecAugment  & 19.91 & 27.73 & 10.93 & 13.25 & 17.96 \\ % WavLM: Combined system Line 104
    \midrule
    ASR5 & ROVER (ASR $\times$ 6 +LM resc.) & 16.42	& 23.71	&9.42	&	11.44	& 15.25 \\
    \bottomrule
  \end{tabular}%
  }
  %\vspace{-2mm}
\end{table}

Table \ref{tab:asr_ablation} shows the tcpWERs of our developed ASR backends when using Oracle diarization and our SE frontend. 
The tcpWERs of all systems were significantly improved compared to those of the baseline. Although the best single ASR system is the Whisper Large v3 (ASR1), the WavLM transducer (ASR4), which is the smallest system, achieved a comparable performance. 
Note that we trained two versions of ASR1 and ASR4 with different training settings, ASR1' and ASR4' in Table \ref{tab:asr_ablation}, respectively. %but only reported the best version in Table \ref{tab:asr_ablation}. 
For ASR1, two models were built with and without curriculum learning, as detailed in Section~\ref{sssec:ASR1}. As seen in Table \ref{tab:asr_ablation}, our proposed curriculum learning improved the recognition performance on all datasets.
For ASR4, two models were trained with and without SpecAugment~\cite{specaugment} during the fine-tuning step using only the CHiME-8 dataset. The two models exhibited close tcpWERs, slightly better when using SpecAugment. %but the application of ROVER affected the improvements. 
ASR5 consists of the combination of 12 hypotheses, which are 1-best hypotheses generated by the six ASR systems with and without LM restoring.

\begin{table}[t]
  \caption{Error rates [\%] ($\downarrow$) for each error type across each ASR system and ROVER. We report errors over all four datasets of the development set. The total error rate indicates tcpWER.}
  %\vspace{-3mm}
  \label{tab:asr_error}
  \centering
  \resizebox{1.0\linewidth}{!}{%
\begin{tabular}{@{}llcccc@{}}
\toprule %\hline
ID   & Model         & Insertion & Deletion & Substitution & Total (tcpWER) \\ %\hline
\midrule
%ASR0 & Nemo Trans. & 4.05& 7.04 &9.19 & 20.29 \\
%\midrule
ASR1 & Whisper-L AED & 4.53 &  \textbf{4.38} &  7.98 & 16.89 \\ % 16.17
ASR1'     & \quad w/o Curriculum Learning & 4.45 & 5.27  &8.05 & 17.77\\
ASR2 & Whisper-M AED & 4.49 & 4.96 &  8.48 & 17.92 \\ %16.84
ASR3 & NeMo Trans.   &  3.71 & 7.10 & 7.75 & 18.56 \\  %17.70
ASR4 & WavLM Trans.  &  4.11 & 5.34 & 8.37 & 17.82 \\ %\hline %16.92
ASR4'     &\quad w/o SpecAugment & 4.23 &5.30&	8.43&17.96 \\

\midrule
%ASR5 & ROVER (ASR 1-4)         &  \textbf{3.56} & 4.96 & \textbf{6.70} & \textbf{15.23} \\ %\hline %15.40 
ASR5 & ROVER (ASR $\times$ 6 +LM resc.) &  \textbf{3.61} & 5.00 & \textbf{6.64} & \textbf{15.25} \\ %\hline %15.40 

\bottomrule
\end{tabular}%
}
\end{table}

The ROVER integration strategy (ASR5) in Table~\ref{tab:asr_ablation} significantly improved the tcpWERs. We analyzed the effect of ROVER in Table~\ref{tab:asr_error} by comparing the insertion, deleting, and substitution error rates of the individual systems before and after ROVER. We report macro error rates on the development set.

We observe that the four ASR models exhibit different error trends. ASR1 achieves a lower deletion error rate compared to all other systems. 
This may be due to the curriculum learning scheme, which filters out unreliable utterances that may include errors in GSS preprocessing, where the speaker's speech is removed but the label remains assigned, resulting in the model improving the recognition of more words.
As for insertion and substitution errors, ASR3 and ASR4 perform better in terms of insertions, while ASR3 and ASR1 do better in terms of substitutions.

In most cases, combining these systems leads to lower error rates than individual systems (the only exception being a lower deletion rate with ASR1).
Overall, ROVER increased deletions by 14.2 \%,  but reduced insertions by 20.3 \% and substitutions by 16.8 \% relatively compared to the best single ASR model (ASR1).

\subsubsection{Impact of LM rescoring on performance}

To evaluate the impact of LM rescoring, we modified the hypotheses used in ROVER and conducted a performance comparison. As discussed in Section \ref{sec:lm_rescoring}, we applied LM rescoring to the recognition results of each ASR model and then performed system combination using ROVER with the original ASR hypotheses and the rescored hypotheses.

To demonstrate the effectiveness of this method, we used the development set and compared three scenarios in Table~\ref{tab:lm_rescoring}: (1) using only the original ASR hypotheses in ROVER, (2) using only the LM-rescored hypotheses in ROVER, and (3) combining both the original and LM-rescored hypotheses in ROVER. In the second and third approaches, which utilized LM rescoring results, we observed a slight performance improvement of 0.1–0.2\% across all tasks, indicating the effectiveness of LM rescoring. 
\textcolor{red}{\sout{Looking at individual tasks, using only the rescored hypotheses for ROVER yielded better performance for CHiME-6 and Mixer 6, while combining both hypotheses in ROVER resulted in better performance for DiPCo and NOTSOFAR-1.}}
Based on the macro average, combining both hypotheses achieved the highest performance, which we adopted for the challenge submission.

\begin{table}[tb]
  \caption{Performance comparison of tcpWER [\%] ($\downarrow$) on the development set when changing hypotheses for ROVER.}
  %\vspace{-3mm}
  \label{tab:lm_rescoring}
  \setlength{\tabcolsep}{2mm}
  \centering
  \resizebox{\linewidth}{!}{%
  \begin{tabular}{@{}lccccccc@{}}
    \toprule
     
     ID & Original hypotheses & LM-rescored & CHiME-6 & DiPCo & Mixer 6 & NOTSOFAR-1 & Macro \\
    \midrule
    (1) & \checkmark &  & 23.25	& 29.30	&9.47	&	17.61	& 19.91 \\
    (2) & & \checkmark & \textbf{22.87}	& 29.49	&\textbf{9.22}	&	\textbf{17.55}	& 19.78 \\
    (3) & \checkmark & \checkmark & 22.99	& \textbf{29.18}	&9.26	&	\textbf{17.55}	& \textbf{19.74} \\
    \bottomrule
  \end{tabular}%
  }
  %\vspace{-2mm}
\end{table}

\bibliographystyle{elsarticle-num} 
\bibliography{refs}

\end{document}